\documentclass[11pt]{article}

\pdfoutput=1

\usepackage{rideshare}
\usepackage{soul}
\usepackage{enumitem}
\usepackage{hyperref}
\hypersetup{
    colorlinks=true,
    linkcolor=blue,
    citecolor=black,
}

\newtheorem{theorem}{Theorem}
\newtheorem{claim}{Claim}

\begin{document}

\title{Spatial Distribution of Supply and the Role of Market Thickness: Theory and Evidence from Ride Sharing}
\author{
    Soheil Ghili and 
    Vineet Kumar\thanks{Yale University. We thank Phil  Haile, Igal Hendel, Nicole Immorlica, Vahideh Manshadi, Larry Samuelson, Yannis Stamatopoulos, and K.  Sudhir. We also thank conference and seminar participants at Lyft  Marketplace Labs, The ACM transactions on Economics and Computation (EC), Marketing Science, and Stanford Institute for  Theoretical Economics (SITE) for helpful comments. We are thankful to Phonkrit Tavanisarut for excellent research assistance. Click \href{https://sites.google.com/view/soheil-ghili/spatialdisrtibution?authuser=0}{here}  for the most current version.} 
}

\maketitle

\begin{abstract}
This paper studies the effects of economies of density in transportation markets, focusing on ridesharing. Our theoretical model predicts that (i) economies of density skew the supply of drivers away from less dense regions, (ii) the skew will be more pronounced for smaller platforms, and (iii) rideshare platforms do not find this skew efficient and thus use prices and wages to mitigate (but not eliminate) it. We then develop a general empirical strategy with simple implementation and limited data requirements to test for spatial skew of supply from demand. Applying our method to ride-level, multi-platform data from New York City (NYC), we indeed find evidence for a skew of supply toward busier areas, especially for smaller platforms. We discuss the implications of our analysis for business strategy  (e.g., spatial pricing) and public policy (e.g., consequences of breaking  up or downsizing a rideshare platform).

\end{abstract}

\section{Introduction}\label{section:Introduction}

Spatial markets are complex:  instead of overall volumes of supply and demand, these markets involve \textit{spatial distributions} of supply and demand. This leads to a number of crucial challenges in studying such markets. Theoretically, it is challenging to characterize the equilibrium spatial distribution of supply and its possible mismatch from that of demand because each supplier, when choosing where to supply, considers not only demand forces, but also externalities  from other suppliers. These externalities can lead to complex phenomena such as agglomeration due to economies of density. Additionally, it is  challenging to determine whether the equilibrium spatial distribution of supply is efficient because, when deciding where to supply, each supplier does not consider the externality her decision may leave on the rest of the market. On the empirical side, it is challenging to measure mismatch between the spatial distribution of supply and that of demand due to the widely acknowledged problem that (unfulfilled) demand is unobservable.

This paper studies these theoretical and empirical questions in the context of ridesharing. Our theoretical model shows that (i) economies of density skew the supply of drivers away from ``less dense'' regions, (ii) the skew will be more pronounced for smaller rideshare platforms, and (iii) a platform's optimal pricing will mitigate but not eliminate the skew. Our empirical analysis provides a method with simple implementation, general application, and limited data requirements to detect spatial skew of supply from demand in spite of the fact that unfulfilled demand is unobserved. Using our empirical method, we find evidence for the implications of our theoretical model regarding the role of economies of density in shaping the distribution of supply. With this evidence in hand, our paper then discusses the implications of the analysis for business strategy (e.g., optimal surge pricing) and public policy (e.g., consequences of breaking up or downsizing rideshare platforms).

Our theoretical model in \cref{section:Theory} examines drivers' decision making on which region to operate in among a set of $I\geq 2$ regions in a spatial market with a monopolist rideshare platform.  Each region has an arrival rate of potential demand. Actual demand is a fraction of potential demand, depending on price in region $i$. Deviating from the literature, our model endows each region with a size rather than considering it a point. Each driver chooses a region $i$ that, given other drivers' choices, will maximize her revenue. Revenue in each region is positively related to the wage per ride in that region  and negatively related to the ``total wait time'' each driver has to wait in the region to give a ride to a passenger. Total wait time consists of (i) ``idle time,'' the time it takes for the driver to be assigned to a passenger requesting a ride, and (ii) ``pickup time,'' the time it takes to arrive at the pickup location after being assigned to a passenger. More drivers operating in  each region $i$  means a higher expected idle time in $i$. This forces the supply of  drivers to geographically  distribute itself proportionally to the distribution of demand. On the other hand, more drivers in region  $i$ means a lower expected pickup time in $i$, forcing drivers to agglomerate. The  interplay between these two forces has a key role in our results. We deliver two sets of results. First, we fix the platform strategy on prices and wages and focus on studying driver behavior. Next, we endogenize platform strategy.

Fixing the platform's prices and wages at spatially uniform values across the market, we deliver two main results.  First, if region $i$ has a higher arrival rate of potential demand per unit of size than region $i'$, then at the equilibrium, region $i$ will also get a higher number of rides even after normalizing by its higher potential demand. In other words, \textit{access} to supply will be spatially skewed toward denser regions. As we will discuss in detail, the intuition for this result is that denser regions enjoy shorter pickup times, giving them economies of density and thereby attracting more drivers. Our second result examines the impact of ``thinning'' the market either on one side  only (a decrease in the  total number of drivers) or on both sides (a proportional decrease in demand in each  region and total number of drivers).  We develop an  inductive technique to prove that while each such thinning preserves the demand ratios, it skews the equilibrium supply ratio between \textit{any} two regions toward the higher-demand region. The basic intuition  is  that the supply of drivers responds to  a ``global thinning'' of the market,  which increases pickup times everywhere, by further agglomerating in regions with ``thicker local markets.''

Our next set of results studies the platform's optimal strategy. We show the efficient spatial distribution of supply from the perspective of the platform \textit{does not involve full elimination} of the skew of supply toward denser areas. This is because the platform, too, suffers from its drivers having to undergo long pickup times. We show, however, that the platform would optimally use regional prices and wages as levers to \textit{mitigate} the skew level that would otherwise arise in equilibrium among drivers. This lack of full alignment between the driver-equilibrium and platform-efficient allocations arises from the fact that each driver, by deciding against operating in a less dense region, makes that region even sparser and, hence, less desirable for other drivers. The platform, unlike the driver herself, is impacted by this externality. In addition to these results, we show that (similar to the case of fixed and uniform prices and wages,) under platform optimal strategy, a thinner market will widen the gap between access to supply in higher and lower density regions.

\cref{section:empiricalanalysis} provides our empirical analysis. In this section, we accomplish two tasks. First, we use individual-ride data on rideshare platforms Uber, Lyft, and Via from New York City (NYC henceforth) in order to test the two main implications of our model: with or without platform intervention, we have (i) geographical inequity in percent fulfillment of potential demand in favor of denser regions and (ii) widened inequity for smaller platforms. Second, we use individual-driver panel data from ``Ride Austin,'' a non-profit rideshare platform in the city of Austin, to directly test the reaction of drivers to pickup times, which is the core ingredient of our model.

With regard to our first objective, empirically testing for  spatial skew of supply from demand is challenging because unfulfilled demand is unobserved. If a platform $k$  has  far fewer rides in region $i$ compared to $i'$, it is not clear how much of this is due to lower demand in $i$ and how much is due lower access to supply (i.e., relatively higher prices and/or wait times in $i$). To overcome this, we develop a method with simple implementation and limited data requirements that is applicable to all passenger-transportation markets irrespective of whether they have a centralized matching system (like rideshare) or a decentralized one (like taxicabs). 

Our method is called ``relative outflows analysis'' and is based on a simple idea that, in our view, has been overlooked in the literature on passenger transportation markets: people move their residences less often than they take rides. Hence, for every trip there is a ``trip back'' by the same person shortly after (otherwise, the trip itself is the ``trip back'' for one that must have happened shortly before). Therefore, if platform $k$ has consistently fewer outgoing rides from region $i$ than it does incoming rides--meaning  $k$ has a low \textit{``relative outflow''} in $i$--then the same population that chooses $k$ to enter $i$ from other regions must have, on average, been more likely to have to choose other options over $k$ to exit. We interpret this as a sign that access to supply of $k$ is lower in  $i$ than outside of it, due to $k$'s high price and/or wait time in $i$. We argue that alternative explanations (such as relatively superior availability of public transit, etc. in $i$) may be ruled out if the relative outflow for $k$ is not as low in other regions with similar characteristics to $i$ or if relative outflows for other rideshare platforms $k'$ are not as low in $i$. In summary, our method deviates from the literature by noting that in order to learn about unfulfilled demand in a region, one can leverage data not only on rides starting in that region, but also on rides \textit{ending} there.


We apply our method to data on rideshare platforms Uber, Lyft, and Via from July 2017 to December 2019 in the proper New York City area (NYC). We first show that the relative outflow of a platform's rides in a region (i.e., outgoing rides per each incoming ride) is strongly positively associated with the regional dropoff density, the number of incoming rides per square mile. This  pattern is counterintuitive because incoming rides are in the numerator of one quantity and the denominator of the other. We argue, however, that this pattern is quite in line with our model prediction of supply being skewed toward denser regions. We show it is robust to controlling for region characteristics such as borough fixed effects, zone-type (e.g., commercial, residential) fixed effects, and other fixed effects, even if they are interacted with each other. In addition, we test the second implication of our theory model regarding the effect of market thickness (i.e., platform size). We show that the gap between relative outflows in more busy and less busy areas is wider for smaller platforms. Again, we document that this pattern is significant and robust to a rich set of controls. Based on all of these results, we conclude that economies of density are indeed playing a role in agglomerating drivers in busier areas and that this is more pronounced for smaller platforms.

Although our NYC analysis tests the main implications of our theory, it does not provide direct evidence for the core ingredient of our model that leads to those implications. That ingredient is drivers' sensitivity to pickup times. To conduct this direct analysis, we turn to an individual-ride level dataset on Ride Austin, with driver identifiers, which allows us to observe when a driver turns her/his app off. Taking advantage of the granularity of this dataset in dealing with potential endogeneity problems, we show that pickup times are indeed crucial to drivers' location decisions.

\cref{section:implications} discusses the implications of our work for platform strategy and public policy. On the platform strategy side, a comparison between our results and the literature on pricing in rideshare suggests that the optimal approach to surge pricing is vastly different when the demand surge is an outcome of a short-run shock (e.g., the end of a sports event) or a long-run recurring one (e.g., rush hours). We argue that in the latter case, unlike the former, the platform may benefit from \textit{reducing the prices and driver wages}. Our second implication for platform strategy is that it our model indeed recommends platforms to have in place a ``pickup-time bonus'' rewarding drivers for choosing to operate in outer regions and suburbs of cities. Our model, however, suggests that the platform should pay for part of this bonus and should consider this payment an investment that will eventually pay off by attracting even more drivers through decreasing pickup times in the region. On the public policy side, we argue that even though breaking up or downsizing a rideshare platform can have positive impacts such as, respectively, increasing competition or reducing congestion, there may be unintended consequences. In particular, such policies will further incentivize drivers to cluster in busier areas, thereby disproportionately hurting the outer regions. We also offer a method for estimating a minimum required platform size for a city so as to avoid skew of supply from outer regions. Applying our method to NYC, we estimate the minimum required size to be between 3.29 and 3.64 million rides/month, which is around the size Lyft reached mid 2018.


\cref{section:conclusion} concludes this paper and discusses avenues for further research. 

\section{Literature Review}\label{section:Literature}
Our paper relates to multiple strands of the literature: (i) the recent and growing literature on the empirical analysis of geographical distribution of supply, and its possible distortion from that of demand, in spatial markets; (ii) the literature on transportation markets (in particular ridesharing); and (iii) the literature that studies the effects of market thickness in two-sided markets.

The empirical literature on the spatial match between supply and demand is new and small. To our knowledge, \cite{buchholz2018spatial,brancaccio2019efficiency} are the only papers directly examining this issue, and papers such as \cite{frechette2018frictions,brancaccio2019geography,brancaccio2019guide} look at related problems. They extend the empirical techniques in the matching literature (see \cite{petrongolo2001looking} for a survey) in order to structurally infer the size of unobserved demand (e.g., passengers searching for rides) in different locations  of  a decentralized-matching market when only the size of supply (e.g., available drivers) and the number of demand-supply matches (e.g., realized rides) are observed. They accomplish this by inverting a matching function that gives the number of rides as a function of searches and vacancies. Our relative-outflows method is complementary. On the one hand, it does not estimate the absolute volume of unfulfilled demand (e.g., failed searches) in each region and, rather, focuses on inequity across regions in percent fulfillment of potential demand. On the other hand, our method (i) is reduced form and easy to implement; (ii) it requires data only on the number of rides rather than rides \textit{and} vacant supply, search time, etc.; (iii) it applies generally to all passenger-transportation markets regardless of whether the matching system is centralized (e.g., rideshare) or decentralized (e.g., taxicabs);\footnote{Another subset of the literature  on  spatial  markets that this paper builds on  is the study of location decisions, resulting in agglomeration. Papers such as \cite{ellison1997geographic,ahlfeldt2015economics,datta2011agglomeration,holmes2011diffusion,miyauchi2018matching} examine agglomeration of firms or residents. We add to this literature by arguing, empirically and theoretically, that agglomeration is also  present  in  transportation markets. In addition, our comparative static theory results, which characterize how the extent of agglomeration is impacted by different factors, may be applied beyond transportation  systems.} and, finally, (iv) our approach detects skew of supply away from a given region $i$ even if in response to short supply,  passengers in $i$ have learned to forego searching (which would make it look like demand is low).

The second strand of the literature to which our paper relates is the set of papers on the functioning of transportation (in particular rideshare) markets. This strand itself can be roughly divided into (at least) two categories. One category is the group of papers focusing on this market as it relates to labor economics.\footnote{For instance, \cite{chen2017value} examine how much workers benefit from the schedule flexibility offered by ridesharing. \cite{cramer2016disruptive} study the extent to which ridesharing, compared to the traditional taxicab system, reduces the portion of time drivers are working but not driving a passenger. \cite{chen2016dynamic} examine the reaction of labor supply to the introduction of ridesharing. \cite{buchholz2018dynamic} estimate an optimal stopping point model to study the labor supply in the taxi-cab industry.} The second category, to which our paper belongs, consists of papers focusing on evaluating the performance of these markets and on market design aspects. Some of those papers, although related to our work in many ways, focus on questions that are inherently not spatial (examples are \cite{levitt2016using,nikzad2018thickness,lian2019optimal,cachon2017role,guda2019your,asadpour2019minimum}). Others study questions that are related to the spatial nature of the market (such as \cite{weyl2017surge,frechette2018frictions}), but they do not examine the spatial distribution of supply and potential mismatches with demand. Many of the papers that do study geographical supply-demand (im)balance in transportation (such as \cite{banerjee2018value,afeche2018ride,castro2018surge}) focus on the short-run, intra-day, aspects. Some other  papers (such as \cite{buchholz2018spatial,lagos2000alternative,lagos2003analysis,bimpikis2016spatial,shapiro2018density,lam2017demand,garg2019driver,ata2019spatial,he2020customer,rosaia2020competing}), however, examine such spatial markets from a long-term perspective. Our paper is complementary to this literature in that it provides a detailed theoretical and empirical investigation of economies of density and market thickness, while abstracting away from some of the phenomena considered in these papers.\footnote{\cite{rosaia2020competing} does examine economies of density as part of his analysis but his analysis is mainly empirical. Additionally, he focuses on part of the time frame of the NYC market in which Uber and Lyft were both large enough and, hence, economies of density turned out to not a significant factor in market equilibrium.}

It is worth noting that a large part of this literature has focused on the ways in which ride-share platforms improve upon the traditional taxi system, in particular due to their flexible pricing and superior matching algorithms (\cite{cramer2016disruptive,buchholz2018spatial,frechette2018frictions,levitt2016using,shapiro2018density,weyl2017surge,castro2018surge,lam2017demand} among others). We  add to this  literature by comparing ride-share platforms to one another. We ask why is it that some rideshare platforms outperform others on  some key issues, such as geographical reach, even though they all have superior technology relative to more  traditional  transportation systems? We conclude that a matching algorithm is not sufficient and that other factors (i.e., adequate platform size) may be needed to ensure geographical reach. This enables  our paper to quantitatively comment on the current policy debate regarding the appropriate sizes of rideshare platforms in NYC and other markets.

The third set of papers to which we relate is a large, mostly theoretical, literature on the impact of market thickness on the functioning of two-sided platforms in general (such as \cite{akbarpour2017thickness,ashlagi2019matching}) and transportation markets in particular (such as \cite{frechette2018frictions,nikzad2018thickness}). This literature, to our knowledge, has not examined how the spatial distribution of supply--and its (mis)alignment with  that of potential demand--responds to a change in market thickness. Our paper focuses on this, both empirically and theoretically.

\section{Theoretical Model}\label{section:Theory}
We develop a model of how the spatial distribution of the supply of a rideshare platform is impacted by density and market thickness. We prove two sets of results. First, we assume prices and wages set by the platform are fixed and uniform across regions so that we can focus on  driver behavior. We show that  the equilibrium spatial distribution of supply is skewed away from that  of demand towards higher density regions. We also show the skew is more intensified for smaller rideshare platforms. Next, we examine the platform's objective and optimal behavior. We first argue that the platform does indeed benefit from some  skew in the geographical distribution of supply. Nevertheless, we show that due to network externalities, the platform's optimal distribution of drivers across regions is less skewed towards denser areas compared to the distribution that arises in the equilibrium among drivers. We then provide multiple results on how the platform should use prices and wages as levers in order to mitigate (but not fully eliminate) the skew in the distribution of supply. Additionally, we also show that the relationship between platform size and skew in the geographical distribution of supply still holds even under platform optimal behavior. That is, the smaller the platform, the more skewed the distribution of supply towards denser areas. The rest of this section is organized as follows. \cref{subsec:setup} sets up the model. \cref{subsec:Drivers} presents our results on driver equilibrium behavior, fixing the platform's strategy. \cref{subsec:Platform} analyzes platform's optimal strategy and how it relates to platform size.

\subsection{Setup}\label{subsec:setup}
We model a market with regions $i\in\{1,...,I\}$ and a monopolist ridesharing platform serving them.
The regions  (which, depending on the  application, one could think of as neighborhoods, boroughs, etc.) are  modeled as circumferences of circles, a la Salop. The circumference of circle $i$ is denoted $t'_i$. The price of a ride in region $i$ is denoted $p_i$. The wage paid to a driver for giving a ride to a passenger in region $i$ is denoted $c_i$. 

In each region, passengers arrive at a rate $\lambda_i(p_i)$ per unit of time. Demand arrival rate is a function of price $p_i$ and take the following form: $$\lambda_i(p_i)=\bar{\lambda}_i f(p_i)$$ where $\bar{\lambda}_i$ is the ``potential demand'' in region $i$ and function $f(\cdot)$ --which is assumed uniform across regions-- captures the fraction of $\bar{\lambda}_i$ that would be willing to pay $p_i$ for a ride. We assume that $f(0)=1$ and that $f(\cdot)$ is decreasing. In some of our results, we assume a functional form for $f(\cdot)$.

Without loss of generality,  we assume that the density of demand in each region is decreasing in its index: $\forall i<j:\, \frac{\bar{\lambda}_i}{t'_i} \geq \frac{\bar{\lambda}_j}{t'_j}$. Also, $\bar{\lambda}$ represents the vector $(\bar{\lambda}_1,...,\bar{\lambda}_I)$. Each arriving passenger's location is uniformly distributed on the circumference of the circle. There is a total mass of $N>0$ drivers who work for the platform. In the first part of our analysis, we treat $N$ as fixed. Later, we endogenize $N$. An allocation of drivers is denoted by vector $n=(n_1,...,n_I)$ such that $\Sigma_{i=1,...,I}n_i=N$. In each region, there is a wait time for drivers before they can provide a ride to a passenger. This total wait time is denoted $W_i(n_i)$ and is a function of $n_i$, the number of drivers present in the region (this notation suppresses the implicit dependence of $W_i$ on prices). Total wait time in each region has two components: idle time and pickup time. Idle time is the time it takes a driver to get assigned to a ride request by a passenger. Idle time in region $i$ is increasing in $n_i$. That is, the more drivers in region $i$, the longer it takes for each of them to get assigned to a passenger. Pickup time is the time it takes a driver, after being assigned to a ride request, to drive and arrive at the passenger's pickup location. Pickup time in region $i$ is decreasing in $n_i$. This is because the more drivers in region $i$, the more densely the region is populated with them. Therefore each driver becomes less likely to be asked to pick up a passenger who is far away in the region. We assume the total wait time for each region $i$ is given by:

\begin{equation}\label{eq:wait}
 \underbrace{W_i(n_i)}_{\text{Total Wait Time}} = \underbrace{\frac{n_i}{\lambda_i(p_i)}}_{\text{Idle Time}} +  \underbrace{ \frac{t'_i}{4n_i} }_{\text{Pickup Time}} = \left(\frac{n_i}{\lambda_i(p_i)} + \frac{t_i}{n_i}\right) 
\end{equation}
where $t_i=\frac{t'_i}{4}$. In the appendix, we provide a micro-foundation for this functional form. From this point on, we refer to $t_i$ (instead of $t'_i$) as the size of region $i$.

\begin{figure}\label{fig: Illustration of W}
\centering
\caption{\scriptsize{Wait time as a function of the number of present drivers, given by \cref{eq:wait}. This is illustrated in a region with $\lambda_i(p_i)=10$ and $t_i=2$. The dashed line is $\frac{n}{\lambda}$, depicting how long the wait time would be if pickup time were zero. }}
\label{fig:circularcity}
\centering
\vspace{-0.02in}
\includegraphics[scale=.6]{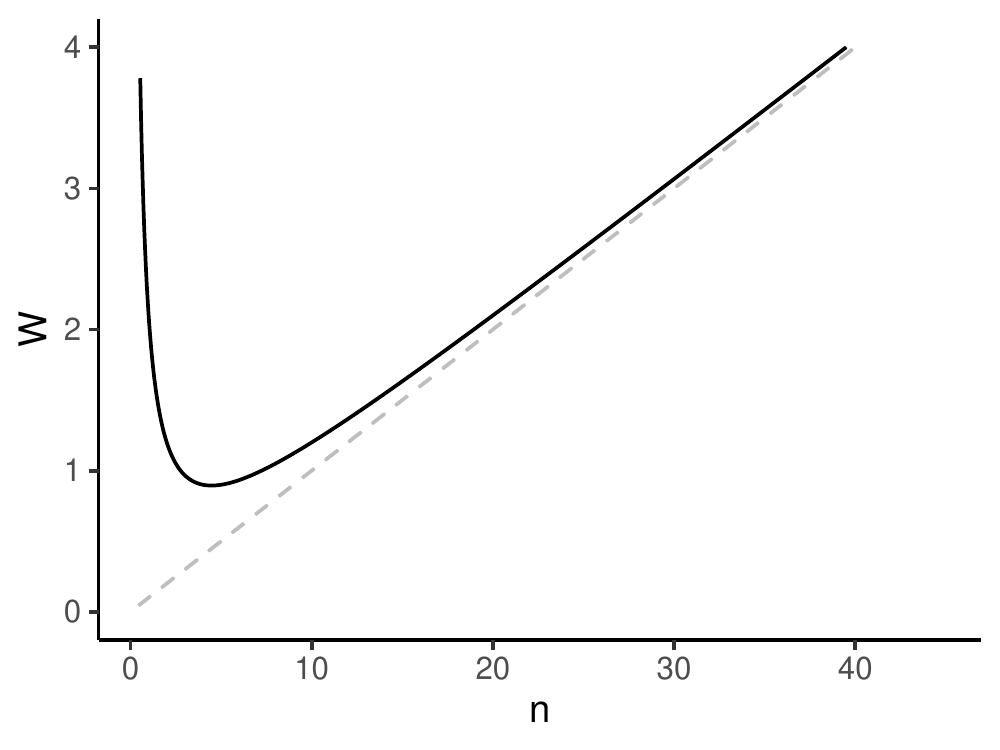}
\end{figure}

The wait time curve given by \cref{eq:wait} is illustrated in \cref{fig: Illustration of W}. As can be seen there, $W_i(\cdot)$  is initially decreasing in $n_i$ because the effect of pickup time is dominant. When $n_i$ is large enough, pickup time becomes less important and $W_i(\cdot)$ becomes increasing in $n_i$ due to the effect of idle time. 

The core of our model is a simultaneous-move game among the drivers in which each one of them chooses one of the $I$ regions to operate in. Each driver seeks to maximize her expected hourly revenue. The hourly revenue in each region $i$ equals the wage per ride in that region multiplied by the frequency of rides given by each driver in the region.   That is, the revenue will be $\frac{c_i}{W_i(n_i)}$.\footnote{Note that this formulation abstracts away from the time it takes to drive a passenger to the dropoff location. This assumption simplifies some of our analysis and we do not expect the results to be sensitive to it.} 

The total number rides given per hour in region $i$, denoted $r_i(n_i)$, is given by the total number of drivers in that region divided by the time each driver has to wait before giving a ride: $r_i(n_i)\equiv\frac{n_i}{W_i(n_i)}$.  We denote ``Access'' to rides in region $i$ by $A_i(n_i)$ and define it as the fraction of the potential demand $\bar{\lambda}_i$ that leads to rides. That is:

\begin{equation}\label{eq:Accecss Plain}
  A_i(n_i)\equiv \frac{r_i(n_i)}{\bar{\lambda}_i}  
\end{equation}

Note that $A_i(n_i)\leq 1$. To see this, observe that:

\begin{equation}\label{eq:Accecss Expanded}
  A(n_i)\equiv \frac{\lambda(p_i)}{\bar{\lambda}_i}\times \frac{r_i(n_i)}{\lambda(p_i)}\equiv \frac{\lambda(p_i)}{\bar{\lambda}_i}\times \frac{1}{1+\frac{t_i \lambda(p_i)}{n_i^2}}
\end{equation}

Both $\frac{\lambda(p_i)}{\bar{\lambda}_i}$ and $\frac{1}{1+\frac{t_i \lambda(p_i)}{n_i^2}}$ are weakly less than 1. These terms show that access to rides could be limited by high price and/or sparsity of drivers in the region. The term $\frac{\lambda(p_i)}{\bar{\lambda}_i}$ captures the effect of price, while $\frac{1}{1+\frac{t_i \lambda(p_i)}{n_i^2}}$ captures the effect of driver sparsity. This latter term will be close to one if the pickup time is low because the region is very small or gets many drivers.\footnote{This term is also close to one if demand is so slow that drivers are able to fully serve it in spite of their non-trivial pickup times.}   One could think of this term as aggregating, in a reduced form, both potential ride requests that are not sent out due to high wait times for passengers and ride requests that are sent out but are accepted by no driver.

We now turn to defining equilibria:

\begin{definition}\label{def_equilibrium}
Under ``market primitives'' $(\bar{\lambda},N,t)$, an allocation $n^*=(n_1^*,...,n_I^*)$ of drivers among the $I$ regions is called an equilibrium if (i) $\Sigma_{i=1,...,I}n^*_i=N$, and (ii) no driver in any location $i$ can strictly increase her total revenue by choosing to drive in another location. Also, we call   $n^*$ an ``all-regions'' equilibrium allocation if it is an equilibrium and if $n_i^*>0$ for all $i$.
\end{definition} 

With this definition in hand, we next turn to our results.

\subsection{Analysis of Driver Behavior}\label{subsec:Drivers}
In this section, we assume that all wages and prices are uniform and fixed. That is, $\exists c,p$ such that   $\forall i:\, c_i=c$ and $p_i=p$. As such, in this section, we suppress the notation on $p_i$ and denote $\lambda_i(p_i)$ simply as $\lambda_i$. Additionally, when $c_i$ is uniform, then the \textit{revenue-maximization objective for drivers boils down to wait-time minimization}. In this section, the ``primitives'' of the market will be $(\lambda,N, t)$.

We provide three main results in this section. Our first result describes the equilibria of the game among drivers. Our second result shows that at the equilibrium, the spatial distribution of supply is skewed toward denser regions due to economies of density. The third result examines how the supply distribution responds to a change in ``market thickness.''

\begin{proposition}\label{prop: driver behavior EQ description}
The following statements are true about the equilibria of the game among drivers.

\begin{enumerate}
    \item An equilibrium $n^*$ always exists. Also, for any subset $J$ of $I$ there is at most one equilibrium $n^*$ under which $n^*_i>0\Leftrightarrow i\in J$.
    \item At any equilibrium $n^*$ such that $n^*_i>0\Leftrightarrow i\in J\subset I$, the total wait time (hence the revenue) is equal across regions: $\forall i,j\in J:\, W_i(n^*_i)=W_j(n^*_j)$.
    \item Suppose ${n}^{*'}$ and $n^*$ are two equilibria such that the set of regions served by $n^*$ (i.e., the set $\{i:n^*_i>0\}$) is a proper subset of the set of regions served by $n^{*'}$. Then driver total wait time is strictly lower (i.e., driver revenue is higher) under $n^{*'}$.
\end{enumerate}
\end{proposition}

The proposition is proven in the appendix. Note that this proposition not only describes the features of the equilibria, but also provides a partial means for equilibrium selection. Part 3, roughly, says that ``larger'' equilibria are more profitable for drivers. In particular, if an ``all-regions'' equilibrium (i.e., $n^*$ with $n^*_i>0$ for all $i$) exists, it is the most desirable equilibrium for \textit{all} drivers.

\begin{proposition}\label{prop: driver behavior econ of density}
Suppose $n^*$ is an equilibrium allocation of drivers under market primitives $(\lambda,N, t)$. Then the following are true:

\begin{enumerate}
    \item For all regions that get positive supply, supply ratios are skewed towards denser areas: $$\forall i<j:\, \frac{n^*_i}{\lambda_i}\geq \frac{n^*_j}{\lambda_j}$$ with equality only if $\frac{\lambda_i}{t_i}=\frac{\lambda_j}{t_j}$.
    \item The same result holds  on access to rides across regions: $$\forall i<j: A_i(n^*_i)\geq A_j(n^*_j) $$ with equality only if $\frac{\lambda_i}{t_i}=\frac{\lambda_j}{t_j}$.
\end{enumerate}
\end{proposition}

  To illustrate, this proposition states that if region $i$ has twice as much demand per unit of size as region $j$, in the equilibrium it may get, say, three times as much supply per unit of size. The basic intuition for why this result is true is the role of pickup times. To see this, consider an allocation $n$ which distributes drivers across regions proportionally to demand arrival rates, meaning $\forall i,j:\, \frac{n_i}{\lambda_i}=\frac{n_j}{\lambda_j}$. It is easy to show that, under such an allocation, lower density areas have higher total wait times. To see this, note that all regions will have the same idle time $\frac{n_i}{\lambda_i}$. However, for any pair of regions $i,j$ with higher demand density at $i$ (meaning $\frac{\lambda_i}{t_i}>\frac{\lambda_j}{t_j}$), the pickup time at $j$ is strictly larger. This is because $\frac{t_i}{n_i}=\frac{t_i}{\lambda_i}\times \frac{\lambda_i}{n_i}$. But $\frac{t_i}{\lambda_i}<\frac{t_j}{\lambda_j}$ and $\frac{\lambda_i}{n_j}=\frac{\lambda_i}{n_j}$ which, together, yield $\frac{t_i}{n_i}<\frac{t_j}{nj}$. Therefore, if we start from the proportional allocation, there will be an incentive for drivers to relocate from less dense areas to denser ones. This, of course, is only the intuition behind the result. The formal proof has multiple extra steps and is provided in the appendix. 
  
  We now turn to studying how the equilibrium spatial distribution of the drivers responds to a change in the platform size (i.e., market thickness). Before that, we define a change in market thickness.
  
  \begin{definition}\label{def:MarketThickness}
  Consider a market with primitives $(\lambda,N,t)$. We call a market with primitives $(\gamma \lambda,\gamma N, t)$ with $\gamma>1$ a ``two-sided thickening'' of $(\lambda,N,t)$. Additionally, $(\lambda,\gamma N, t)$ is a ``one-sided'' thickening of  $(\lambda,N,t)$.
  
  \end{definition}

Intuitively, a two-sided thickening increases both the total number of drivers and the demand arrival rate in each area by the same factor $\gamma>1$. A one-sided thickening only increases the total number of drivers. It is crucial to note that both of these changes \textit{preserve the demand ratios between any two regions}. Nevertheless, as our next result shows, making a market thinner will \textit{skew the supply ratio between any two regions towards the denser one.} 

\begin{proposition}\label{prop:MarketThickness_fixedN}
Suppose $n^*$ is an equilibrium allocation of drivers under market primitives $(\lambda,N,t)$. Also assume $(\lambda',N',t')$ is a one- or two-sided thickening of $(\lambda,N,t)$. Then there exists an equilibrium allocation $n^{*'}$  under $(\lambda',N',t')$, which satisfies the following:

\begin{enumerate}
    \item $\{i:\, n^*_i>0\}= \{i:\, n^{*'}_i>0\}$.
    \item For all $i,j$ with $n^*_i>0$ $$\forall i<j:\, \frac{A_j(n^*_j)}{A_i(n^*_i)}\leq \frac{A_j(n^{*'}_j)}{A_i(n^{*'}_i)} \leq 1$$ 
    
    where both inequalities are strict when $\frac{\lambda_i}{t_i}>\frac{\lambda_j}{t_j}$.

    \item There will be equitable access to rides as the market gets sufficiently thick: $$\forall i,j:\, \lim_{\gamma\rightarrow \infty} \frac{A_j(n^{*'}_j)}{A_i(n^{*'}_i)}=1$$

\end{enumerate}

\end{proposition}

The underlying intuition for the result is that as the market gets thicker (i.e., the platform gets larger,) \textit{all regions get denser} with drivers. As such, the importance of pickup times relative to idle times decreases in drivers' decision making, leading to a supply distribution that is more balanced with demand. The proof of this proposition is also given in the appendix.  The main technique used to carry out the proof is strong induction in the number of regions. A crucial part of the proof in the induction is to show that, when the market gets thicker, so does any ``sub-market'' consisting of any arbitrary subset of all the $I$ regions (this will be necessary for the induction step). That is, as the global market gets thicker, the distribution of drivers does shift towards less dense areas \textit{but not so much as to make some of the denser ``sub markets'' thinner relative to before the global increase in market thickness.} For details, see appendix.

To sum up, in this subsection we show that, in order to avoid longer pickup times, drivers tend to disproportionately locate in regions with higher demand densities. We also show that this may lead to some regions not being served at all. Additionally, we prove that this supply-demand imbalance dwindles as the platform grows. All of these, however, were analyzed under the assumption that the platform chooses uniform and fixed wages across the regions. The next natural question is, what will happen when the platform optimally uses prices and wages as levers? Does the platform's optimal strategy involve ``going along'' with the supply-demand distribution mismatch? Or does it involve some corrections? Next subsection is dedicated to the analysis of these questions.

\subsection{Analysis of Optimal Platform Strategy}\label{subsec:Platform}
In this subsection, we analyze the platform's optimal strategy regarding prices and/or wages. Formally, we assume prices and/or wages are set to maximize the platform profit per hour which is given by: $$\pi(p,w)=\Sigma_{i=1,...,I}(p_i-c_i)\times r_i$$

where $r_i$ is the number of rides per hour given in region $i$ and is a function of wages and prices among other things.

Before studying the platform's optimal strategy, we give a result about the difference between the equilibrium distribution of drivers and the platform-optimal distribution of drivers under uniform wages and prices.

\begin{proposition}\label{prop:Platform vs EQ}
Suppose that allocation $n^*$ is an equilibrium under market primitives $(\lambda,N,t)$. Then, for any two regions $i,j$ with $\frac{\lambda_i}{t_i}>\frac{\lambda_j}{t_j}$, if the platform were to optimally reallocate the $n^*_i+n^*_j$ drivers between the two regions, it would choose $n^{**}_i$ and $n^{**}_j$ (subject to $n^{**}_i+n^{**}_j=n^*_i+n^*_j$) such that: $$\frac{A_j(n^*_j)}{A_i(n^*_i)}<\frac{A_j(n^{**}_j)}{A_i(n^{**}_i)}<1.$$

\end{proposition}

That is, the platform would desire some inequity in access across regions but not as much as the equilibrium allocation among drivers naturally gives rise to. The intuition for this result is that the platform dislikes its drivers having to do long pickups. Therefore, it also prefers some level of geographical ``imbalance'' between supply and demand. However, the platform internalizes the externalities that drivers leave on each other when deciding where to locate. These externalities come mainly from the fact that when, at the equilibrium, a driver chooses a dense region $i$ over  the less dense $j$, she makes $j$ even sparser which increases the pickup in $j$. Of course her joining $i$ does slightly decrease the pickup time in $i$ but the effect on $i$ is not as large compared to $j$, given the diminishing sensitivity of pickup times to the number of drivers present in a region. All of this impacts other drivers and, hence, the platform. 

\cref{prop:Platform vs EQ} shows the platform would modify the distribution of drivers if it could do so at no cost. The next natural question is whether the platform would optimally take potentially costly actions of  changing local prices and wages  in order to smooth out the distribution of supply across the regions of a market. The rest of this subsection is devoted to this analysis. The key required change in the model is to assume that regional prices $p_i$ and wages $c_i$ are not uniform and exogenous anymore, but are, rather,  determined by the platform optimally. In addition, endogenizing the wages will require us to endogenize the total number of drivers $N$. We assume the ``equilibrium'' number of drivers will be such that each driver's earning per hour is equal to an exogenously given reservation value $\bar{c}$. A flexibile $N$ means the new market primitives will now be $(\bar{\lambda},t)$ instead of $(\bar{\lambda},N,t)$. As such, making the market ``thicker'' when $N$ is flexible can only take place in one form: scaling up the demand and going from primitives $(\bar{\lambda},t)$ to $(\gamma\bar{\lambda},t)$ where $\gamma>1$. In other words, the one-sided versus two-sided distinction between ways of making the market thicker no longer exists.

The game in this section of the paper has, therefore, two stages. First, the platform optimally decides all regional prices and/or wages. In the second stage, potential drivers from an infinitely large pool simultaneously decide whether to enter the market and, if so, which region to operate in. As such, an equilibrium of the game will involve a platform strategy as well as a driver distribution.

We now turn to the formal analysis of platform optimal strategy. We provide three main results. We first endogenize wages, while keeping prices fixed and uniform. Next, we endogenize prices while keeping wages fixed and uniform. Finally, we endogenize both wages and prices. In each of these three results, we speak both to the characterization of the equilibrium spatial distribution of supply, and to how this distribution changes in response to a changed market thickness.

\begin{proposition}\label{prop:endogenous Wages}
Suppose that prices are fixed at $p_i=p$ for all $i$ and that market primitives are $(\bar{\lambda},t)$. Also suppose the pair of vectors $(c^*,n^*)$ is an equilibrium.  Then, the following are true: 
\begin{enumerate}
    \item $n^*$ is unique. Also $c^*_i$ for any $i$ with $n^*_i>0$ is unique. In other words, if $(c'^{*},n'^{*})$ is another equilibrium, we have $n^*=n'^{*}$, and for any $i$ with $n^*_i>0$, we have $c^*_i=c'^{*}_i$.
    \item Regions that are not supplied are those with lowest demand densities: $$\exists \mu:\,\text{s.t.}\, \forall i:\, \frac{\bar{\lambda}_i}{t_i}<\mu \Leftrightarrow n^*_i=0$$
    \item For any two regions $i,j$ with $n^*_i\neq 0\neq n^*_j$, we have: $$\frac{\bar{\lambda}_i}{t_i}\geq \frac{\bar{\lambda}_j}{t_j} \Rightarrow A_i(n^*_i)\geq A_j(n^*_j)$$
    where the latter comparison holds with equality only if the first one does.
    \item For any regions $i,j$ with $n^*_i\neq 0\neq n^*_j$, we have:
    $$\frac{\bar{\lambda}_i}{t_i}\geq \frac{\bar{\lambda}_j}{t_j} \Rightarrow c^*_i\leq c^*_j$$
    where the latter comparison holds with equality only if the first one does.
\end{enumerate}

In addition, consider a thickening of the market from $(\bar{\lambda},t)$ to $(\gamma\bar{\lambda},t)$ where $\gamma>1$. Suppose that $(c^{*'},n^{*'})$ is an equilibrium under the new primitives $(\gamma\bar{\lambda},t)$. Then, the following are true:

\begin{enumerate}
    \item For any $i$: $n^*_i>0\Rightarrow n^{*'}_i>0$.
    \item For any $i,j$: $$\frac{\bar{\lambda}_i}{t_i}\geq \frac{\bar{\lambda}_j}{t_j} \Rightarrow \frac{A_j(n^{*}_j)}{A_i(n^{*}_i)}\leq
    \frac{A_j(n^{*'}_j)}{A_i(n^{*'}_i)}\leq 1$$
    and the latter two inequalities hold strictly only if the first one does.
    \item There will be equitable access to rides as the market gets sufficiently thick: $$\forall i,j:\, \lim_{\gamma\rightarrow \infty} \frac{A_j(n^{*'}_j)}{A_i(n^{*'}_i)}=1$$
\end{enumerate}

\end{proposition}

This proposition has three main messages. First, if wages are flexible, the platform's optimal strategy will involve wage incentives for drivers to operate in areas with lower densities of potential demand. This is, again, in line with the intuition that even though platforms, like drivers, find long pickup times undesirable, they would still like to intervene and make the distribution of drivers across regions less skewed toward busier areas. In other words, the platform will optimally try to ``build economies of density'' in sparser regions. The second message is that in spite of the platform's intervention, the equilibrium will still exhibit lower access to supply in less dense areas compared to denser ones. The third message is that, similar to the case of exogenous and uniform wages, here too an increase in market thickness will lead to a more balanced supply.

In the first glance, the result in \cref{prop:endogenous Wages} may seem at odds with previously established results in the literature that the optimal response to high demand in a region is a wage increase in that region in order to encourage drivers to relocate to the said region and meet the demand \citep{castro2018surge}. Note, however, that the result in \cite{castro2018surge} has to do with a \textit{short term} local demand shock which could only be met if drivers in other regions are incentivized  to incur the costs of relocation. Our model, however, is complementary in that it captures steady-state distribution of supply in the market and how it is impacted by economies of density, abstracting away from short run shocks. As such, driver location choice in our model should be thought of as a driver's general strategy for where to drive on a regular basis rather than a (costly) relocation from a region to another. Therefore, our result and the result in \cite{castro2018surge} are not inconsistent with each other. They are, rather, complements to each other, each shedding light on a different aspect of spatial pricing in the market.

Our next proposition keeps wages fixed and allows the platform to optimally decide the regional prices. For this proposition, we make a functional form assumption on function $f(\cdot)$ in equation $\lambda_i(p_i)=\bar{\lambda}f(p_i)$. For this proposition and the next one, we assume $f(p_i)\equiv 1-\alpha p_i$ with $\alpha>0$.

\begin{proposition}\label{prop:endogenous Prices}
Suppose that wages are fixed at $c_i=c$ for all $i$ and that market primitives are $(\bar{\lambda},t)$. Also suppose the pair of vectors $(p^*,n^*)$ is an equilibrium.  Then, the following are true: 
\begin{enumerate}
    \item $n^*$ is unique. Also $p^*_i$ for any $i$ with $n^*_i>0$ is unique. In other words, if $(p'^{*},n'^{*})$ is another equilibrium, we have $n^*=n'^{*}$, and for any $i$ with $n^*_i>0$, we have $p^*_i=p'^{*}_i$.
    \item Regions that are not supplied are those with lowest demand densities: $$\exists \mu:\,\text{s.t.}\, \forall i:\, \frac{\bar{\lambda}_i}{t_i}<\mu \Leftrightarrow n^*_i=0$$
    \item For any regions $i,j$ with $n^*_i\neq 0\neq n^*_j$, we have:
    $$\frac{\bar{\lambda}_i}{t_i}\geq \frac{\bar{\lambda}_j}{t_j} \Rightarrow p^*_i\geq p^*_j$$
    where the latter comparison holds with equality only if the first one does.
\end{enumerate}

In addition, consider a thickening of the market from $(\bar{\lambda},t)$ to $(\gamma\bar{\lambda},t)$ where $\gamma>1$. Suppose that $(p^{*'},n^{*'})$ is an equilibrium under the new primitives $(\gamma\bar{\lambda},t)$. Then, the following are true:

\begin{enumerate}
    \item For any $i$: $n^*_i>0\Rightarrow n^{*'}_i>0$.
    \item There will be equitable access to rides as the market gets sufficiently thick: $$\forall i,j:\, \lim_{\gamma\rightarrow \infty} \frac{A_j(n^{*'}_j)}{A_i(n^{*'}_i)}=1$$
\end{enumerate}

\end{proposition}

Broadly, \cref{prop:endogenous Prices} has similar messages to those of \cref{prop:endogenous Wages}. There are, however, two additional points worth noting about \cref{prop:endogenous Prices}. First, unlike \cref{prop:endogenous Wages}, \cref{prop:endogenous Prices} does \textit{not} claim that as the market gets thicker, access to rides becomes more balanced across regions. One could construct a counter-example for such a claim in the case of fixed wages and endegenous prices. One can show that the comparative static result does hold if the platform is large enough (i.e., if $\gamma$ is large enough).\footnote{We skip the provision of a counter-example as well as the proof for large $\gamma$ (both  would be available upon request). Instead, we provide some intuition for why the result does not always hold for small $\gamma$ values. To see the intuition, consider a region $i$ that is just dense enough to attract non-zero supply under the platform's optimal pricing strategy and with a fixed wage $c$. Now consider the effect of an increase in the region's potential demand arrival rate from $\bar{\lambda}_i$ to  $\gamma\bar{\lambda}_i$ with $\gamma>1$. In response to this demand increase, the platform can decide either (i) to do very little price increase and enjoy the extra volume of demand which also helps with attracting more drivers, or (ii) to increase the price substantially and focus on the margin instead of the volume. If, under the original $\bar{\lambda}_i$, the price elasticity is low at $p^*_i$, then the platform will choose the latter strategy in response to a scale-up to $\gamma\bar{\lambda}_i$. This could lead to a decrease in access to rides. If the fixed wage $c$ is small enough, indeed the market can get formed at a $p^*_i$ low enough so that it induces a low price elasticity. We have confirmed this intuition using numerical simulations of the market.} 

The second, and more crucial, point about \cref{prop:endogenous Prices} is how it should be understood as it relates to surge pricing. \cref{prop:endogenous Prices} suggests that prices should be higher regions with higher densities of potential demand. This outcome may seem in-line with what one would naturally expect in this spatial market and with results already established in the literature \citep{castro2018surge}. Nevertheless, our result holds \textit{only because of the network externalities that arise from pickup times.} That is, regions with higher demand density can attract more drivers which leads to lower pickup times which, in turn, further helps sustain the number of present drivers. Given the diminishing sensitivity of pickup times to the number of drivers in the region, the effect of a price increase in dense regions will be restricted to a decrease in demand. However, in regions with lower demand densities and, hence, longer pickup times, the demand decrease arising from a price increase will have a more substantial adverse consequences through network effects: with every driver leaving the region as a result of the lower demand, the pickup time for the remaining drivers increases, further encouraging drivers to leave. As a consequence, a permanent price increase is a safer action for the platform in denser regions compared to less dense ones. This mechanism is different from \cite{castro2018surge} who focus on short-run demand shocks. In fact, in our model, if we abstract away from pickup time frictions (by setting vector $t$ to zero), the optimal price will be uniform across regions even though some regions have higher demand densities than others.

Next, we study the platform's optimal strategy when both wages and prices are flexible. Propositions \ref{prop:endogenous Wages} and \ref{prop:endogenous Prices}  suggest that there are multiple economic forces governing the platform's optimal behavior, and that these forces may pull the optimal strategy in different directions.  To see this, consider the following scenarios. First suppose the platform, in line with \cref{prop:endogenous Wages}, offers higher wages in less dense regions. In this case, when we make the prices also flexible, the platform would have some incentive to also increase the prices in less dense areas because the platform has a higher marginal cost (due to driver wages) in those areas. As a second scenario, suppose the platform, in line with \cref{prop:endogenous Prices}, has decided to offer lower prices in less dense areas. In this case, when wages also become flexible, there is some incentive for the platform to offer lower wages in less dense areas, passing at least part of the regional price cut on to drivers. Third, and last, the platform could adhere to the results from both \cref{prop:endogenous Wages} and \cref{prop:endogenous Prices} and offer both lower prices and higher wages in less dense regions. It is not obvious which one of these scenarios (or what combination of them) would prevail when prices and wages are both flexible. Put differently, the platform's incentive to build economies of density pulls the prices down and wages up in sparser areas. But lower prices themselves push wages down; and, likewise, higher wages themselves push prices up. The overall outcome of the interaction among these multiple forces is not clear. Our next result speaks to this question:

\begin{proposition}\label{prop:endogenous Wages and Prices}
Suppose prices and wages are both flexible and that market primitives are $(\bar{\lambda},t)$. Also suppose the triple $(c^*,p^*,n^*)$ is an equilibrium.  Then, the following are true: 
\begin{enumerate}
    \item $n^*$ is unique. Also $c^*_i$ and $p^*_i$ for any $i$ with $n^*_i>0$ are unique. In other words, if $(c'^{*},p'^*,n'^{*})$ is another equilibrium, we have $n^*=n'^{*}$, and for any $i$ with $n^*_i>0$, we have $c^*_i=c'^{*}_i$ and $p^*_i=p'^{*}_i$.
    \item Regions that are not supplied are those with lowest demand densities: $$\exists \mu:\,\text{s.t.}\, \forall i:\, \frac{\bar{\lambda}_i}{t_i}<\mu \Leftrightarrow n^*_i=0$$
    \item For any two regions $i,j$ with $n^*_i\neq 0\neq n^*_j$, we have: $$\frac{\bar{\lambda}_i}{t_i}\geq \frac{\bar{\lambda}_j}{t_j} \Rightarrow A_i(n^*_i)\geq A_j(n^*_j)$$
    where the latter comparison holds with equality only if the first one does.
    \item For any regions $i,j$ with $n^*_i\neq 0\neq n^*_j$, we have:
    $$\frac{\bar{\lambda}_i}{t_i}\geq \frac{\bar{\lambda}_j}{t_j} \Rightarrow  \begin{cases} c^*_i\leq c^*_j  \\ 
    p^*_i\leq p^*_j  \\
     p^*_j-c^*_j \leq p^*_i-c^*_i \end{cases} $$
    where the latter three comparisons hold with equality only if the first one does.
\end{enumerate}

In addition, consider a thickening of the market from $(\bar{\lambda},t)$ to $(\gamma\bar{\lambda},t)$ where $\gamma>1$. Suppose that $(c^{*'},p^{*'},n^{*'})$ is an equilibrium under the new primitives $(\gamma\bar{\lambda},t)$. Then, the following are true:

\begin{enumerate}
    \item For any $i$: $n^*_i>0\Rightarrow n^{*'}_i>0$.
    \item For any $i,j$: $$\frac{\bar{\lambda}_i}{t_i}\geq \frac{\bar{\lambda}_j}{t_j} \Rightarrow \frac{A_j(n^{*}_j)}{A_i(n^{*}_i)}\leq
    \frac{A_j(n^{*'}_j)}{A_i(n^{*'}_i)}\leq 1$$
    \item There will be equitable access to rides as the market gets sufficiently thick: $$\forall i,j:\, \lim_{\gamma\rightarrow \infty} \frac{A_j(n^{*'}_j)}{A_i(n^{*'}_i)}=1$$
\end{enumerate}

\end{proposition}

\cref{prop:endogenous Wages and Prices} has similar messages to the previous two results. It shows that (i) the platform uses prices and wages to make the distribution of supply more even, that (ii) the platform's optimal strategy involves only mitigating (rather than eliminating) the uneven distribution of supply at the equilibrium, and that (iii) the unevenness will be more pronounced for smaller platforms.

In addition to the above messages, \cref{prop:endogenous Wages and Prices} also resolves the ambiguity regarding the overall effect of the aforementioned opposing economic forces on regional prices and wages. \cref{prop:endogenous Wages and Prices} says that it is the higher wages implied by \cref{prop:endogenous Wages} for lower density areas that pushes the prices up in those regions (rather than the lower prices implied by \cref{prop:endogenous Prices} pushing the regional wages down). One possible intuition for this result is the more direct effect of wages (compared to prices) on establishing a network of drivers in a region. A lower price $p_i$ in region $i$ reduces the idle time in the region but does not directly change the pickup time (it will impact the pickup time only in equilibrium once more drivers are attracted to the region). A higher wage $c_i$, however, acts as if both the pickup time and the idle time have been reduced.  

\subsection{Inter-region rides and testable implications}

An abstraction in our model so far was that it assumes rides take place only within regions. This subsection does two things. First, it shows that similar versions of propositions \ref{prop:Platform vs EQ} through \ref{prop:endogenous Wages and Prices} hold when we allow some of the rides to be cross-regional. Second, it derives testable implications of our analysis based on ``relative outflows" of rides for different groups of regions. We start by the basic notations and definitions.

Let $\bar{\lambda}_{ij}$ denote the potential demand for rides from $i$ to $j$. By construction, we have $\bar{\lambda}_i=\Sigma_j \bar{\lambda}_{ij}$. Define realized demand $\lambda_{ij}(p_i)$ in a similar way: $\lambda_{ij}(p_i)=\bar{\lambda}_{ij}f(p_i)$. Define $r_{ij}$ as the realized number of hourly rides from $i$ to $j$. Assume that when access to rides $A_i$ is less than one, the proportion of unfulfilled demand does not depend on the destination. That is: $\forall j:\frac{r_{ij}}{\bar{\lambda}_{ij}}=\frac{r_{i}}{\bar{\lambda}_{i}}=A_i$. This assumption has also been made in \cite{bimpikis2016spatial}.

Realized $r_{ij}$ values lead to ``internal flows'' of drivers among regions. By internal, we mean flows of drivers who are already in the market. In addition to that, we now also allow for an external (negative or positive) net flow of drivers $p_i$ out of each region $i$. In this new environment, we need a new notion of equilibrium which is given by \cref{def: stable EQ}.

\begin{definition}\label{def: stable EQ}
The pair of vectors $(n^*,\rho^*)$ is an equilibrium if it satisfies the following two conditions:

\begin{enumerate}
    \item \textbf{Steady State:} $\forall i: \rho^*_i+\Sigma_j r^*_{ij}=\Sigma_j r^*_{ji}$
    \item \textbf{Optimality:} No small perturbation in the rates $\rho^*$ should be able to strictly improve driver revenue in any region $i$.
\end{enumerate}
\end{definition}

We now show that the results provided in the previous section (i.e., without inter-region rides) all hold with inter-region rides as well. \cref{prop: inter-region equivalence} formalizes this.

\begin{proposition}\label{prop: inter-region equivalence}
Suppose market primitives $(\bar{\lambda},t,\bar{c})$ and platform strategy $(c,p)$ are given. Then for any $(n^*,\rho^*)$ with $n^*>>0$, the pair $(n^*,\rho^*)$ is a driver equilibrium according to Definition \ref{def: stable EQ} if and only if $n^*$ is  a driver equilibrium  in the no-inter-region-ride version of the problem.
\end{proposition}

The proof of this proposition is straightforward and is provided in the appendix. This result shows that as long as we focus on all-region equilibria, all of our results from Section \ref{subsec:Platform} should hold. The intuition is simple. If there are too many incoming rides into a region $i$ with overall low payoff, that will lead to many drivers opting to exit the market and pursue an outside option upon dropping a passenger in $i$. Similarly, attractive payoff in a region $i$ with low inflow of rides will garner local driver entry. Note that this does \textit{not} mean drivers' entry/exit decisions are directly influenced by cross-region flows of rides. Rather, cross-region flows of rides influence the regional payoffs which in turn impact entry/exit flows.

Next, we turn to providing testable implications of our model. In the context of a two-region version of our model, we show that even though potential demand values $\bar{\lambda}$ are unobservable, one may still infer something about how access to rides compares across different regions by looking at the flows of realized rides. To this end, define the ``relative outflow'' of rides in each region $i$ as $RO_i=\frac{r^\rightarrow_i}{r^\leftarrow_i}$ where $r^\rightarrow_i$ is the realized number of rides exiting $i$, meaning $r^\rightarrow_i=\Sigma_{j\neq i}r_{ij}$ and $r^\leftarrow_i$ is the realized number of rides entering $i$.


\begin{proposition}\label{prop: testable implications 1}
Suppose $N=2$ and $(n,\rho)$ is a  steady-state (not necessarily equilibrium) driver allocation under $(c,p,\bar{\lambda},t)$. Also suppose potential demands for rides are ``balanced:''  $\bar{\lambda}_{12}=\bar{\lambda}_{21}$. Then we have: $$\frac{A_1(n)}{A_2(n)}=RO_1$$
\end{proposition}

In other words, the ratio between access to rides in the two regions can be measured without observing $\bar{\lambda}$ values, and just by looking at the ``relative outflow'' of rides. Under its balancedness assumption, and combined with our previous propositions,  this result predicts that regions with higher access to rides will have more outflows of realized rides than inflows.

\begin{proposition}\label{prop: testable implications 2}
Suppose $N=2$, and $(n,\rho)$ is a steady state allocation under $(c,p,\bar{\lambda},t)$. Also suppose $(n',\rho')$ is a stable allocation under possibly different potential demands and platform strategies but in the same market $(c',p',\bar{\lambda}',t)$. Assume neither of the two potential demands $\bar{\lambda}$ or $\bar{\lambda}'$ has to be balanced but they are ``similarly unbalanced'': $\frac{\bar{\lambda}_{12}}{\bar{\lambda}_{21}}=\frac{\bar{\lambda}'_{12}}{\bar{\lambda}'_{21}}$. Then we have: $$\frac{A_1/A_2}{A'_1/A'_2}=\frac{RO_1}{RO'_1}$$

\end{proposition}

In this result, the two different primitives $(c,p,\bar{\lambda},t)$ and  $(c',p',\bar{\lambda}',t)$  may represent two different platforms in the same city or the same platform in the same city in two different times. This result says that even without a balancedness assumption, as long as potential demands are ``similarly unbalanced,'' one can still use ride flows for comparing unobservable access levels. Combined with our previous propositions, this result predicts that as a platform gets larger, its relative outflow in busier areas decreases.



\section{Empirical Analysis}\label{section:empiricalanalysis}
In this section, we provide empirical evidence for our theory. We do this in two parts. First, \cref{subsec: NYC} analyzes the model prediction on relative outflows using data on Uber, Lyft, and Via from New York City. Next, \cref{subsec: Austin} provides direct evidence for the role of pickup times on driver location decisions using data from Austin.

\subsection{Empirical Analysis of Rideshare in NYC}\label{subsec: NYC}
The purpose of this section is to empirically test the relevance and validity of the model through testing some of its main implications. More specifically, we would like to test (i) whether access to supply decreases as density of potential demand decreases and (ii) whether the gap between access to supply in higher and lower density areas is wider for smaller platforms. These are two implications of the theoretical model which were robust to whether the rideshare platform is deliberately intervening to change the spatial distribution of supply. We see these tests as tests of whether our model and insights and recommendations are empirically relevant.

Direct empirical tests of spatial inequity in access to rides are impossible. This is because in any region  $i$, access to rides $A_i$ is defined as $\frac{r_i}{\bar{\lambda}_i}$, the fraction of potential demand $\bar{\lambda}_i$ that end up with rides. The number of rides $r_i$ is observable in some datasets but potential demand $\bar{\lambda}_i$ is unobservable even by rideshare platforms themselves. The closest thing to $\bar{\lambda}_i$ on which data is available (usually only to rideshare platforms) is the number of people who opened the rideshare app regardless of whether they ended up with a ride. Note, however, that this does not provide a reliable measure of $\bar{\lambda}_i$ given that, in the long run, those who need rides in region $i$ may respond to persistently high prices and/or wait times in the region by \textit{not going on the app in the first place}. In other words, in the long run, only the portion of $\bar{\lambda}_i$ that expects to find a ride may go on the app. Therefore, $r_i$ divided by the total frequency of app sessions in region $i$ may be fairly homogeneous even if there is substantial heterogeneity in $\frac{r_i}{\bar{\lambda}_i}$ across regions $i$. This unobservability of $\bar{\lambda}_i$ makes a direct test of heterogeneity in $\frac{r_i}{\bar{\lambda}_i}$ challenging to carry out.

Our solution to this challenge is to devise an indirect tests based on Propositions \ref{prop: testable implications 1} and \ref{prop: testable implications 2} of our theoretical analysis. To this end, we leverage a feature of passenger transportation markets which (i) we believe is a powerful tool in empirically identifying spatial supply-demand mismatches, and which (ii) has been overlooked in the literature. In passenger transportation markets, it is reasonable to assume that \textit{for every trip, there is a ``trip back'' by the same person.} Roughly, this suggests that if passengers consistently use platform $k$ less often to exit region $i$ than they do to enter it, then it means \textit{the same population} that chooses platform $k$ over other options to enter $i$ is systematically more likely to end up having to choose other options over platform $k$ to exit.\footnote{Crucially, this gap between the inflow and the outflow is not impacted by whether passengers try and fail to find outgoing rides or they have learned to not even go on the app to search.} Under assumptions that we will be more specific about later in this section, and after controlling for some possible confounds, we conclude from such a data pattern that access to (inter-region) rides with platform $k$ is lower in region $i$ than outside of it.\footnote{Note that our theoretical model does not capture flows of rides across regions. Consequently, it does not capture relative outflows directly. Nevertheless, we do not see this as a weakness of the model. Our theory is focused on how to \textit{explain} the impacts of economies of density and market thickness on spatial distribution of supply, whereas our empirical analysis will be focused on how to \textit{identify} those impacts. Relative outflows have a key role in the identification; therefore they are at the heart of our empirical analysis. They do not, however, have a crucial role in the underlying mechanism. Therefore, our theoretical model abstracts away from them. Put in simpler terms, relative outflow patterns are informative consequences of economies of density rather than key antecedents of it.}

Based on the strategy above, we devise regression analyses to test both whether there is geographical heterogeneity in access to rideshare across regions and whether this heterogeneity is exacerbated when the platform is smaller (i.e., the market is thinner). Before getting to the regressions, however, we discuss our data and provide several visualizations that convey the main intuition behind our subsequent empirical analysis.

\subsubsection{Data and Summary Statistics}\label{subsec:Data}
We have multiple sources of data. For our main analysis, we use ride-level data on all rides with three  platforms Uber,  Lyft, and Via within the New York City proper area from July   2017 to December 2019.  Uber, is the largest platform whose total number of rides given in NYC per month grew from 8.7M in the beginning of our dataset to 15.6M. Lyft is the second largest platform with the highest growth rate. Its size, over the same time period, grew from 2.2M rides/month to 5.2M rides/month in NYC. Via is the smallest platform with a size that oscillated around 0.9M rides/month over the course of our data.

\begin{figure}
    \centering
    \caption{\scriptsize{Population densities of the five boroughs  of NYC as of April 2019 in thousands/sq mile. The color scale is logged.} }
    \label{figure:popDensities}
    \includegraphics[scale=.4]{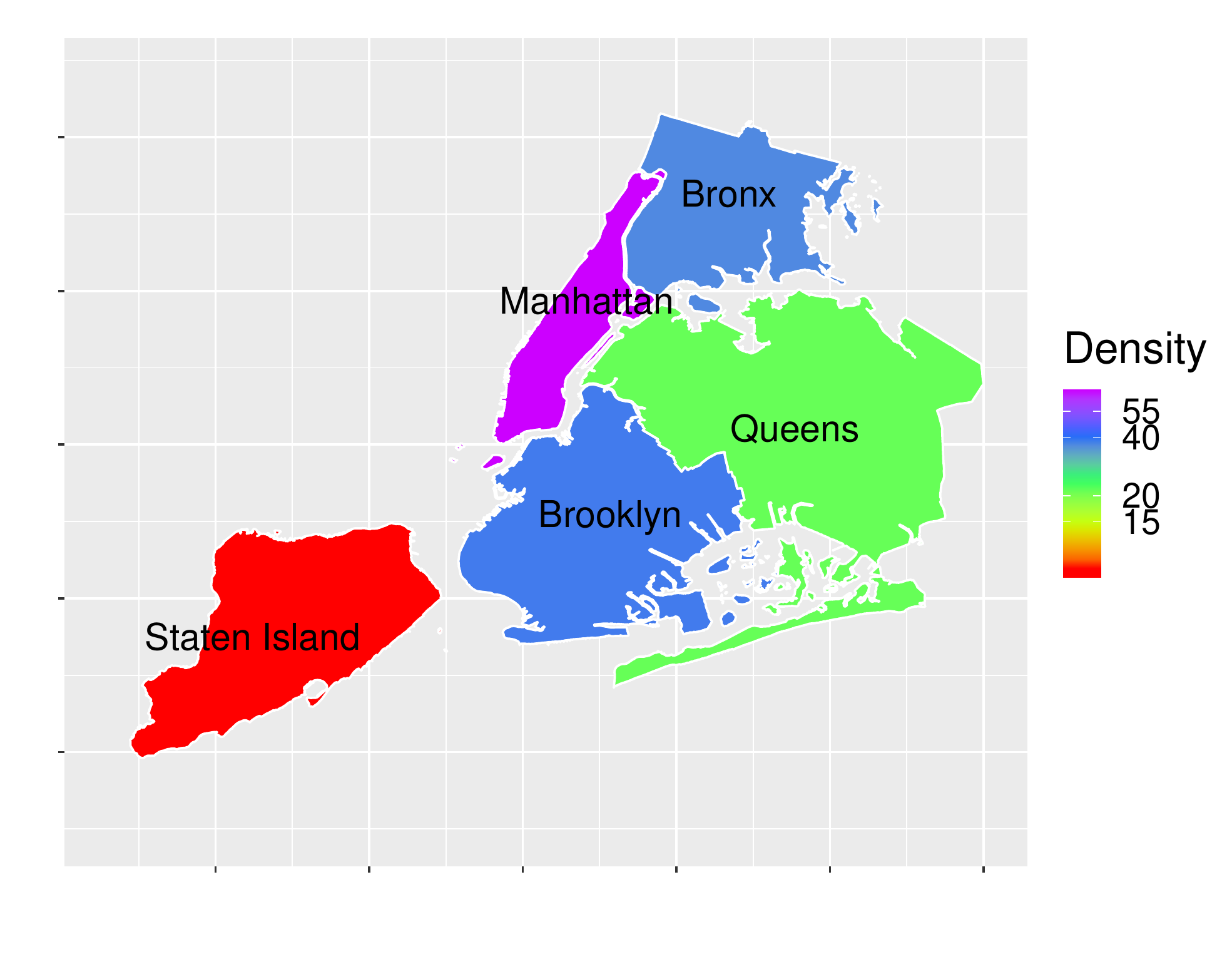}
\end{figure}

For each ride from these three platforms, we observe the exact date, time, and location  both for the pickup and the dropoff. We augment. The ``location'' in our  data is one of 263 ``taxi zones'' that partition NYC. Each such zone belongs to one of the five ``boroughs'' of NYC: Manhattan, Brooklyn, The Bronx, Queens, and Staten Island. These boroughs and their population densities are shown in \cref{figure:popDensities} . In addition, we leverage a zoning districts data (provided by NYC Planning Labs) which contains information on the ``zoning type'' for each taxi zone. The possible types are: residential, commercial, park, and manufacturing. We use these data to provide empirical tests of our theory.

There are a number of reasons why we chose NYC as the setting for our empirical analysis. First, it is the only city from which data on \textit{both pickup and dropoff locations} is available from \textit{multiple rideshare platforms.} Both of these features play key roles in our analysis. Second, NYC is one of the densest cities in the world. Therefore, if frictions from low density impact the spatial distribution of supply in NYC, it is reasonable to conclude they must be relevant in other markets too. Third, the NYC government provides a number of additional datasets which are useful for our analysis (such as data on taxicabs and data on districting of the city into residential, commercial, manufacturing, and park zones).

Before getting to the main empirical analysis, we visualize some data patterns that speak to the tests we will be carrying out in the subsequent sections.

\begin{figure}
\centering
 \caption{\scriptsize{Relative outflows for Lyft, Uber, and Via. Panel (a) is July 2017 and Panel (b) is  July 2018. The figure depicts two important patterns: (i) relative outflow is highest in Manhattan and decreases as we move towards the outer areas; (ii) the gap between relative outflows of Manhattan and outer boroughs is wider for smaller platforms. Both patterns are in line with our theory model. See appendix for the same figure at the borough level.
 \textbf{Note:} Via was not operating in Bronx and Staten Island during July 2017.}}\label{figure:relOutflows1718_zones}
\subfigure[]{\includegraphics[width=0.48\textwidth]{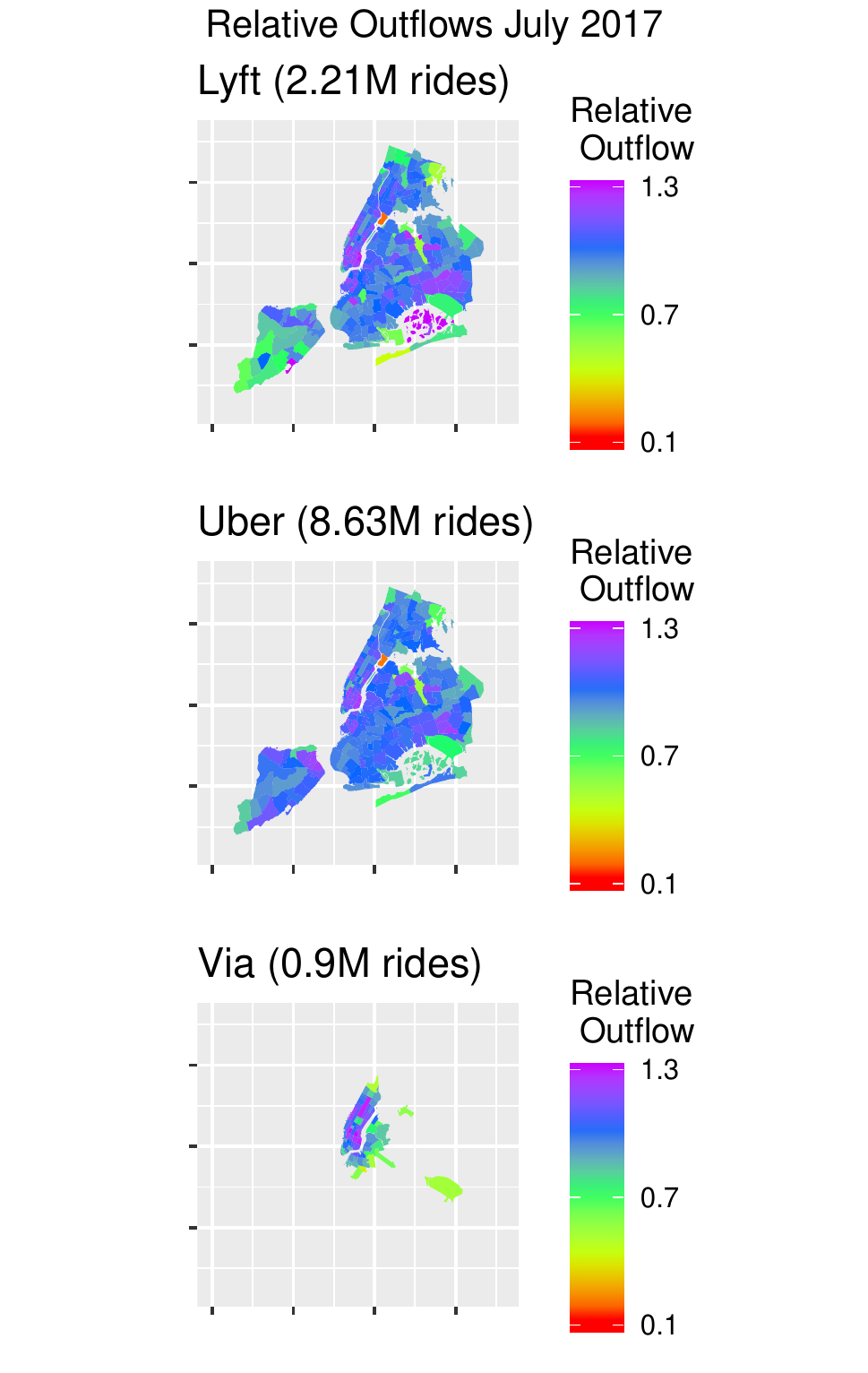}} 
\subfigure[]{\includegraphics[width=0.48\textwidth]{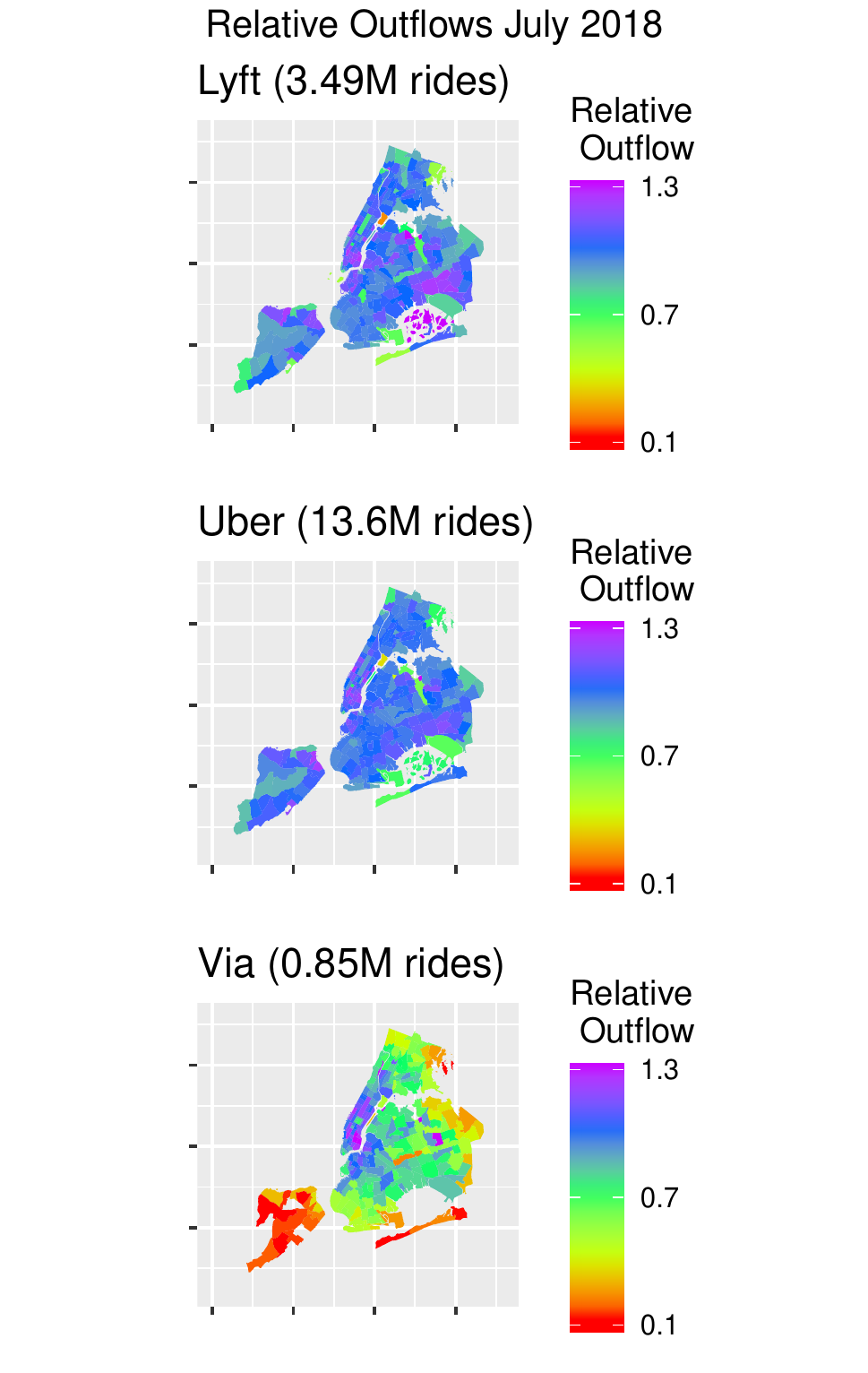}} 
\end{figure}

\cref{figure:relOutflows1718_zones} represent the ``relative outflows'' across different taxi zones in NYC for Uber, Lyft, and Via. Panel (a) corresponds to July 2017 and Panel (b) depicts one year later, July 2018. The relative outflow for any platform $k$ in region $i$ during some period $d$ is defined by the number of rides with $k$ that exit $i$ during period $d$ divided by the number of rides with $k$ entering $i$ during the same period.  Two patterns are noteworthy in these figures. First, the relative outflow tends to be higher in busier areas of the city, and it decreases as we move towards the outer, less busy regions. Second, the gap between the relative outflows of busier and less busy areas is wider for smaller platforms than it is for larger ones (\cref{figure:relOutflows1718_zones} visualizes this by showing that the heat maps of relative outflows for smaller platforms are more ``colorful''). For instance, the relative outflow for Lyft in Staten Island is was close to 60\% during July 2017, suggesting that out of every 100 passengers who chose Lyft over other options to enter that regions, close to 40 had to use other options to leave. The same gap between Staten Island and Manhattan, however, is not observed for Uber during the same period. Neither is it observed for Lyft in July 2018 when Lyft was a larger platform. A much wider such gap is observed for Via which was much smaller in size than both Uber and Lyft. See appendix for \cref{figure:relOutflows1718_boroughs} which is similar to \cref{figure:relOutflows1718_zones} but shows the relative outflows at the borough level instead of the taxi zone level. There we also discuss some of the less intuitive patterns (such as higher relative outflows for some zones in Staten Island and Queens).

In subsequent sections we will formally argue that that the observed relative-outflow gap among regions has to do with economies of density. We do so by controlling for a variety of possible confounds. Before getting to that, more formal, analysis, however, we show additional data patterns in this section that are informative about the nature of the observed gap between the relative outflows in different areas. 

\begin{figure}
\centering
\caption{\scriptsize{Hourly flows in absolute and relative terms during July 2017 for Lyft from  three zones zones. The hourly relative-outflows pattern in Alphabet City Manhattan (residential) is similar to that of Riverdale in Bronx (residential) and different from Greenwhich Village in Manhattan (commercial). Nevertheless, the overall relative outflow in Alphabet City is high, unlike Riverdale but  similar to (even slightly higher than) Greenwhich Village. }}\label{figure:hourlyRelOutflowsNeighborhoods}
\subfigure[]{\includegraphics[width=0.32\textwidth]{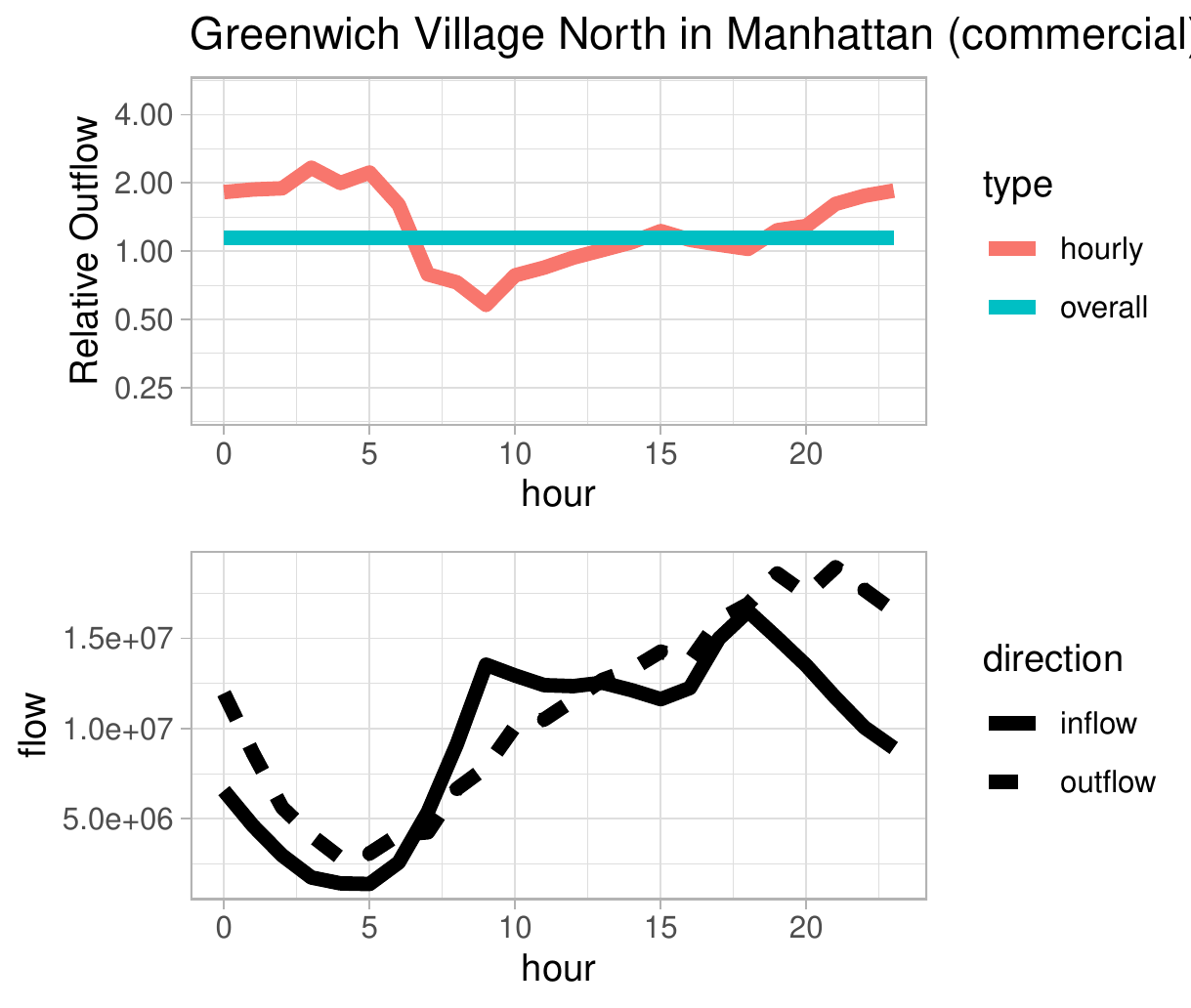}} 
\subfigure[]{\includegraphics[width=0.32\textwidth]{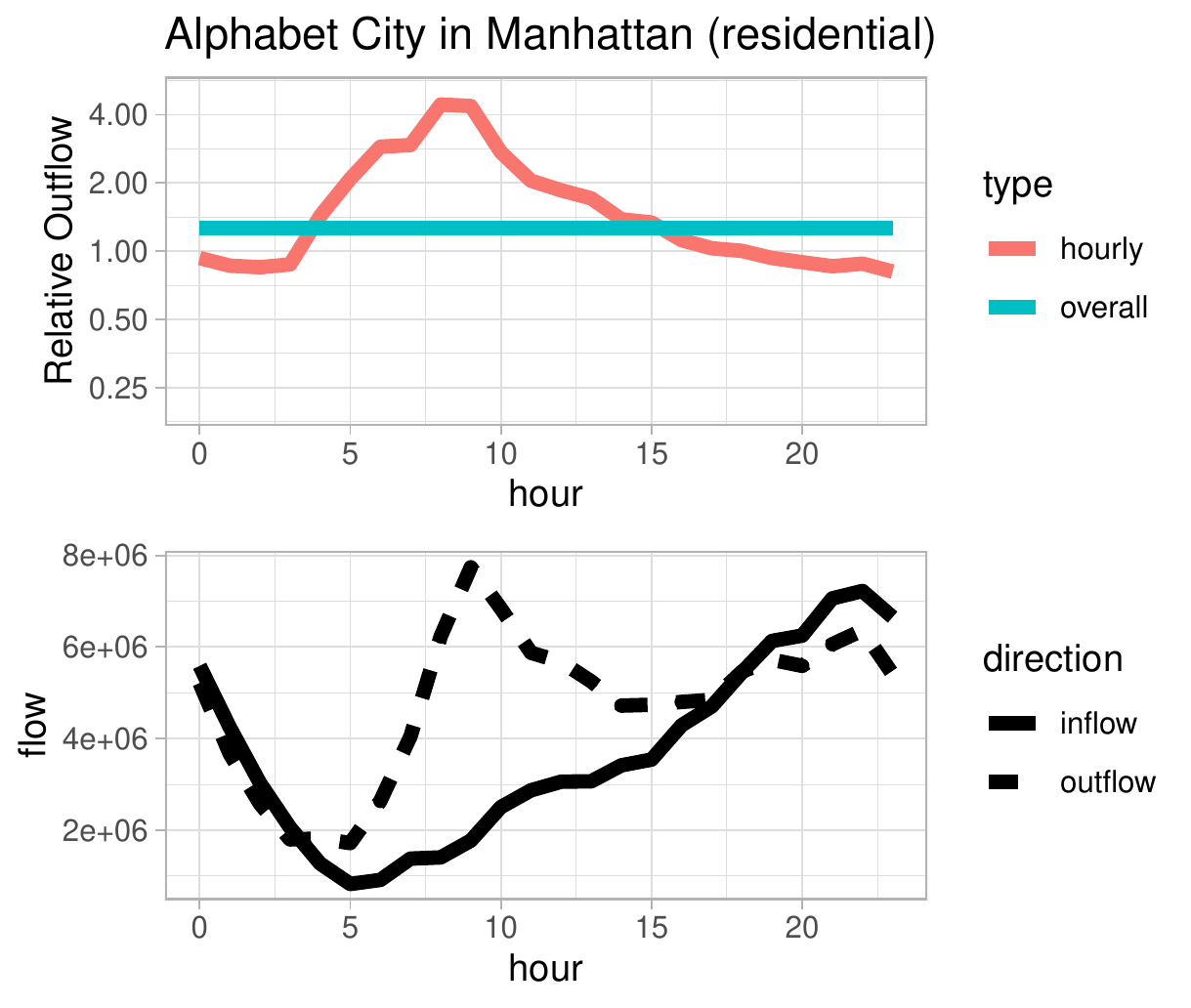}} 
\subfigure[]{\includegraphics[width=0.32\textwidth]{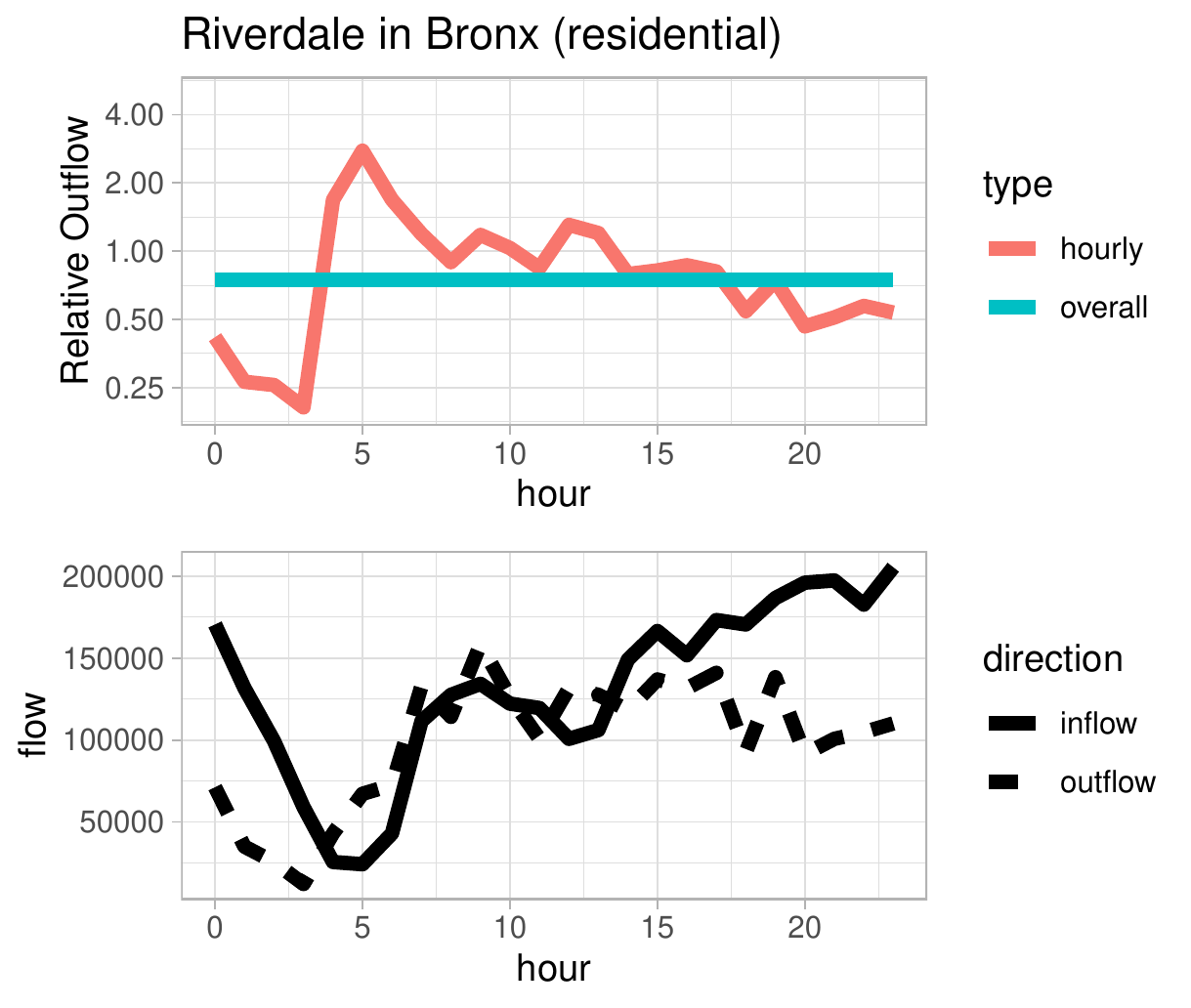}} 
\end{figure}

\cref{figure:hourlyRelOutflowsNeighborhoods} depicts the hourly patterns of relative outflows during July 2017 for Lyft from three different zones. Panel (b) is Alphabet City, a residential area in lower Manhattan. Similar to other residential areas (such as Riverdale in the Bronx, depicted in panel (c)), the relative outflow in Alphabet City peaks in the morning and then gradually decreases. This is the opposite of the pattern observed for a commercial area like Greenwhich Village North, also in Manhattan. In spite of having a similar hourly pattern to Riverdale, however, Alphabet City has a high relative outflow, much more similar to Greenwhich Village North than to Riverdale. \cref{figure:hourlyRelOutflowsNeighborhoods}, hence, is suggestive that the overall high relative outflow in Manhattan is not merely an outcome of the concentration of commercial areas in Manhattan. If anything, we will show later, that the relative outflow in Manhattan is high \textit{in spite of} (rather than because of) its commercial areas.

\begin{figure}
\centering
\caption{\scriptsize{Cross-platform hourly comparisons of relative outflows in Staten Island. During July 2017 --panel (a)-- Lyft's relative outflow is consistently smaller than Uber's. During July 2018 (when Lyft had grown substantially in size) the two platforms' relative outflows are very close to each other throughout the day --panel (b). Via (which started operating in Staten Island in 2018,) on the other hand, has a consistently smaller relative outflow compared to Uber and Lyft. This figure suggests the cross platform differences in relative outflows are unlikely to be an artifact of a spike that takes place during a certain, small, time window. }}\label{fig: hourly RO Staten Island}
\subfigure[]{\includegraphics[width=0.44\textwidth]{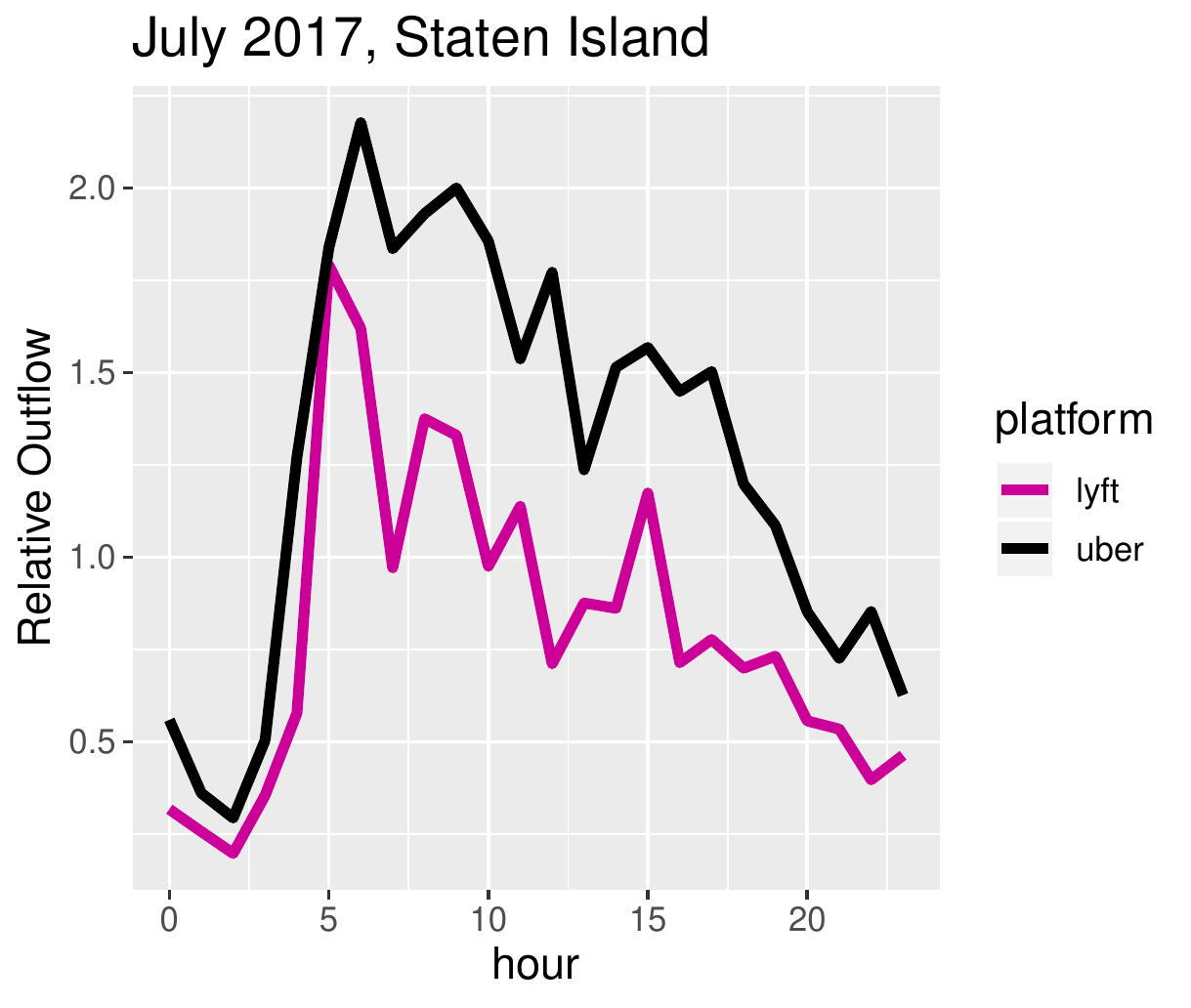}} 
\subfigure[]{\includegraphics[width=0.44\textwidth]{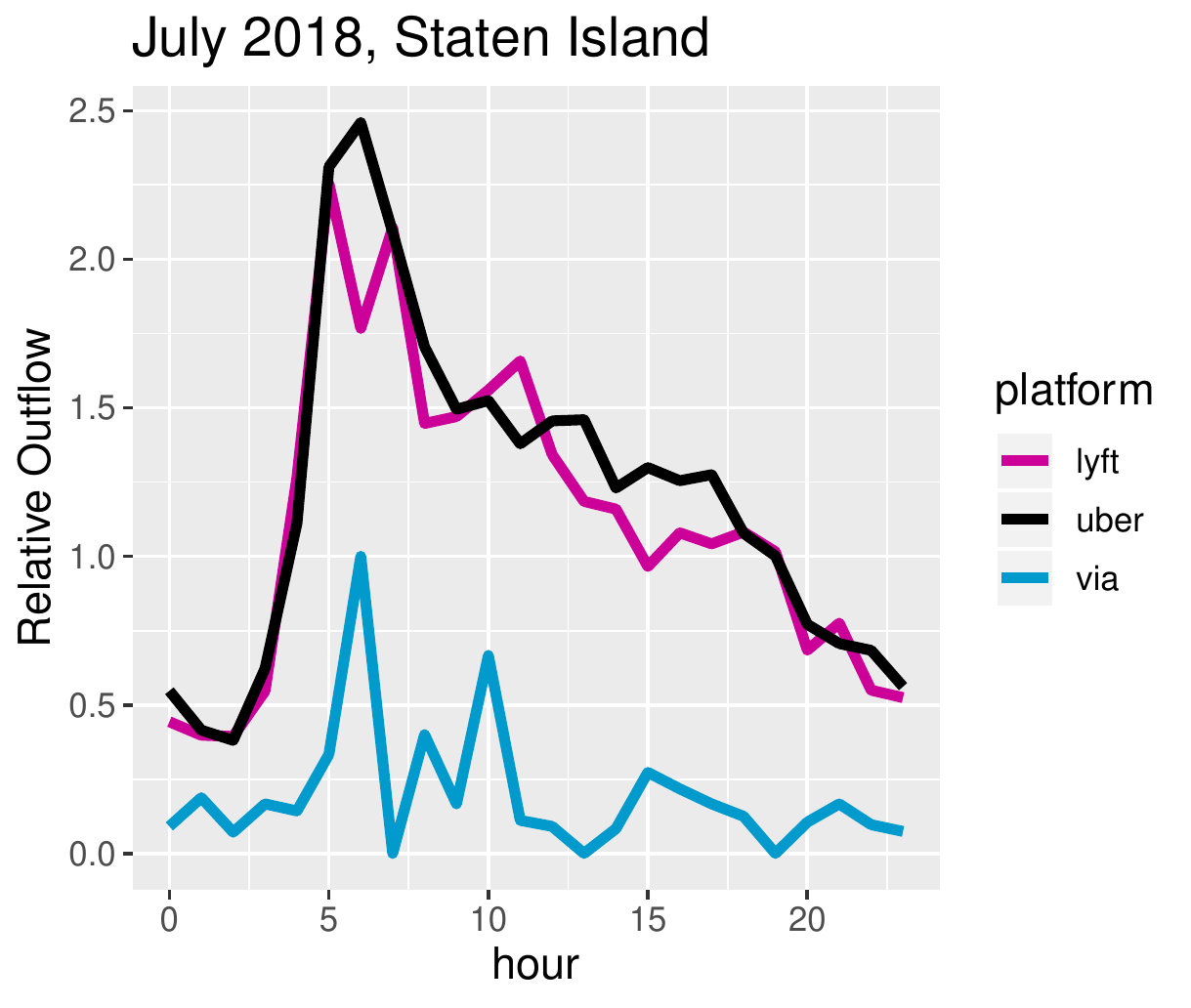}} 
\end{figure}

\cref{fig: hourly RO Staten Island} offers a different illustration by, this time, fixing the location and varying the platform. This figure focuses on Staten Island and exhibits relative outflows for Uber and Lyft during July of 2017 (the first month of our data) and those for Uber, Lyft, and Via one year later. As can be seen from this figure, smaller platforms have relative outflows (in Staten Island) that are \textit{consistently} lower than those of larger platforms. This happens even though the non-rideshare outside options are the same for the passenger of all three platforms. This figure suggests that the persistent differences across relative outflows of rideshare platforms are because of systematic differences in ``access'' rather things such as the schedules of passengers, or outside options.

Next, we turn to carrying out the empirical tests.

\subsubsection{Testing for Economies of Density}
If all of our quantities of interest were observed,  we would ideally directly test our theoretical result: access to rides $A_i=\frac{r_i}{\bar{\lambda}_i}$ is higher in regions that have higher densities of potential demand $D_i=\frac{\bar{\lambda}_i}{t_i}$.  The challenge, as mentioned before, is that potential demand $\bar{\lambda}_i$ is essentially unobservable.


Our solution, inspired by the last two propositions in the theory section and the discussion in the beginning of the empirical section, is to turn to relative outflows and devise a test that approximates, as closely as possible, a test of a positive association between $A_i$ and $D_i$. Recall that we defined the relative outflow in region $i$ as:

\begin{equation}\label{eq:DefRelOutflow}
    RO_i\equiv \frac{r^{\rightarrow}_i}{r^{\leftarrow}_i}
\end{equation}

where $r^{\rightarrow}_i$ is the rate of outgoing rides from region $i$ and $r^{\leftarrow}_i$ the rate of incoming ones. Similarly, we define the density of dropoffs in region $i$ by $D^\leftarrow_i\equiv \frac{r^\leftarrow_i}{t_i}$ where $t_i$ is the size of region $i$ in sqaure miles.

 
\cref{prop: testable implications 1} suggests that we should expect higher $RO_i$ values for denser regions. The regression specification that we use for testing our hypothesis is as follows:

\begin{equation}\label{eq:Regression Test econ of density}
    \log(RO_{ikd})=\alpha \log(D^\leftarrow_{ikd})+\beta X_{ikd}+\epsilon_{ikd}
\end{equation}

Here, $RO$ and $D^\leftarrow$ have been indexed not only by region $i$ but also by platform $k$ and time duration $d$. In this setting, $i$ denotes a ``taxi zone'' as described previously. Also $d$ in this regression is a year-month combination. The coefficient of interest is $\alpha$, which according to the predictions of our model, is expected to be positive and significant. Finally, $X_{ikd}$ is a set of some controls that we use in order to deal with possible confounds and aid the interpretation of the test results as a sign of economies of density. 

Before presenting the results of this regression, we make two points regarding their interpretation. First, \cref{prop: testable implications 1} may leave the impression that in order in order to interpret $RO_i$ as a measure of access to rides in $i$ relative to outside of $i$, it is necessary that potential demands for rides be balanced. That is:

\begin{equation}\label{eq:Balanced Potential Demand}
   \bar{\lambda}_{ikd}^\rightarrow\equiv \bar{\lambda}_{ikd}^\leftarrow 
\end{equation}

where $\bar{\lambda}_{ikd}^\rightarrow$ is the potential demand for rides with platform $k$ that exit $i$ during period $d$ (a month in this case) and $\bar{\lambda}_{ikd}^\leftarrow$ is the potential demand for rides entering it.

Of course \cref{eq:Balanced Potential Demand} would be too strong of a formalization for our assumption that ``for every ride there is a ride back by the same person shortly before or after.'' But \cref{prop: regression equivalance 1} shows that, with the right controls, \cref{eq:Balanced Potential Demand} is \textit{not necessary} for our intended interpretation of the results from regression \ref{eq:Regression Test econ of density}.
 
\begin{proposition}\label{prop: regression equivalance 1}
Suppose that vector $Q$ is a partition of all $ikd$ observations in the data based on some characteristics.\footnote{To illustrate, if $Q$ partitions the data based on borough and platform, it means $\forall\, ikd\, \&\, i'k'd':\, Q_{ikd}=Q_{i'k'd'}$ if and only if  $k=k'$ and regions $i$ and $i'$ are within the same borough.} Also suppose that $\frac{\bar{\lambda}^\rightarrow_{ikd}}{\bar{\lambda}^\leftarrow_{ikd}}$ depends only on characteristics $Q$. That is, for some function $g$, we have: $$\forall ikd:\, \frac{\bar{\lambda}^\rightarrow_{ikd}}{\bar{\lambda}^\leftarrow_{ikd}}=g(Q_{ikd})$$

Under these conditions, the following regression will lead to the exact same estimated $\alpha$ as regression \ref{eq:Regression Test econ of density} if controls in $X_{ikd}$ include fixed effects at the level of $Q$ or finer.

\begin{equation}\label{eq:Regression Test econ of density ideal data}
    \log(\frac{A^\rightarrow_{ikd}}{A^\leftarrow_{ikd}})=\alpha \log(D^\leftarrow_{ikd})+\beta X_{ikd}+\epsilon_{ikd}
\end{equation}

\end{proposition}

The proof can be found in the appendix. \cref{prop: regression equivalance 1}  shows that $X_{ikd}$ helps not only with controlling for omitted factors that may be impacting access to rides, but also with correcting for possible error in measurement of $\frac{A^\rightarrow_{ikd}}{A^\leftarrow_{ikd}}$ using $RO_{ikd}$. In these ways, the controls help rule out a number of possible alternative hypotheses to the effect of density on access. For instance, it is in principle possible (though not likely) that Manhattan  provides worse public transit options than the outer boroughs, prompting passengers to turn to rideshare. This can lead to $\bar{\lambda}_{ikd}^\rightarrow> \bar{\lambda}_{ikd}^\leftarrow $ when $i$ is in Manhattan, implying that a higher relative outflow in Manhattan is just an artefact of higher demand for (rather than better access to) rideshare there. This issue, as \cref{prop: regression equivalance 1} shows, can be taken care of with borough fixed effects if we assume the gap between $\bar{\lambda}_{ikd}^\rightarrow$ and $ \bar{\lambda}_{ikd}^\leftarrow $ can be explained by boroughs. 

The second point we need to make about the interpretation before proceeding to results is that our ideal regressor, as mentioned before, would be $D_{ikd}=\frac{\bar{\lambda}_{ikd}}{t_i}$ but it is essentially unobservable (even by platforms.) In choosing a proxy for it, we decided to use the density of incoming rides  $D^\leftarrow_{ikd}$. We made this choice for two reasons. First, as mentioned before, those who enter region $i$ must have a need to exit shortly before or after. As a result, incoming rides seem like a reasonable proxy for potential demand. Second, working with incoming rides means we are testing the following hypothesis: regions with more incoming rides per square mile have more outgoing rides per incoming rides. That is, $r^\leftarrow_{ikd}$ is in the numerator on the right-hand-side and in the denominator on the left-hand-side. This mimics what we would have if we were able to directly observe $A_{ikd}$ and $D_{ikd}$ which would involve $\bar{\lambda}_{ikd}$ in the numerator on the right-hand-side and in the denominator on the left-hand-side. In both of these cases, a positive association between the two objects would be counter-intuitive but possible to explain using economies of density. That said, we did try using other proxies than $D^\leftarrow_{ikd}$ for $D_{ikd}$ and the results were robust.\footnote{We tried using $D^\rightarrow_{ikd}$ and the results were robust. We also tried using $\Sigma_k D^\leftarrow_{ikd}$ which means using the total number of incoming rides across all platforms as a proxy for potential demand for each of them. The results were robust again. } We now proceed to presenting the results.

\textbf{Results:} In different specifications, we allow $X_{ikd}$ to  capture a variety of fixed effects such as borough fixed effects, platform fixed effects, zone type fixed effects, year-month fixed effects, and interactions among the above. Results of this regression analysis have been reported in \cref{tab:zoneLevelRegression}.

\begin{table}[!htbp] \centering 
  \caption{\scriptsize{Effect of dropoff density in a region on its relative outflow (i.e., pickups per dropoff) is always positive and significant. This is robust to a rich set of fixed effects specifications.}} 
  \label{tab:zoneLevelRegression} 
\begin{tabular}{@{\extracolsep{5pt}}lccccc} 
\\[-1.8ex]\hline 
\hline 
 & \multicolumn{5}{c}{\textit{Dependent variable: log relative outflow}} \\ 
\cline{2-6} 
& (1) & (2) & (3) & (4) & (5)\\ 
\hline \\[-1.8ex] 
 log dropoff density & 0.072$^{***}$ & 0.081$^{***}$ & 0.069$^{***}$ & 0.075$^{***}$ & 0.074$^{***}$ \\ 
  & (0.001) & (0.001) & (0.001) & (0.001) & (0.001) \\ 

  Fixed Effects$^{\dagger}$ & Constant & B & P & Z & B$\times$P$\times$Z \\ 
\hline \\[-1.8ex] 
Observations & 21,357 & 21,357 & 21,357 & 21,357 & 21,357 \\ 
R$^{2}$ & 0.299 & 0.378 & 0.385 & 0.369 & 0.495 \\ 
Adjusted R$^{2}$ & 0.299 & 0.377 & 0.385 & 0.369 & 0.494 \\ 

\hline 
\hline \\[-1.8ex] 
  & \multicolumn{5}{r}{$^{*}$p$<$0.1; $^{**}$p$<$0.05; $^{***}$p$<$0.01} \\ 
 \multicolumn{6}{l}{$\dagger$: P:Platform, B:Borough, Z:Zone-type}\\
\end{tabular} 
\end{table}

\cref{tab:zoneLevelRegression} shows that the coefficient of interest, $\alpha$, is always positive and significant under multiple fixed effects specifications. The simplest column to interpret in this table is column (1) which pertains to the specification with no fixed effects. The positive and significant estimate for $\alpha$ in this column simply means $D^\leftarrow_{ikd}$ and $RO_{ikd}$ are positively correlated across observations.

As mentioned before, the controls $X_{ikd}$ have two roles. First, they help deal with the issue that $RO_{ikd}$ may not directly measure $\frac{A_{ikd}^\rightarrow}{A_{ikd}^\leftarrow}$ due to unbalancedness in potential demand. In addition,  these controls also help deal with alternative hypotheses to the role of economies of density, even if $\bar{\lambda}_{ikd}^\rightarrow= \bar{\lambda}_{ikd}^\leftarrow $. The variations that our robustness columns (i.e., (2) through (5)) in \cref{tab:zoneLevelRegression} leverage should help alleviate concerns about such alternative hypotheses.

 For instance, a model with borough fixed effects would estimate $\alpha$ only based on within-borough variation in relative outflows as a function of dropoff density, which would not pick up differences between Manhattan and the outer boroughs. This rules out the alternative hypothesis that the relative outflows results arise from heterogeneity across boroughs in their provision of alternative transportation options.  Likewise, a zone-type fixed effects model would estimate $\alpha$ from the variation in dropoff density among regions that are of the same type (e.g., residential). To illustrate, recall \cref{figure:hourlyRelOutflowsNeighborhoods}. A regression with zone-type fixed effects compares overall relative outflows between the two residential zones Alphabet City (panel b) and Riverdale (panel c) to each other in order to infer $\alpha$, the effect of dropoff density; but it does \textit{not} compare either of the two to Greenwhich Village (panel a) which is mostly commercial. The robustness of our results to this specification rules out the possibility that the observed relative outflows are only an artefact of differential access to outside options across different zone types.\footnote{We would like to add that in fact Manhattan has a higher relative outflow \textit{in spite of} having a concentration of commercial areas rather than because of it. To see this, we note that in column (4) of \cref{tab:zoneLevelRegression}, the fixed effects coefficient on commercial areas (not reported in the table) comes back smaller than those of the other zone types (i.e., residential, manufacturing, and park). This means all else (including density of dropoffs) equal, a commercial area is expected to have a lower relative outflow for rideshare compared to other zone types. We believe this should not be surprising given the relative abundance of non-rideshare options in commercial areas.}

More sophisticated controls help alleviate concerns about more sophisticated alternative hypotheses. For instance, it is conceivable that the overall balance of flows of rides on a \textit{daily or monthly level} gets impacted not by differential densities but by the \textit{hourly level} complexities in the movement of passengers within the city. As an example, in the morning there is a net flow of passengers from the outer boroughs toward Manhattan when individuals commute to work. In the afternoon, the flow is reversed when they return home. Thus, if supply of rideshare is more limited in the morning compared to the evening (e.g., perhaps because some rideshare drivers are at their ``first jobs'' in the morning), then this can explain why access to rideshare --and, hence, the relative outflow of rides-- is higher in Manhattan independently of density. Which controls help with rule out this alternative depends on our assumptions. If we assume this hour-level complexity impacts the relative flow of Manhattan compared to other boroughs homogeneously across platforms, then it is ruled out simply by the borough fixed effects in column (2) of \cref{tab:zoneLevelRegression}. If we believe, however, that the effect of this mechanism differs across platforms, we will need $B\times P$ or, like column (5) of the table, $B\times P\times Z$.   
 

\textbf{Other robustness checks.} The fixed effects specifications shown in \cref{tab:zoneLevelRegression} are only a subset of those we examined. In particular, we examined interacting all of the fixed effects columns in that table with year-month (YM) fixed effects. The most general case (i.e., B$\times$P$\times$Z$\times$YM) would have about 1600 separate fixed effects. Under all of our additional regressions, the  result was robust. the results also seem robust to functional form assumptions (we tried using the relative outflow itself instead of the log and all results were robust). Finally, we ran the model with only subsets of the full data by taking one platform off each time. Again, $\alpha$ always came back positive and significant.

Before closing this section, we test another prediction by the model. Our model predicts that the positive effect of density $D_i$ on access $A_i$ diminishes as $D_i$ grows large enough. Again directly testing this is not feasible given $A_i$ and $D_i$ are both unobservable due to unobservability of $\bar{\lambda}_i$. Therefore, again, we test this in the context of the relationship between $RO_i$ and $D^\leftarrow_i$. To this end, we change the regression in \cref{eq:Regression Test econ of density} by adding $\log(D^\leftarrow_i)^2$ as another regressor:

\begin{equation}\label{eq:Regression Test Diminishing returns}
    \log(RO_{ikd})=\alpha_1 \log(D^\leftarrow_{ikd})+\alpha_2 \log(D^\leftarrow_{ikd})^2+\beta X_{ikd}+\epsilon_{ikd}
\end{equation}

If our model's prediction of diminishing sensitivity to density is empirically relevant, we should expect a positive $\alpha_1$ and a negative $\alpha_2$ in the estimation results.\footnote{A more precise test of diminishing sensitivity of access to density would use $D^\leftarrow_{ikd}$ and $D^{\leftarrow2}_{ikd}$ in the regression rather than $\log(D^\leftarrow_{ikd})$ and $\log(D^\leftarrow_{ikd})^2$. Nevertheless, we used the latter to keep \cref{eq:Regression Test Diminishing returns} compatible with \cref{eq:Regression Test econ of density}. We have examined the regression with $D^\leftarrow_{ikd}$ and $D^{\leftarrow2}_{ikd}$ and the results were robust. } As \cref{tab:zoneLevelRegression_2} shows, the empirical results are robustly in line with the model prediction.

\begin{table}[!htbp] \centering 
  \caption{\scriptsize{A square density term added to regressions in \cref{tab:zoneLevelRegression}. The effect of the square density term is robustly negative and significant, in line with model predictions.}} 
  \label{tab:zoneLevelRegression_2} 
\begin{tabular}{@{\extracolsep{5pt}}lccccc} 
\\[-1.8ex]
\hline 
\hline 
 & \multicolumn{5}{c}{\textit{Dependent variable: log relative outflow}} \\ 
\cline{2-6} 
& (1) & (2) & (3) & (4) & (5)\\ 
\hline 
 log dropoff density & 0.327$^{***}$ & 0.311$^{***}$ & 0.297$^{***}$ & 0.331$^{***}$ & 0.104$^{***}$ \\ 
  & (0.007) & (0.007) & (0.007) & (0.007) & (0.009) \\ 
 (log dropoff density)$^2$ & $-$0.008$^{***}$ & $-$0.007$^{***}$ & $-$0.007$^{***}$ & $-$0.008$^{***}$ & $-$0.001$^{***}$ \\ 
  & (0.0002) & (0.0002) & (0.0002) & (0.0002) & (0.0003) \\ 
 
  Fixed Effects$^{\dagger}$ & Constant & B & P & Z & B$\times$P$\times$Z \\ 
\hline \\[-1.8ex] 
Observations & 21,357 & 21,357 & 21,357 & 21,357 & 21,357 \\ 
R$^{2}$ & 0.342 & 0.408 & 0.415 & 0.406 & 0.495 \\ 
Adjusted R$^{2}$ & 0.342 & 0.407 & 0.415 & 0.406 & 0.494 \\ 

\hline 
\hline \\[-1.8ex] 
\textit{Note:}  & \multicolumn{5}{r}{$^{*}$p$<$0.1; $^{**}$p$<$0.05; $^{***}$p$<$0.01} \\ 
\multicolumn{6}{l}{$\dagger$: P:Platform, B:Borough, Z:Zone-type}\\
\end{tabular} 
\end{table}

Next, we turn to testing for the role of market thickness in determining the spatial distribution of supply.

\subsubsection{Testing for the Role of Market Thickness (i.e., Platform Size) in Economies of Density}
A second crucial prediction of our model was that the gap between access to rides in busy areas and access to rides in less busy areas is wider for smaller platforms. To this end, we run a regression on rideshare platforms' rides in NYC at the borough level. Specifically, we analyze the following specification:

\begin{equation}\label{eq_reg_outflow_size}
RO_{ikd}=\alpha_0+\alpha_1\log(\rho_i)+\alpha_2 \log(S_{kd})+\alpha_3 \log(S_{kd})\log(\rho_i)+\nu_{ikd}    
\end{equation}
where $RO_{ikd}$ is the relative outflow for platform $k$ at borough $i$ on date $d$ (note that this regression, compared to previous ones, is coarser on $i$ but finer on $d$). Also $\rho_i$ is the population density of borough $i$.\footnote{In this section, we use population density instead of dropoffs density. Population densities allow us to define busier and less busy regions in a way that is constant across platforms and time. This allows us to focus on cross-platform (and within-platform over time) comparisons in size as they impact the spatial distribution of relative outflows. } Finally, $S_{kd}$ is the size of platform $k$ on date $d$, which is measured by the total number of rides given by that platform in NYC during the month in which date $d$ occurs. Tables (\ref{table:outflowsizeregressionsimpleFE}) and (\ref{table:outflowsizeregressioncomplexFE}) report the results from this regression. The first table reports results when we either do not include any fixed effects in the regression or we do have fixed effects but they are not interacted (that is, platform fixed effects, year-month fixed effects, or borough fixed effects).  \Cref{table:outflowsizeregressioncomplexFE} incorporates a  richer set of fixed effects. It starts, in its first three columns, with fixed effects on (i) platform interacted by year-month, (ii) borough interacted by platform, and (iii) borough interacted by year-month. It then incorporates these three pairs into one single regression. Finally, the last column has interaction fixed effects among boroughs, platforms, and years (not year-month in this column).

The coefficient of interest is $\alpha_3$, the interaction coefficient between platform size and borough population density. Based on the predictions of our theoretical model, we would expect a negative estimated value for $\alpha_3$, indicating that as a platform gets smaller, access to its supply in lower density areas falls further behind that in denser areas. As can be seen from tables (\ref{table:outflowsizeregressionsimpleFE}) and (\ref{table:outflowsizeregressioncomplexFE}), this is exactly the result that we get from the empirical analysis and it is robust the controls.

\begin{table}[!htbp] \centering 
\ra{1.0}  \caption{\scriptsize{Relative Outflow regression with single fixed effects. The coefficient of interest is the interaction coefficient which is robustly negative and significant.}}  
\label{table:outflowsizeregressionsimpleFE} 
\tabcolsep=0.05cm
\begin{tabular}{lcccc} 
\\[-1.8ex]\hline 
\toprule
 & \multicolumn{4}{c}{\textit{Dependent variable: Relative Outflow}} \\ 
\cline{2-5} 
& (1) & (2) & (3) & (4)\\ 
\midrule
 log(population density) & 2.154$^{***}$ & 2.222$^{***}$ & 2.141$^{***}$ & -\\ 
  & (0.041) & (0.040) & (0.041) & \\ 
 log(size) & 0.492$^{***}$ & 0.483$^{***}$ & 0.490$^{***}$ & 0.448$^{***}$ \\ 
  & (0.009) & (0.014) & (0.009) & (0.008) \\ 
  log(population density)$\times$\\ log(size) & $-$0.126$^{***}$
  & $-$0.130$^{***}$ & $-$0.125$^{***}$ & $-$0.113$^{***}$ \\ 
  & (0.003) & (0.003) & (0.003) & (0.002) \\ 
  Fixed Effects$^{\dagger}$ & Constant & P & YM & B\\
\midrule
Observations & 7,709 & 7,709 & 7,709 & 7,709 \\ 
R$^{2}$ & 0.595 & 0.624 & 0.598 & 0.725 \\ 
\bottomrule
\textit{Note:}  & \multicolumn{4}{r}{$^{*}$p$<$0.1; $^{**}$p$<$0.05; $^{***}$p$<$0.01} \\ 
\multicolumn{5}{l}{$\dagger$:  P:Platform, B:Borough, YM:Year-Month}\\
\end{tabular} 
\end{table}

\begin{table}[!htbp] \centering 
\ra{1.0} 
 \caption{\scriptsize{Relative Outflow regression with interaction Fixed Effects. The coefficient of interest is the interaction coefficient which is robustly negative and significant.}}
  \label{table:outflowsizeregressioncomplexFE}
\tabcolsep=0.05cm
\begin{tabular}{lccccc} 
\toprule
 & \multicolumn{4}{c}{\textit{Dependent variable: Relative Outflow}} \\ 
\cline{2-5} 
& (1) & (2) & (3) & (4) \\ 
\midrule

 log(population density) & 2.182$^{***}$ & - & - & - \\ 
  & (0.040) &  &   &  \\ 
 log(size) & - & 0.596$^{***}$ & 0.444$^{***}$ & 0.402$^{***}$ \\ 
  & & (0.036) & (0.008) &  (0.061) \\ 
 log(population density)$\times$\\ log(size) & $-$0.128$^{***}$ & $-$0.167$^{***}$ & $-$0.112$^{***}$  & $-$0.110$^{***}$ \\ 
  & (0.003) & (0.011) & (0.002)  & (0.018) \\   
  Fixed Effects$^{\dagger}$ & P$\times$YM & B$\times$P & B$\times$YM  & B$\times$P$\times$Y  \\
\midrule
Observations & 7,709 & 7,709 & 7,709 & 7,709 \\ 
R$^{2}$ & 0.629 & 0.829 & 0.738 & 0.835 \\ \bottomrule
\textit{Note:}  & \multicolumn{4}{r}{$^{*}$p$<$0.1; $^{**}$p$<$0.05; $^{***}$p$<$0.01} \\ 
\multicolumn{5}{l}{$\dagger$:  P:Platform, B:Borough, YM:Year-Month, Y:Year}\\
\end{tabular} 
\end{table}

We end this subsection by emphasizing (without introducing a formal proposition) that, similar to regression \cref{eq:Regression Test econ of density}, regression \cref{eq_reg_outflow_size} does not require the assumption in \cref{eq:Balanced Potential Demand} for $RO_{ikd}$ to be interpreted as a representation of access to rides in $i$ relative to outside of $i$. A similar proposition to \cref{prop: regression equivalance 1} can be proven showing that, with the right controls, $\alpha_3$ remains intact if one replaces $RO_{ikd}$ in the left hand side of \cref{eq_reg_outflow_size} with $\frac{A^\rightarrow_{ikd}}
{A^\leftarrow_{ikd}}$. Moreover, note that our coefficient of interest in this section is different from that in the previous section. Here we are interested in $\alpha_3$, the \textit{interaction} coefficient between platform size and borough population density. Therefore, even without controls, we do not expect this coefficient to be impacted by the possibility that $\frac{\bar{\lambda}^\rightarrow_{ikd}}
{\bar{\lambda}^\leftarrow_{ikd}}$ can be region-specific as long as it is not region-platform specific. In other words, even if things such as availability of public transportation options can create an imbalance on the flow of rides to and from a region $i$, it should not impact $\alpha_3$ (even if we do not use additional controls) as long as public transportation options are assumed to create the same amount of flow imbalance in $i$ for all platforms.

To see how these controls help rule out  alternative hypotheses, we review two such alternatives. First, in principle it might be that Lyft has lower relative outflows in the outer regions because Lyft drivers are more likely to live in busier areas and, hence, prefer to driver there. Column 2 of \cref{table:outflowsizeregressioncomplexFE} rules this out by exploiting only within-platform variation in each borough. Such variation can be seen in \cref{figure:relOutflows1718_zones} as well: Lyft's relative outflows become more balanced geographically as it grows in size. A second, more important, possible issue is the fact that Via is a much different platform in structure than Uber and Lyft; therefore, its more concentrated relative outflows may be a consequence of its different structure rather than its size. Again, our controls that exploit only within-platform (such as columns 1 and 2) variation in each borough show that Via's possible differences from Uber and Lyft is not the main driver of our results.

Finally, in addition to the arguments above, \cref{fig: hourly RO Staten Island} should be useful in illustrating that our results are not an artifact of a differences in the hourly transportation patterns of the users of these platforms. As can be seen in the figure, comparisons among different platforms' relative outflows in Staten Island is remarkably consistent over time of day, both during July 2017 and a year later. This rules out the possibility that, for instance, Lyft had a lower relative outflow in Staten Island in July 2017 because Lyft users tended to need to exit that region at certain times of the day when rideshare was less available relative to other options.

We finish this section with two further notes, before moving to an analysis of data from Austin.

\textbf{Note on possible role of competitive forces.} We finish this section by briefly arguing against the hypothesis that the observed cross-platform geographical differences in access to supply is an artifact of platforms' strategic spatial positioning to avoid competition from one another (i.e., tacit collusion). There are multiple arguments against the role of competitive forces in shaping the observed spatial differences. First, if competitive forces pushed the platforms to each focus on some areas and be less accessible in other areas, then we would see a different ``area of focus'' for each platform. However, in the data, \textit{all platforms} are more accessible in busier regions compared to the outer areas; it is just the extent of the difference between more and less busy regions that differs by platform. Second, if the platforms were dividing the market among themselves, it would be more natural to expect larger  and older platforms such as Uber to claim more attractive markets such as Manhattan. However, Via's supply is showing the highest degree of concentration in the Manhattan area.

\textbf{Note on comparison to taxicabs.} \cref{section:discussion} provides a relative-outflows analysis of the NYC Yellow Taxicab rides and shows that economies of density, perhaps unsurprisingly, is a substantially more pronounced problem in that market compared to rideshare.


\subsection{Evidence from Individual Driver Behavior in Austin}\label{subsec: Austin}

The previous subsection provided evidence for economies of density in the rideshare market. Nevertheless, it did not provide \textit{direct evidence} for the mechanism that our model puts forth for economies of density: drivers' location choice in response to pickup times.\footnote{In particular, one can propose a model of economies of density based solely on demand; and that model's implications would not be empirically distinguishable from our model if one were to solely use relative outflows data. That said, we have performed simulations that show demand side economies of density has similar implications to its supply-side counterpart which are put forth in this paper: access to rides gets skewed toward denser regions, and the platform would want to mitigate but not eliminate this skew using prices and wages.} This section provides such direct evidence. We use data from ``Ride Austin," a non-profit rideshare platform from June 2016 till April 2017. During this period, Ride Austin was the sole rideshare platform operating in the city due to a one-year exit by Uber and Lyft. Using this data, we show that drivers pay substantial attention to pickup times when deciding whether to stop operating in an area.

\textbf{Data Description and Summary Statistics.} Our data encompasses all of the rides given by Ride Austin from early June 2016 till mid April 2017. For each ride, we observe the exact data and time the ride request was sent through the app and the driver was assigned, the time the driver arrived at the passenger's location, and the time s/he dropped the passenger off at the destination. We also observe the corresponding locations. Moreover, we observe the surge multiplier applied to that ride by the platform. Finally, in addition to a driver ID that is constant for each driver for the duration of our data, we have a driver ID in our dataset which resets every time the driver turns her/his app off and and back on.

Note that using the above dataset, important variables for our analysis can be constructed: One can construct the idle time for the driver by calculating the difference between when s/he is dispatched and the last dropoff s/he made. Pickup time is constructed using the difference between when the driver was dispatched and when s/he picked the passenger up. Finally, whether can observe when a driver turned her/his app off after a ride, by looking at whether her/his ID was reset. Table (\ref{tab: Austin Summary Stats}) provides a summary.

\begin{table}[ht]
\centering
\begin{tabular}{lcc|cc}
  &  \multicolumn{2}{c|}{Ride Level} & \multicolumn{2}{c}{District Level}\\
variable & mean & std & mean & std \\ 
  \hline
Pickup Time (minutes) & 6.33 & 4.34 & 7.15 & 0.96 \\ 
  Idle Time (minutes) & 9.55 & 18.04 & 12.74 & 5.88 \\ 
  Surge Factor & 1.10 & 0.39 & 1.07 & 0.04 \\ 
  App Turnoff & 0.40 & 0.49 & 0.42 & 0.03 \\ 
   \hline
\end{tabular}
\caption{Summary statistics of variables of interest from the Ride Austin Data. Ride level statistics are calculated using the full, ride level data (820K observations). For the district level statistics, the data is first aggregated at the district level (11 observation) and then means and variances are calculated.}
\label{tab: Austin Summary Stats}
\end{table}

\textbf{Analysis of Individual Driver Behavior.} We next turn to analyzing whether pickup times impact drivers' decision making on whether to turn their app off in a region after they drop a passenger. We do this while controlling for idle times and prices as well as various fixed effects. This is implemented using the following logit regression:

\begin{equation}\label{eq: app turnoff regression}
    O_{\eta}=\alpha_0+\alpha_1 p_\eta + \alpha_2 W^{idle}_\eta + \alpha_3 W^{pickup}_\eta + \varepsilon_\eta
\end{equation}

In this equation, $\eta$ indexes an individual ride. Variable $O_\eta$ is 1 if the driver turned her/his app off after finishing the ride. Variables $p_\eta$, $W^{idle}_\eta$, and $W^{pickup}_\eta$ are, respectively, the  surge multiplier, idle time, and pickup time that the driver \textit{expects} to have for her/his next ride if s/he keeps her app on and keep searching for rides upon finishing $\eta$. We use expected values instead of actual values for two reasons. First, in the case of rides $\eta$ after which the driver does turn off her/his app (i.e., $O_\eta=1$), one cannot observe what the surge factor and pickup and idle times would have been had s/he decided to keep searching. Second, even in the case of rides $\eta$ after which the driver keeps searching in the area (i.e., $O_\eta=0$), it is reasonable to assume s/he only does not exactly know these quantities for her next rides with full precision. The expectations for each of these quantities are calculated by taking the average among all  Ride Austin rides started at the same ``time-of-week'' and in the same ``tile location'' as ride $\eta$ ended. Our definition of time of weak is a slight modification of that in \cite{chen2017value}; and our tiles are a set of 1-square-mile squares partitioning the city of Austin and its greater area.

Table \ref{tab: app turnoff regression results} presents the results of analyzing regression \ref{eq: app turnoff regression} using different fixed effects specifications.  We use a combination of area fixed effects and time-of-week fixed effects. Area fixed effects divide the city into 11 parts: 10 official districts and what we consider the outskirt which is any location not belonging to one of the 10 districts. Time-of-weak indicators were described in the previous paragraphs. More details about these two fixed effects can be found in Figure \ref{fig:Austin FE}.

\begin{table}[ht]
  \begin{varwidth}[b]{0.6\linewidth}
    \centering
    \begin{tabular}{c|c}

\begin{tabular}[c]{@{}c@{}}bucket\\ name\end{tabular}            & description                                                                                      \\
\hline
Weekday Regular                                                  & Mon-Fri: 10am-4pm                                                                                \\
\hline
\begin{tabular}[c]{@{}c@{}}Weekday\\ Morning Rush\end{tabular}   & Mon-Fri: 8am-10am                                                                                \\
\hline
\begin{tabular}[c]{@{}c@{}}Weekday\\ Afternoon Rush\end{tabular} & Mon-Fri: 4pm-9pm                                                                                 \\
\hline
Friday Night                                                     & \begin{tabular}[c]{@{}c@{}}Fri: \textgreater{}4pm,\\ Sat: \textless{}2am\end{tabular}            \\
\hline
Saturday Morning                                                 & Sat: 2am-1pm                                                                                     \\
\hline
Saturday Rush                                                    & \begin{tabular}[c]{@{}c@{}}Sat: \textgreater{}1pm,\\ Sun \textless{}2am\end{tabular}             \\
\hline
Quiet                                                            & \begin{tabular}[c]{@{}c@{}}All other times\end{tabular}
\end{tabular}
    \caption{Student Database}
    \label{table:student}
  \end{varwidth}%
  \hfill
  \begin{minipage}[b]{0.4\linewidth}
    \centering
    \includegraphics[width=80mm]{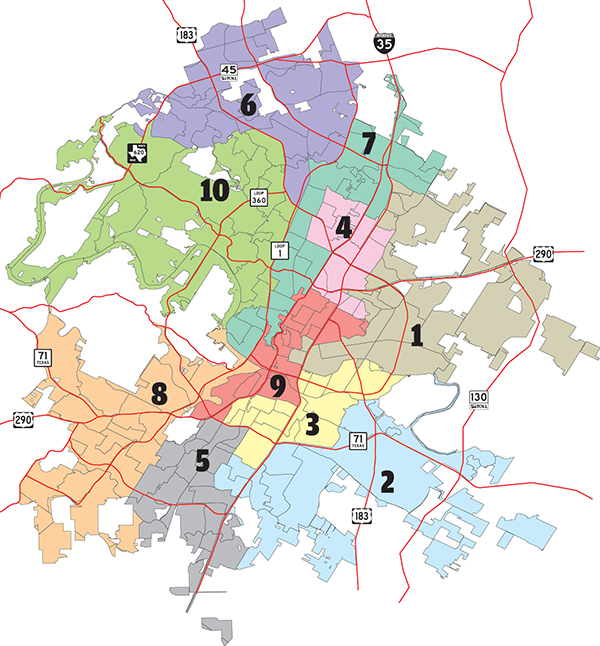}
    
    \label{fig:Austin FE}
  \end{minipage}
\end{table}

\begin{table}[!htbp] \centering 
  \caption{Results from regressing app turnoff decisions on pickup times, controlling for prices, idle times, and different fixed effects.} 
  \label{tab: app turnoff regression results} 
\begin{tabular}{@{\extracolsep{5pt}}lccccc} 
\\[-1.8ex]\hline 
\hline \\[-1.8ex] 
 & \multicolumn{5}{c}{\textit{Dependent variable: App Turned Off}} \\ 
\cline{2-6} 

\\[-1.8ex] & (1) & (2) & (3) & (4) & (5)\\ 
\hline \\[-1.8ex] 
 Pickup Time & 0.004$^{***}$ & 0.004$^{***}$ & 0.012$^{***}$ & 0.007$^{***}$ & 0.007$^{***}$ \\ 
  & (0.001) & (0.001) & (0.001) & (0.001) & (0.001) \\ 
  
 Idle Time & 0.003$^{***}$ & 0.006$^{***}$ & 0.001$^{***}$ & 0.001$^{***}$ & 0.001$^{***}$ \\ 
  & (0.0001) & (0.0001) & (0.0001) & (0.0002) & (0.0002) \\ 
  
 Surge Factor & $-$0.289$^{***}$ & $-$0.228$^{***}$ & $-$0.123$^{***}$ & $-$0.050$^{***}$ & $-$0.056$^{***}$ \\ 
  & (0.010) & (0.010) & (0.013) & (0.014) & (0.016) \\ 
  
 Constant & 0.657$^{***}$ & - & - & - & - \\ 
  & (0.012) &  & &  &  \\ 
  Fixed Effects$\dagger$ & N & L & T & T+L & T$\times$L \\
\hline \\[-1.8ex] 
Observations & 820,343 & 820,343 & 820,343 & 820,343 & 820,343 \\ 
Log Likelihood & $-$575,179 & $-$574,801 & $-$572,880 & $-$572,693 & $-$572,618 \\ 
Akaike Inf. Crit. & 1,150,367 & 1,149,630 & 1,145,784 & 1,145,429 & 1,145,421 \\  

\hline 
\hline \\[-1.8ex] 
\textit{Note:}  & \multicolumn{5}{r}{$^{*}$p$<$0.1; $^{**}$p$<$0.05; $^{***}$p$<$0.01} \\ 
\multicolumn{6}{l}{$\dagger$: N: None,  L: Location, T: Time}\\
\end{tabular} 
\end{table}

\textbf{Interpretation and Robustness.} Regression results shown in table \ref{tab: app turnoff regression results} all have the expected signs. All else equal, drivers are more likely to turn off their apps when surge factor is lower or wait time (idle time or pickup time) is higher. Most specifications suggest drivers dislike pickup times more than idle times. This is natural: pickup times are essentially idle times plus fuel costs. Finally, it is not surprising to see that once controlling for time and location fixed effects, the effect of price is attenuated.

It is also worth noting that the individual-driver nature of the dataset is reassuring when it comes to possible endogeneity problems. This is because we do not need to worry about reverse causality: no individual driver's behavior should change the expected prices and wait times in regions. The only possible concern would be omitted variables which we control for using fixed effects based on fine partitioning of time and space.

According to the results from the column (4) of table \ref{tab: app turnoff regression results}, drivers would like a 7-minute reduction in pickup times almost as much as they would like a surge factor of 2. According to this column, and comparing with district-level summary statistics from table \ref{tab: Austin Summary Stats}, drivers care about a one-standard deviation change in pickup times almost four times as much as they do about a one-standard-deviation change in the surge factor. The magnitude of the effect according to other columns of the table is smaller. Nevertheless, in all of them, pickup times have a non-trivial impact on drivers' decision making on where (not) to operate.

Though the use of multiple fixed effects specification along with the individual nature of the data is reassuring on the front of robustness, we do perform another check. We restrict the dataset to observations from only the month of March in 2017 (the latest month from which we have full data) and run the same regression. We do this in order to make sure our results are not confounded by underlying macroeconomic or demographic changes during the (almost) one-year duration of our data. Results from this additional analysis are broadly similar to those from table \ref{tab: app turnoff regression results}. To the extent that there is any difference, it slightly strengthens the role of pickup times. Table \ref{tab: app turnoff March 2017} shows the results.

This concludes our empirical analysis of Ride Austin. Appendices \ref{appx: austin RO} and \ref{appx: further Austin} provide further analysis in support of economies of density for Ride Austin. \cref{appx: austin RO} carries out relative outflows analysis for Austin and \cref{appx: further Austin} shows evidence that drivers move toward busier regions both when they have their app off and when they have it on.

\begin{table}[!htbp] \centering 
  \caption{Repeating the logit regression analysis of app turnoff decisions, restricting the data to March 2017 only} 
  \label{tab: app turnoff March 2017} 
\begin{tabular}{@{\extracolsep{5pt}}lccccc} 
\\[-1.8ex]\hline 
\hline \\[-1.8ex] 
 & \multicolumn{5}{c}{\textit{Dependent variable: App Turned Off}} \\ 
\cline{2-6} 
\\[-1.8ex] & (1) & (2) & (3) & (4) & (5)\\ 
\hline \\[-1.8ex] 
 Pickup Time & 0.009$^{***}$ & 0.004$^{**}$ & 0.015$^{***}$ & 0.007$^{***}$ & 0.006$^{***}$ \\ 
  & (0.001) & (0.002) & (0.001) & (0.002) & (0.002) \\ 
  
 Idle Time & 0.003$^{***}$ & 0.004$^{***}$ & 0.001$^{***}$ & 0.001$^{***}$ & 0.002$^{***}$ \\ 
  & (0.0002) & (0.0003) & (0.0003) & (0.0004) & (0.0004) \\ 
  
 Surge Factor & $-$0.173$^{***}$ & $-$0.118$^{***}$ & $-$0.101$^{***}$ & $-$0.041 & $-$0.032 \\ 
  & (0.022) & (0.023) & (0.030) & (0.032) & (0.037) \\ 
 
 Constant & 0.520$^{***}$ & - & - & - & - \\ 
  & (0.027) &  &  &  &  \\ 
  Fixed Effects$\dagger$ & N & L & T & T+L & T$\times$L \\
\hline \\[-1.8ex] 
Observations & 167,838 & 167,838 & 167,838 & 167,838 & 167,838 \\ 
Log Likelihood & $-$118,836 & $-$118,762 & $-$118,628 & $-$118,579 & $-$118,521 \\ 
Akaike Inf. Crit. & 237,681 & 237,553 & 237,279 & 237,201 & 237,224 \\ 

\hline 
\hline \\[-1.8ex] 
\textit{Note:}  & \multicolumn{5}{r}{$^{*}$p$<$0.1; $^{**}$p$<$0.05; $^{***}$p$<$0.01} \\ 
\multicolumn{6}{l}{$\dagger$: N: None,  L: Location, T: Time}\\
\end{tabular} 
\end{table}

\section{Implications}\label{section:implications}

Our empirical analysis in \cref{section:empiricalanalysis} tests the relevance of our theoretical model in \cref{section:Theory}. Based on the empirical results, we indeed find the theory and its possible implications empirically relevant. Thus, this section discusses those implications for business strategy and public policy.

\subsection{Business Strategy}
Regarding business strategy implications of our works, we make three notes. First, we discuss the implications for surge pricing. Next, we discuss how our analysis relates to the optimal spatial incentive scheme by rideshare platforms. 

\textbf{Surge Pricing.} Comparing our theoretical results to other papers on surge pricing such as \cite{castro2018surge} leads to an important implication for surge pricing as it relates to recurrent versus temporary shocks to demand. Under a temporary positive shock to demand --such as the end of a sports event-- in region $i$, the platform should raise the regional price and driver wage in order to encourage drivers to relocate to $i$. \cite{castro2018surge} even recommend that the platform reduce wages in some other locations to speed up the process of relocation to $i$.  Our model abstracts away from temporary shocks. Recurring demand shocks such as rush-hours, however, are different in that it is reasonable to assume drivers (i) know about them and (ii) can predict that other drivers respond to them. Under these assumptions, our model would suggest that platforms \textit{should not} do surge pricing during rush hours. In fact, our model points out that high demand leads to economies of density which is attractive to drivers due to shorter pickup times. Consequently, platforms can even \textit{reduce} driver wages during rush hours, effectively charging drivers for the provision of economies of density. Platforms can then pass on some of this cost saving to passengers in order to get even more demand and strengthen the local economies of density. Of course  this recommendation hinges on at multiple  assumptions that our model in \cref{section:Theory} is making. First, we  abstract away from congestion. Second, as mentioned before, our model assumes drivers know about the lower pickup times in region $i$.  Third, our model assumes that higher demand only means a scale-up in the demand curve and does \textit{not} involve a change in the price elasticity. If either of these two assumptions is violated, then it may still be in a rideshare platform's interest to raise the prices and wages during rush hours. Fourth, and perhaps most importantly, we assume  $\bar{c}$ is homogeneous \textit{across drivers}. If drivers are substantially heterogeneous in their reservation values $\bar{c}$, then it may make sense to increase wages during rush hours in order to encourage drivers with higher $\bar{c}$ to join the market.

\textbf{Optimal Spatial Pricing Scheme.} \cref{prop:endogenous Wages and Prices} characterizes the optimal spatial price and wage policy by the platform. It describes the overall form of the optimal spatial strategy in the presence of opposing economic forces presented by \cref{prop:endogenous Wages} and \cref{prop:endogenous Prices}. Our \cref{prop:endogenous Wages and Prices} states  that the platform's optimal strategy would entail offering higher driver wages in sparser regions (unless they are too sparse). This is in line with the ``pickup time bonus'' policy (i.e., paying drivers for longer pickups) which some platforms have recently adopted in some (but not all) markets. To our knowledge, platforms pass on the full bonus to passengers. \cref{prop:endogenous Wages and Prices}, however, recommends that \textit{the platform should pay for part of this bonus} (i.e., pass-through should be less than 1). Footing part of this bill, according to our model, would be an investment by the platform into building economies of density in sparser regions: any driver who chooses to drive in a less busy area will slightly decrease the pickup time in that area, thereby encouraging more drivers to locate there without requiring additional bonus from the platform.

\subsection{Public Policy}
Containment of the growth of rideshare platforms has been a  ongoing policy debate  in the city of New York. Multiple policies --such as direct caps on the number of rideshare driver licences issued, a minimum wage, or congestion taxes-- have proposed and there has been opposing reactions from different stakeholders (these proposals and reactions to them may be found in numerous media articles on this issue. Examples are \cite{NYT_2019_SuicideCharge,NYT_2019_DriverSuicide,NYT_2019_CongestionTaxFate,washingtonPost_2018,wired_2019,theHill_2019,theVerge_2018_Vote,techCrunch_2018_Vote,techCrunch_2019}). While there are possible benefits from controlling the sizes of rideshare platforms --such as less congestion--, our analysis in this paper points out that there is a potential unintended consequence from such policies that has to do with the role of market thickness and that has been overlooked: a smaller platform size (i.e., thinner market) incentivizes drivers to avoid the outer areas.\footnote{It might seem at first that ``multi-homing'' (i.e., the phenomenon  of drivers working for multiple platforms \citep{bryan2019theory}) might mitigate the excess clustering of supply of small platforms in busier areas, because drivers working for multiple platforms are, in effect, working for one large rideshare system. We note, however, that, for  multi-homing to mitigate agglomeration, it must be that  the matching  systems across platforms are fully integrated. This would imply, for instance,  that a Lyft driver would not get asked to pick up a passenger  who is far away, if there is an Uber driver in the vicinity of that passenger. We believe that in reality, the integration of matching systems is substantially less than perfect, rendering multi-homing less impactful on the extent of agglomeration. Indeed, if multi-homing could eliminate agglomeration, it should have balanced out the relative outflows of the platforms, which is not what we observe in Figures \ref{figure:relOutflows1718_zones}, \ref{figure:hourlyRelOutflowsNeighborhoods}, \ref{fig: hourly RO Staten Island}, and \ref{figure:relOutflows1718_boroughs}.}

Related to this issue, we answer an interesting question motivated by our theoretical analysis. Our theoretical results show that geographical inequity in access to supply diminishes as platform size becomes infinitely large due to the fact that pickup times lose their importance against idle times. In a sense, this implies that if the platform size is ``large enough,'' then geographical inequity will not be a first order concern. A practical question is how large is this ``large enough'' size? To find out, we modify regression equation (\ref{eq_reg_outflow_size}), replacing the log function applied to platform size by a function that satiates to an upper limit as the platform size increases. 

We implement this by using $\log (\min(a_{Max},\#Rides))$ instead of $\log (\#Rides)$, where $a_{Max}$ is the parameter capturing the adequate size and is to be estimated (one could interpret $a_{Max}$ as the size at which the impact of size on the geo-distribution of relative outflows becomes small enough so that it cannot be distinguished from noise). We choose this way of capturing the adequate size over adopting a functional form that converges smoothly as size grows. The reason behind this choice is  that we want the identification of the adequate size to come mainly from the data points at  which  relative outflows stop responding to platform size, as opposed to the data points at which the platform size is well  below the upper limit.\footnote{The identification comes from the size at which  Lyft's spatial distribution of relative outflows (i) substantially slows down in its response to Lyft's growth in size and (ii) satiates to a distribution that looks very close to that of Uber. } The regression equation implementing  this notion is  very similar to the earlier regression \Cref{eq_reg_outflow_size} on relative outflows, with the difference being the inclusion of $a_{Max}$. Equation (\ref{reg_size_upperLimit_log}) describes this regression:
\begin{equation}\label{reg_size_upperLimit_log}
RO_{ikd}=\alpha_0+\alpha_1\log(\rho_i)+\alpha_2 \log(\min(a_{Max},S_{kd}))+\alpha_3 \log(\min(a_{Max},S_{kd}))\log(\rho_i)+\nu_{ikd}  
\end{equation}

In order to make  sure that the functional form of log is not substantially impacting our estimate of $a_{Max}$, we  also estimate a version in which the size itself,  as opposed to its natural log, is used. Equation (\ref{reg_size_upperLimit_linear}) represents this:
\begin{equation}\label{reg_size_upperLimit_linear}
RO_{ikd}=\alpha_0+\alpha_1\log(\rho_i)+\alpha_2 \min(a_{Max},S_{kd})+\alpha_3 \min(a_{Max},S_{kd})\log(\rho_i)+\nu_{ikd}  
\end{equation}

Regressions (\ref{reg_size_upperLimit_log}) and (\ref{reg_size_upperLimit_linear}) are estimated using non-linear least squares, and the results are reported in Table (\ref{table:adequateSize}). The adequate size parameter, $a_{Max}$ is estimated at 3.65M rides/month using regression (\ref{reg_size_upperLimit_log}) and at 3.30M rides/month using regression (\ref{reg_size_upperLimit_linear}). Both estimates are statistically very significant. They are also fairly close to each other, suggesting the robustness of $a_{Max}$ to the model specification, as we expected. The values estimated for $a_{Max}$ are in the range of the size tha Lyft reached mid 2018.

These results suggest that NYC needs to use caution if it were to downsize Lyft and, especially, Via. Uber, on the other hand will not face distorted geographical supply distribution if downsized. We note that given a similar dataset to what we used here, the method we laid out in this section can help identify $a_{Max}$ in any other metropolitan area. These estimates are in line with the analysis by \cite{rosaia2020competing} which finds that economies of density was not large in 2019 for Uber and Lyft.

\begin{table}[!htbp] \centering 
\ra{1.0}  \caption{\scriptsize{Results of regressions (\ref{reg_size_upperLimit_log}) and (\ref{reg_size_upperLimit_linear}). The main coefficient of interest is $a_{Max}$, the adequate size for a rideshare platform to contain geographical inequity in supply.}}
  \label{table:adequateSize} 
\begin{threeparttable}
\begin{tabular}{lcccc} 
\toprule
 & \multicolumn{4}{c}{\textit{Dependent variable: Relative Outflow}} \\ 
\cline{2-5} 
& (1) & (2)\\ 
\midrule
Regression & Equation (\ref{reg_size_upperLimit_log}) & Equation (\ref{reg_size_upperLimit_linear})\\
\midrule
$\alpha_0$ & $-$15.07$^{***}$ & $-$1.449$^{***}$  \\ 
  & (0.2829) & (0.027) \\ 
$\alpha_1$ & 4.129$^{***}$ & 6.408$^{***}$ \\ 
  & (0.079) & (7.604e-03) \\ 
$\alpha_2$ & 1.030$^{***}$ & 5.876e-07$^{***}$  \\ 
  & (0.019) & (1.254e-08) \\ 
$\alpha_3$  & $-$0.264$^{***}$ & $-$1.505e-07$^{***}$ \\ 
  & (0.005) & (3.428e-09) \\ 
$a_{Max}$ & 3.648e+06$^{***}$ & 3.295e+06$^{***}$ \\
  & (7.120e+04) & (3.916e+04) \\ 
\midrule
\hline \\[-1.8ex] 
Observations & 7,709 & 7,709  \\ 
\bottomrule
\textit{Note:}  & \multicolumn{4}{r}{$^{*}$p$<$0.1; $^{**}$p$<$0.05; $^{***}$p$<$0.01} \\ 
\end{tabular} 
\end{threeparttable}

\end{table}

\section{Conclusion}\label{section:conclusion}

This paper examined the effects of economies of density and market thickness on spatial distribution of supply. We focused on ridesharing and analyzed this market both theoretically and empirically. Theoretically, we showed that, at the equilibrium, regions with higher densities of potential demand get more supply--even after normalizing by their higher demand--compared to lower density regions. We showed that this ``spatial skew'' of supply from demand is more intense for smaller platforms. Finally, we showed that a rideshare platform's optimal pricing strategy would involve mitigating, but not fully eliminating, the spatial mismatch between supply and demand. The lack of full alignment between the equilibrium driver distribution (if the platform does not intervene) on the one hand and the platform's optimal distribution on the other comes from externalities that drivers leave on each other which the platform internalizes.

On the empirical side, we devised a method with simple implementation and limited data requirements to detect a mismatch between supply and demand in passenger transportation markets. Our method, ``relative outflows analysis,'' can be applied to all passenger transportation markets regardless of whether the matching technology is centralized (e.g., rideshare) or decentralized (e.g., taxicabs). We applied our method to the rideshare market in NYC and tested the implications of our theory model. We found evidence that access to supply is indeed skewed toward higher density areas and that the skew is more pronounced for smaller platforms. In addition to these tests of our model's implications, we also provided direct tests of the main model ingredient  (i.e., the role of pickup times in drivers' location decisions) through an analysis of app turn-off decisions by drivers of a rideshare service in Austin.

We also discussed the implications of our model for business strategy and public policy. We argued that, according to our model, surge pricing should not be done for recurring demand shocks (such as rush hours) in the same way it is done for temporary shocks (such as the end of a sporting event). We also argued that platforms should give bonuses to drivers for driving in sparser regions  but should avoid passing on the full bonus to passengers. Finally, we argued that breaking up or downsizing a rideshare platform can have the unintended consequence of further incentivizing drivers to operate in busier areas, thereby disproportionately hurting outer areas and suburbs.

The analysis in this paper can be extended in multiple directions. On the theoretical side, a next step would be to extend the model to markets with decentralized matching such as the taxicab market.  Also, it would add value to investigate the role of platform competition in shaping the spatial distribution of supply.

On the empirical side, an important next step would be to quantify the optimal (for a firm or social planner) price and wage strategy in the presence of economies of density. Our theoretical model shows that the platform's optimal policy involves some incentives for drivers to operate in less dense regions. But our paper does not empirically determine the exact magnitude of the price and wage differences across regions. An empirical analysis providing such  empirical recommendations  would be of value especially if it captures economies of density alongside other mechanisms that could bring about inefficiencies in the spatial distribution of drivers. Several papers that are complementary to this paper point out mechanisms leading to inefficiencies in the distribution of supply. For instance, \cite{buchholz2018spatial} studies frictions arising from intra-day dynamics, and \cite{brancaccio2019efficiency} examine ``pooling externalities.'' A unified analysis comparing the magnitudes among the efficiency effects of these mechanisms and economies of density could certainly enrich our understanding of spatial markets.

\bibliographystyle{apalike}
\bibliography{sample}

\appendix
\section*{Appendices}
\setcounter{lemma}{0}
\renewcommand{\thelemma}{A\arabic{lemma}}
\setcounter{lproof}{0}
\renewcommand{\thelproof}{A\arabic{lproof}}

\renewcommand{\thesection}{\Alph{section}}

This section provides six appendices. \cref{subsection:anecdotalEvidence} provides anecdotal evidence for the relevance of our model. \cref{appx:micro foundation} provides a micro foundation based on a circular city model for the total wait time formula that we use throughout the paper. \cref{appx: proofs 1} gives the proofs for those propositions in which the platform prices and wages are considered exogenous and uniform (i.e., \cref{prop: driver behavior EQ description} through \cref{prop:Platform vs EQ}). \cref{appx: proof 2} supplies proofs for those propositions in which platform pricecs and wages are determined optimally (i.e., \cref{prop:endogenous Wages} through \cref{prop:endogenous Wages and Prices}). \cref{appx: proof empirical} provides the proof to our proposition in the empirical section (i.e., \cref{prop: regression equivalance 1}). Finally, \cref{sec:Appx:RelOutfows Borough} visualizes relative outflows at the borough level and discusses similarities to/differences from the zone level relative outflows. \cref{section:discussion} does a relative outflows analysis of the Yellow Taxicab system in NYC and argues that economies of density is a substantially larger problem there compared to rideshare platforms in NYC. \cref{appx: austin RO} provides a relative-outflows analysis of Ride Austin. \cref{appx: further Austin} delivers further evidence of economies of density using individual-level driver behavior analysis of Ride Austin.

\section{Anecdotal Evidence from Media and Online Forums}\label{subsection:anecdotalEvidence}
{
This appendix points to a list of anecdotal pieces of evidence (by no means exhaustive) from online rideshare forums on how drivers complain about Lyft's far pickups in suburbs and how they recommend responding to it. The explanations in brackets within the quotations are from us.

\begin{itemize}
    \item \textbf{From the online forum ``Uber People,''\footnote{In spite of what the name suggests, this is a general ridesharing forum, not exclusively about Uber.} a thread in the Chicago section:} The title of the thread is ``To those who drive Lyft in the suburbs.'' The thread was started on Dec 19 2016. The first post says ``Are the ride requests you get on Lyft always seem to be far away from you location? Seems like they are always 5 miles or more for the pickup location. I got one for 12 miles last night. I drive in the Schaumburg/Palatine area [two northwestern suburbs of Chicago about 30mi away from downtown].''
    \item \textbf{From the same thread:} ``iDrive primarily in Palatine. about two out of every five ride requests are for more than 10 minutes away. I ignore those''.
    \item \textbf{From the same thread:} ``I was a victim of that once. Never again I take a ping more than 10 minutes away in the burbs''.
    \item \textbf{From the same thread:} ``Yesterday was my 1st day on Lyft. Was visiting in Homer Glen [a village about 30mi southwest of downtown Chicago] \& decided to try Lyft for the first time. First ping was 18 minutes away. Dang, I could make it 1/2 way downtown in that time! I ignored the ride request. 2nd ping was also 18 minutes away. Lyft app complained my acceptance rate is too low. I ignored the 2nd ping \& went off-line.''
    \item \textbf{From the same forum, a thread titled ``First 3days of Lyft'':} ``If your area is spread out...and you have to take those > 10 minute requests, well...I might look for another job.''
    \item \textbf{From the same thread:} ``Yeah, another (mostly) Lyft-specific problem, especially when working in the suburbs, is you sometimes (fairly frequently, actually) receive trip requests that are not close to your current location. I've received requests from passengers 20 miles away.''
    \item \textbf{From Chicago Tribune article titled ``Lyft takes on Uber in suburbs'':} Jean-Paul Biondi, Chicago marketing lead for Lyft is quoted to explain the reason for Lyft's planned expansion into surburbs as follows ``The main reason is we saw a lot of dropoffs in those areas, but people couldn't get picked up in those areas.'' Which is in line with our reasoning that small relative-outflow is a sign of potential demand which does not get served due to under-supply.
    \item \textbf{From the rideshare website ``Become a Rideshare Driver'':} It says successful Lyft drivers use the following strategy:
    \begin{itemize}
        \item ``The drivers usually run the Lyft app exclusively when they are in the busy downtown or city areas.''
        \item ``Usually in the suburbs, Uber is busier than Lyft, and in such areas, the drivers run both the Uber and Lyft apps.''
    \end{itemize}
\end{itemize}
}

\section{Micro Foundation for the Total Wait Time Formula}\label{appx:micro foundation}
This section provides a simple micro-foundation for the total wait time formula \cref{eq:wait} using a circular city model ala Salop. The circular model of regions is illustrated in \Cref{fig:circularcity}. The $n_i$ drivers (in the figure, $n_i=5$) are placed on equidistant locations on the circumference. \footnote{While we do not model the locational choice of drivers within the region, it is fairly easy to see that equidistant positioning location from neighbors is an equilibrium. While there might be other locational choices that might also be equilibria, we focus on the equidistant positioning equilibrium.} Drivers are matched to arriving passengers via a centralized matching system. Each driver's ``range'' or ``catchment area'' will be the arc consisting of all the points on the circle that are closer to that driver than they are to any other driver in region $i$. Given that the total arrival rate in the region is $\lambda_i$, the arrival rate in each driver's catchment area is $\frac{\lambda_i}{n_i}$. Each driver picks up the first passenger that arrives within that driver's catchment area. In practice, ridesharing platforms implement a similar matching rule (\cite{frechette2018frictions} use a similar approach to model a centralized matching market).

Suppose the time it takes a driver to travel a full circumference to pick up a passenger is $t'$. The platform allocates an arriving customer to the closest driver.  Because drivers are situated at equidistant points on the circumference of the region, their catchment areas include half the distance to their nearest neighbors on both sides. The idle time expected for a customer to arrive in the driver's  area is $\frac{n_i}{\lambda_i}$.  The distance between drivers is $\frac{l}{n_i}$ where $l$ is the circumference. Since consumer location is uniform, the distance a consumer will be from the driver along the arc is distributed $d\sim U[0,\frac{l}{2n_i}]$, implying that the expected distance is $E[d] = \frac{l}{4n_i}$. Thus, the expected pickup time is $\frac{t'}{4n_i}$.

\begin{figure}
\centering
\caption{Illustration of Circular
Model of each Region $i$ and Driver Allocation}
\label{fig:circularcity}
\centering
\vspace{-0.02in}
\includegraphics[scale=.3]{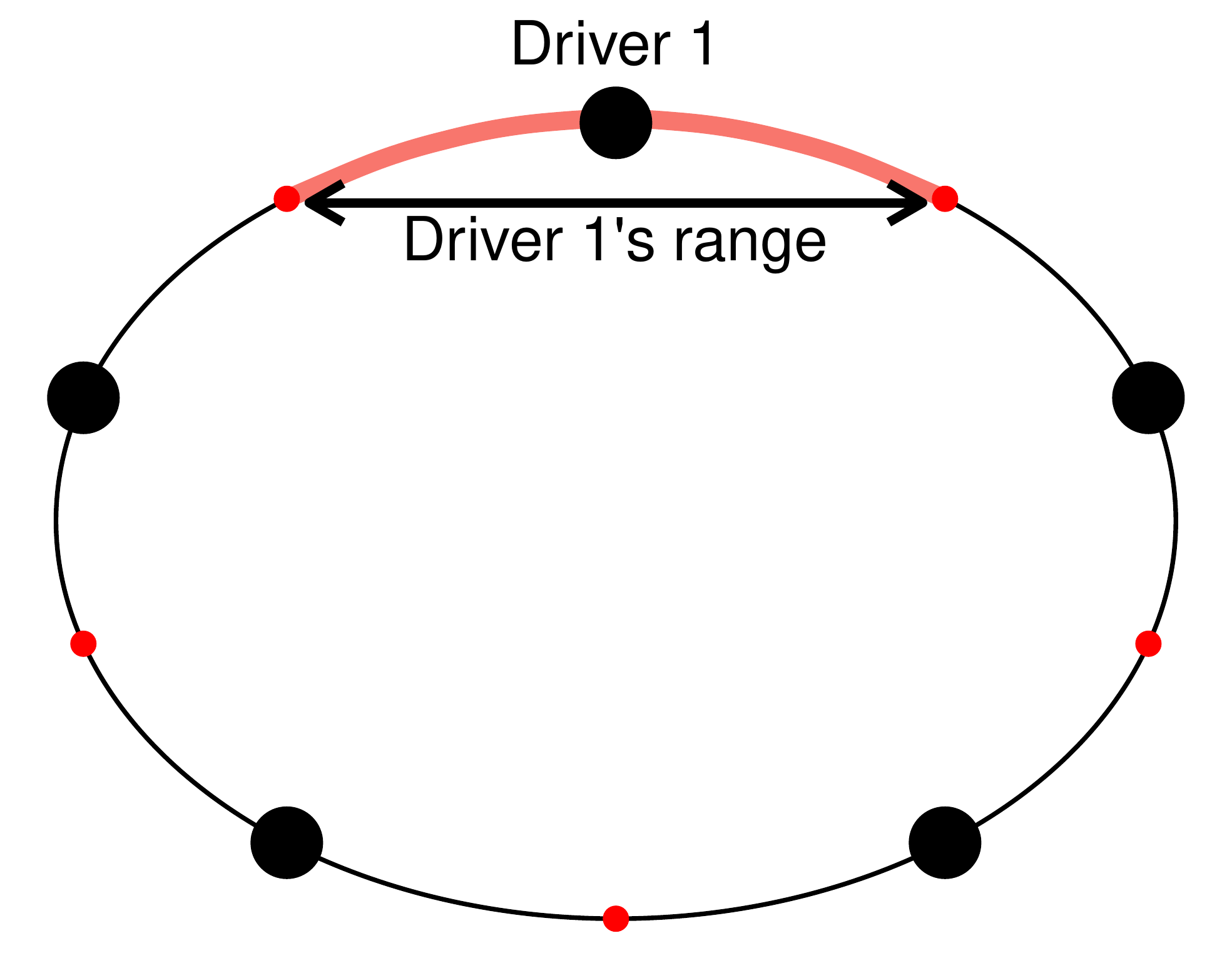}
\end{figure}

Based on the above, the expected total wait time $W_i(n_i)$ defined from the driver's perspective as:
\begin{equation}\label{eq:waitAppendix}
 \underbrace{W_i(n_i)}_{\text{Total Wait Time}} = \underbrace{\frac{n_i}{\lambda_i}}_{\text{Idle Time}} +  \underbrace{ \frac{t'}{4n_i} }_{\text{Pickup Time}} = \left(\frac{n_i}{\lambda_i} + \frac{t}{n_i}\right) 
\end{equation}
where $t=\frac{t'}{4}$. This is exactly the same as \cref{eq:wait} except the notation on $\lambda_i(p_i)$ is suppressed to $\lambda_i$.

\section{Proofs for the results with fixed and uniform prices and wages}\label{appx: proofs 1}

This section is organized as follows. In \cref{subsec:preliminary results} we build up a set of results that will later be used in proving our propositions in the main text (under fixed and uniform prices and wages). These preliminary results have two shortcomings. First, they are ``all-regions equilibria,'' meaning equilibria in which each region gets a positive number of drivers. Second, they are developed assuming size $t_i$ is homogeneous across all regions. After building up these preliminary results, our next step in \cref{subsec: proofs when region size homogeneous} is to extend the analysis to multiple equilibria and provide proofs of the propositions in the main text  regarding driver behavior under fixed and uniform prices and wages  under homogeneous $t_i$. Finally, in \cref{subsec: full proofs of driver behavior}, we show that once our propositions are true for homogeneous $t_i$, they are also true for heterogeneous $t_i$.


\subsection{Preliminary results}\label{subsec:preliminary results}
This section studies the market under fixed prices and wages and fixed $N$. In these circumstances, for each region $i$, we have $\lambda_i(p)=\bar{\lambda}_i f(p)$ where $p$ is the uniform price across regions. Therefore, we will suppress the notation on prices and represent the arrival rate of demand in region $i$ by $\lambda_i$ rather than $\lambda_i(p_i)$. This makes the primitives of the market $(\lambda,N,t)$ where $\lambda$ is the vector of all $\lambda_i$ demand arrival rates, $N$ is the fixed total number of drivers, and $t$ is the uniform region size which is a scalar in this subsection but will be heterogeneous in \cref{subsec: full proofs of driver behavior}. 

\begin{definition}\label{def_undersupply_Theory}
We say allocation $n$ is under-supplied in region  $j$, relative to region $i$, if we have $A_j(n_j)<A_i(n_i)$. Under fixed prices and wages, this is equivalent to:

$$\frac{n_j}{\lambda_j}<\frac{n_{i}}{\lambda_{i}}$$

The ``degree of under-supply'' in region $j$ relative to region $i$ is defined by $\kappa_{ji}=\frac{A_i(n_i)}{A_j(n_j)}= \frac{\frac{n_{i}}{\lambda_{i}}}{\frac{n_j}{\lambda_j}}$.

\end{definition}

In general, when developing the preliminary results, we do not use the notation on access and work directly with $n_i$ and $\lambda_i$. In \cref{subsec: proofs when region size homogeneous} and \cref{subsec: full proofs of driver behavior}, we will see how these results imply our results on access in the main text.

\subsubsection{Statements of Preliminary Results when there are Only Two Regions}\label{subsec: statements two regions}
 
 In this section, we present our results for the case of $I=2$.  We will present two results. First, if the demand arrival rate in region 1 is strictly larger than that of region 2, then in any all-regions equilibrium, region 2 will be strictly under-supplied. Second, we show that the under-supply problem in region 2 is mitigated as the size of the platform increases, holding fixed the ratio between $\lambda_1$ and $\lambda_2$.

First we give a result that helps to visually understand an all-regions equilibrium.
 
\begin{proposition}\label{proposition:focal}
At any all-regions equilibrium, the wait times in the two regions are equal. Also the wait time for each region is locally increasing in the number of drivers present in that region.
\end{proposition}
\textbf{Proof.} If $W_1(n_1)\neq W_2(n_2)$, then,  given the wait time functions are  continuous, a small mass of drivers can relocate from the region with the higher wait time  to the  region with the lower wait time and  be strictly  better  off. Thus, at  equilirium allocation $n^*$, we have  $W_1(n^*_1)= W_2(n^*_2)$. Next, if at equilibrium, the wait time curve in region $i$ is strictly decreasing, then a small mass  of drivers from  region $j$ can relocate to $i$ and become strictly  better off.$\blacksquare$

Next, we introduce a result that speaks to the existence and uniqueness of an all-regions equilibrium.
 
 \begin{proposition}\label{proposition:multi_region_existence}
 There is exactly one all-regions equilibrium if assumptions (A1) to (A3) hold. Otherwise, there is no all-regions equilibrium.
 \begin{description}
\item[(A1)] $N \ge \sqrt{\lambda_1 t} + \sqrt{\lambda_2 t}$
\item[(A2,A3)] $2 \sqrt{\dfrac{t}{\lambda_j}} \le \dfrac{N-\sqrt{\lambda_j t}}{\lambda_i}	+ \dfrac{t}{N-\sqrt{\lambda_j t}}$ for $j=1,2$ and $i=3-j$
\end{description}
 \end{proposition}
 
\begin{figure}[!h]
\centering
\caption{Wait Time and Driver Allocation. An all-regions equilibrium exists only in  panel (c)}\label{fig:equilibria}
\centering
\vspace{-0.2in}
\subfigure[No All-Regions Equilibrium]{\includegraphics[width=0.48\textwidth]{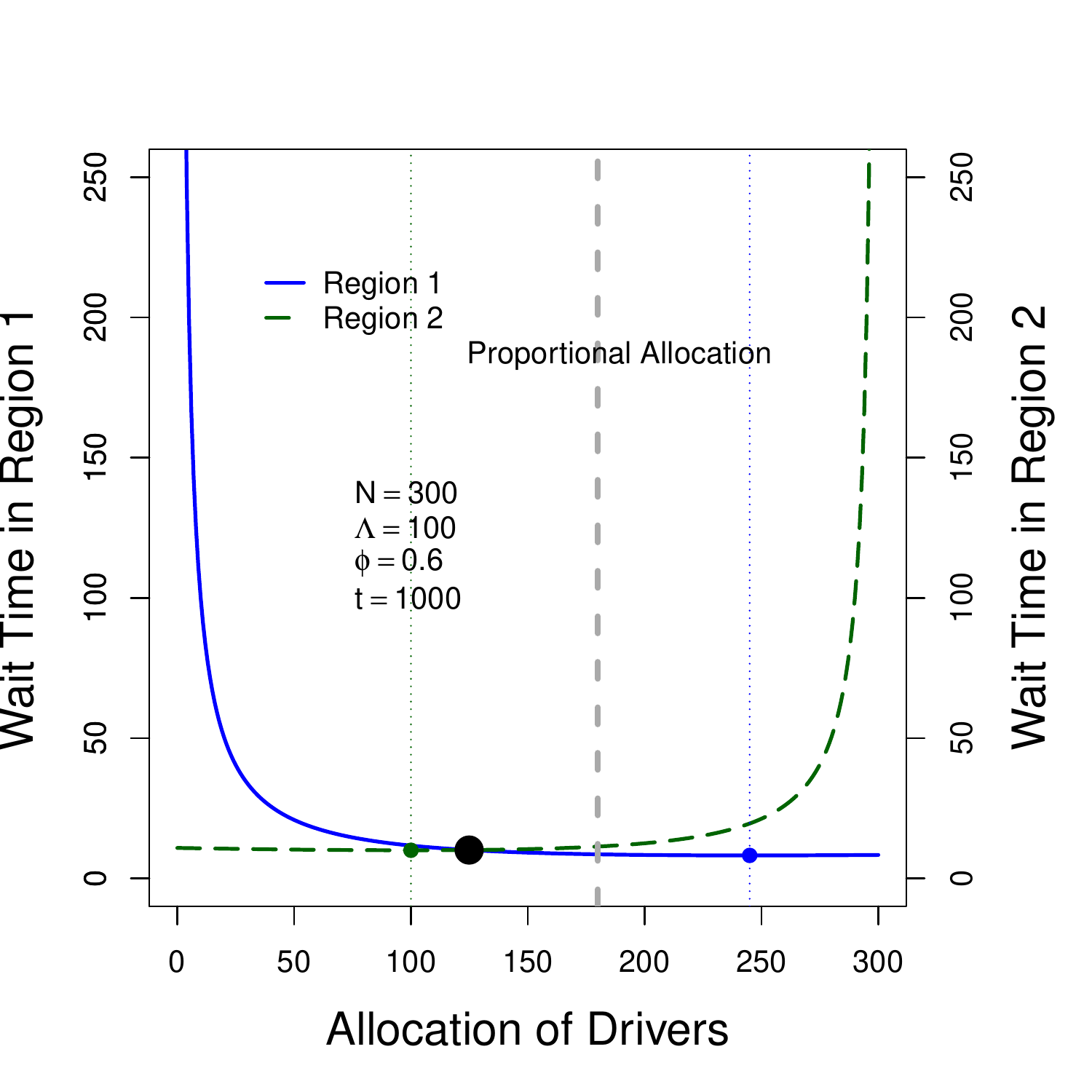}}\label{fig:eqC}
\subfigure[No All-Regions Equilibrium]{\includegraphics[width=0.48\textwidth]{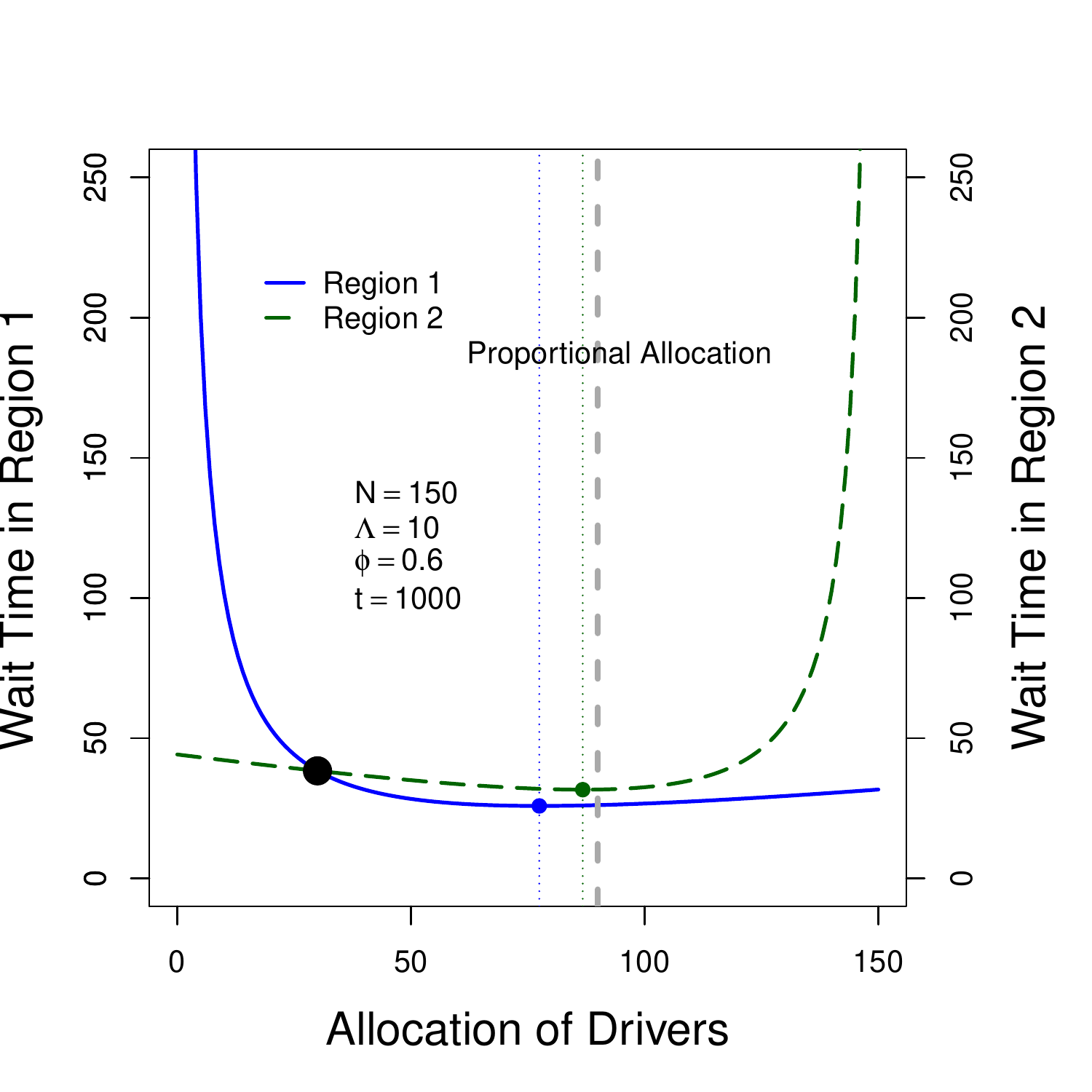}}\label{fig:eqB}\\
\subfigure[Unique All-Regions Equilibrium]{\includegraphics[width=0.48\textwidth]{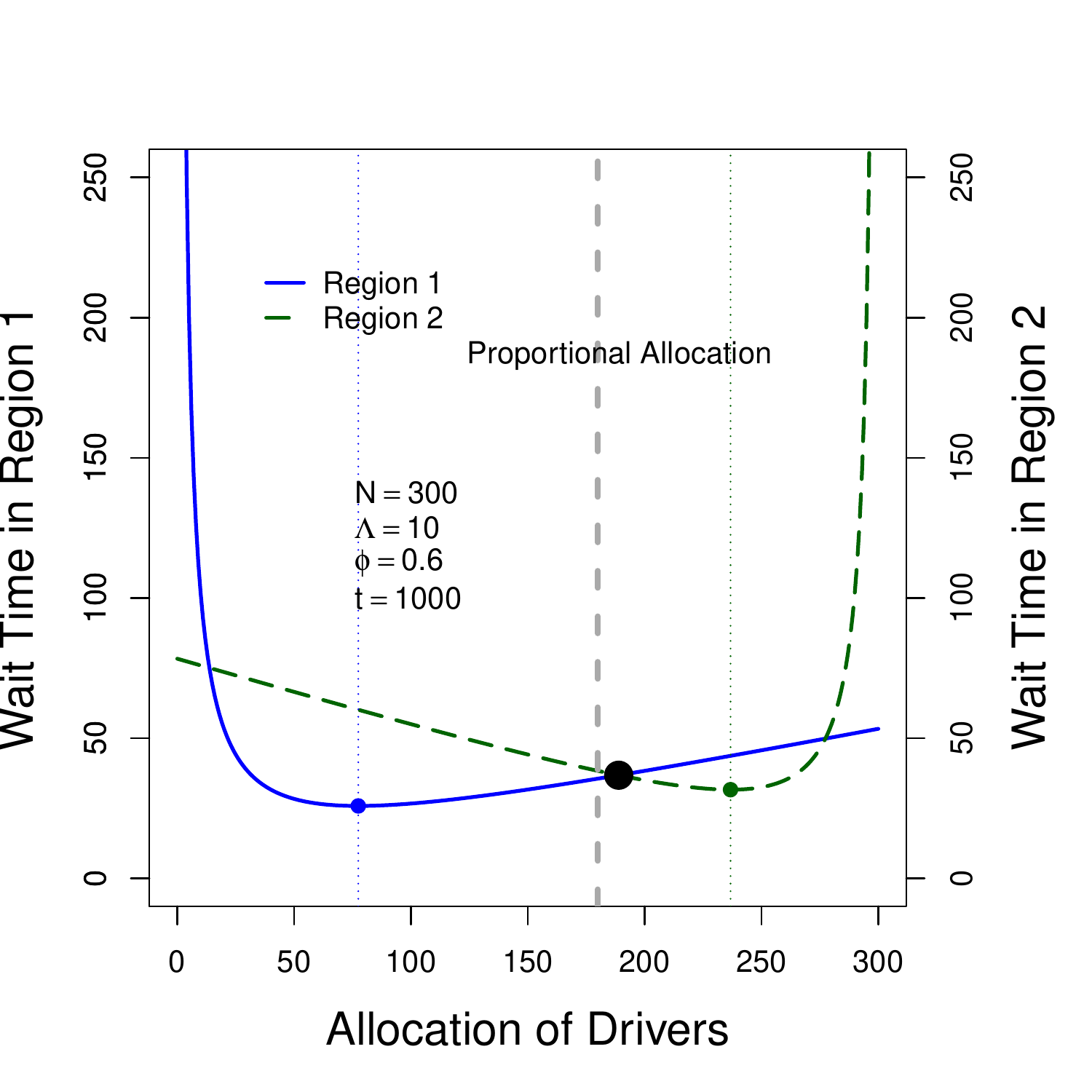}}\label{fig:eqA}
\end{figure}

 \Cref{fig:equilibria} visually illustrates Propositions \ref{proposition:focal} and \ref{proposition:multi_region_existence}. In each  panel, the wait time curves for the two regions are plotted opposite from  each other. In each region,  the wait time is initially decreasing in the total number  of drivers present in that region due to  the decrease  it causes in pickup  times. But as the region  gets more drivers,  the effect on  pickup  time dwindles and  overall wait time increases  due to increased  idle time for drivers.\footnote{Total wait time curves being U-shaped has been mentioned in other studies (such as \cite{weyl2017surge}). To our knowledge, this curve and the U-shaped assumption on it are used by ride-share platforms  in the determination of various strategies including  surge pricing.} Each point on the horizontal axis of the graph corresponds to a driver allocation between the two regions. One such  point is the ``demand-proportional'' allocation which satisfies $\frac{n_1}{\lambda_1}=\frac{n_2}{\lambda_2}$. This allocation  is shown in  the figure  by a dashed vertical gray line. At each  point,  the solid blue line shows the total wait time in region 1, and the dashed green line gives the total  wait time in region 2.
 
 Translated to these graphical  terms, \cref{proposition:focal} states that an all-regions equilibrium is a point of intersection between the two wait-time curves, at which  both curves are increasing. Among  the  panels of \cref{fig:eqA}, such  equilibrium only exists in panel (c).\footnote{One can verify that in all panels of \cref{fig:eqA}, allocations that put all drivers in one of the two regions are in fact equilibria. To illustrate why, note that under  allocation $(n_1,n_2)=(N,0)$, the wait time at region 2 is $\infty$ due to high pickup time. Thus,  no driver has an incentive to move from region 1 to region 2.} \cref{proposition:multi_region_existence} explains why. In order for an all-regions equilibrium to exist, there should exist allocations for which the total wait time at each region is increasing in the number of drivers present in that region. This is what assumption  \textbf{(A1)} requires. Graphically, the trough for the wait time curve in region 2 (emphasized by a green circle) should be to the right of the trough of the wait time in region 1 (blue circle). Panel (a) in \cref{fig:eqA} lacks this feature and, hence, also lacks an all-regions equilibrium.  In addition to  \textbf{(A1)}, the existence of an all-regions equilibrium  would also require that the  two wait-time  curves do intersect over  the range in  which they are  both increasing. In  order for this to happen, we  require assumptions  \textbf{(A2)}  and \textbf{(A3)}. They require that, under the allocation that minimizes  the total wait time in  region 1, the total wait time  in  region 2  be higher than that  in  region 1. They impose a similar  condition  on  the allocation   that minimizes the total wait  time  in region 2. Graphically, they require that the total-wait-time curve for region 2 (the green dashed line) be above the trough of the wait-time  curve in  region 1  (the blue circle), and vice versa. Panel (c) satisfies both \textbf{(A2)} and  \textbf{(A3)} and, hence, has an all-regions equilibrium given by the intersection between  the two  wait-time  curves, emphasized by a large black circle. Panel (b), although satisfying \textbf{(A1)}, has  the wait time curve for  region  1 pass below the trough of the  wait  time curve in region 2. Therefore, there is no  all-regions equilibrium  in  panel  (b).
 
 The reason why different panels in \cref{fig:eqA} differ in terms of having an all-regions equilibrium is that they pertain to different market primitives $(\lambda,N,t)$ (in the figure as well as some of the proofs in the appendix, instead of its components $\lambda_1$ and $\lambda_2$, the vector $\lambda$ is represented by total demand $\Lambda=\lambda_1+\lambda_2$ and share of region 1 from demand $\phi=\frac{\lambda_1}{\Lambda}$). The figures are already suggestive of what affects the existence of an  all-regions  equilibrium (e.g., a  large enough $N$ is necessary) or where the all-regions  equilibrium is located when it  exists (to the right of the gray  dashed line --i.e., the demand-proportional allocation-- instead  of on it  due to  agglomeration  of  drivers in region  1). Our next  results in  this  section formalize and generalize such observations  from  the figure and add other    results describing the role of market thickness.



\begin{proposition}\label{prop:clustering}
Suppose that $\lambda_1>\lambda_2$ and that an all-regions equilibrium $(n_1^*,n_2^*)$ exists. In that case, the all-regions equilibrium is strictly under-supplied in region 2:

$$\frac{n_1^*}{\lambda_1}>\frac{n_2^*}{\lambda_2}$$
\end{proposition}

 To illustrate, if region 1 has 80\% of the demand, then, in equilibrium, 90\% of the drivers might prefer to drive in region 1. This result coincides with our empirical observation that the relative outflow was greater in busier areas than in less busy areas.  This can be  graphically seen  in  \cref{fig:eqC} panel (c): the equilibrium is  to the right of  the  gray dashed line representing the proportional  allocation.

We now turn to our second result which speaks to the impact of market thickness. We prove that ``making the market thicker'' will decrease the extent of geographical inequity in supply. The next two results show this, respectively, for thickening the market on both sides (increasing  the number of drivers and all-regional demand arrival rates) and thickening it on one side (increasing the number of drivers only).


\begin{proposition}\label{prop_impactOfSize}
Suppose that $\lambda_1>\lambda_2$, and $(n_1^*,n_2^*)$ is the all-regions EQ under $(\lambda_1,\lambda_2,N,t)$. Consider scaling up the platform size by $\gamma>1$ to $(\lambda'_1,\lambda'_2,N',t)=(\gamma\lambda_1,\gamma\lambda_2,\gamma N,t)$. Under these new primitives, an all-regions equilibrium exists and under-supply in region 2 decreases with scaling up, i.e., $\frac{n_1'^*}{N'}<\frac{n_1^*}{N}$
In particular, as $\gamma\rightarrow\infty$,the relative under-supply in region 2, $\kappa_{21}$, tends to zero.
\end{proposition}

Our next proposition proves similar results to those shown in  \cref{prop_impactOfSize}, but this time for thickening  the market only  on  one  side.

\begin{proposition}\label{prop_impactOfSize_SupplyOnly}
Suppose that $\lambda_1>\lambda_2$ and $(n_1^*,n_2^*)$ is the all-regions EQ under $(\lambda_1,\lambda_2,N,t)$. If we scale up to $(\lambda_1,\lambda_2,N',t)$ for some $N'>N$, then an all-regions equilibrium still  exists.  Also, the new equilibrium shows less  under-supply of rides in region 2.
In particular, as $N'\rightarrow\infty$, under-supply in region 2 (and in region 1) tends to zero.
\end{proposition}

\subsubsection{Statements of Preliminary Results when there are $I \geq 2$ Regions}\label{subsec_results more regions}

Our main result in this section extends all of the results presented so far from two regions to any number of regions $I\geq 2$. This theorem is powerful in that it provides, among other results, a description of how the market responds to a changed thickness, at the most granular level. That is, it describes what happens to the supply ratio between \textit{any} two regions $i,j$. As formalized below, the proposition shows that the market responds to a ``global thinning'' by further agglomerating the supply at the thickest ``local markets.'' 

\begin{theorem}\label{prop_extensionKregions}
In the general version of the game (i.e., $I\geq2$), the following statements are true:
\begin{enumerate}
    \item For an all-regions equilibrium, the total wait time is equal across all $I$ regions. Also, at the equilibrium allocation, the total-wait-time curve for any region is strictly increasing in the number of drivers present in that region.
    \item Any all-regions equilibrium $n^*=(n^*_1,...,n^*_I)$ is unique.
    \item At any all-regions equilibrium, for any $i<j$, we have $\frac{n^*_i}{\lambda_i}\geq\frac{n^*_j}{\lambda_j}$. The inequality is strict if and only if $\lambda_i>\lambda_j$.
    \item Suppose an all regions equilibrium $n^*=(n^*_1,...,n^*_I)$ exists under primitives $(\lambda,N,t)$ where $\lambda=(\lambda_1,...,\lambda_I)$. Then, if supply and demand both scale up, that is, under new primitives $(\gamma\lambda,\gamma N,t)$ with $\gamma>1$, we have:
    \begin{itemize}
        \item An all-regions equilibrium $n^{*'}=(n^{*'}_1,...,n^{*'}_I)$ exists.
        \item The new equilibrium $n^{*'}$ shows less geographical supply inequity than $n^*$ in the sense that for any $i<j$, we have $1\leq \frac{\frac{n^{*'}_i}{\lambda_i}}{\frac{n^{*'}_j}{\lambda_j}}\leq \frac{\frac{n^*_i}{\lambda_i}}{\frac{n^*_j}{\lambda_j}}$. Both inequalities are strict if and only if $\lambda_i>\lambda_j$.
        \item All $\frac{\frac{n^{*'}_i}{\lambda_i}}{\frac{n^{*'}_j}{\lambda_j}}$ tend to 1 as $\gamma\rightarrow\infty$
    \end{itemize}
    \item The same statement is true if instead of proportionally scaling up both $\lambda$ and $N$, we scale up only $N$.
\end{enumerate}
\end{theorem}



\subsubsection{Proofs of Preliminary Results}\label{subsec_Extension}
\textbf{Proof of \cref{proposition:multi_region_existence}.}
First, we prove the necessity part, and then sufficiency and uniqueness. 
\begin{description}
\item[Necessity:] 
We prove necessity of (A1)-(A3) by contradiction.
First, if (A1) is not satisfied, then we show that the wait time curves can only intersect when at least one of them  is decreasing. To see  this, note that taking the first order condition on \cref{eq:wait} shows the wait time  curve in  each region $i$ is minimized at $n_i^{min}=\sqrt{\lambda_i  t}$. Thus, condition  (A1) simply requires that $N\geq n_1^{min}+n_2^{min}$. Without (A1), there would be  no possible allocation  of  drivers under which  the total wait time in  each region  is  increasing in  the number  of drivers present  in  that region. Therefore, by \cref{proposition:focal}, there would be no all-regions equilibrium. Next, suppose condition (A2) were not true. Thus, at the minimum wait time for region $1$, i.e. at the allocation $(n_1=n_1^{min},n_2=N-n_1^{min})$, the wait time for region $1$ is higher than region $2$. Thus, the wait time curves can only intersect in the decreasing region for market $1$, which we know cannot be an all-regions equilibrium.
The necessity of (A3) is similar to (A2).

\item[Sufficiency:] Observe that when (A2) is true, $W_1(n_1^{min})< W_2(N-n_1^{min})$. Similarly from (A3), we have $W_2(n_2^{min})< W_1(N-n_2^{min})$. We know that for $n_1>n_1^{min}$, $W_1$ is an increasing function, and similar is the case for $W_2$. Since we have a reversal in relative magnitude for $W_1$ and $W_2$, and since the two curves are continuous, we must have an intersection of the curves between $n_1^{min}$ and $n_2^{min}$, when both wait time curves are increasing. Such an intersection permits no profitable deviation by switching to the other market for any driver, and is thus an all-regions equilibrium.

\item[Uniqueness:]  Both wait time curves are monotonic for the region $n_1>n_1^{min}$ and $n_2=(N-n_1)>n_2^{min}$, implying that there can only be one intersection between the curves when they are both increasing.
\end{description}

Together, these conditions are proven equivalent to existence and uniqueness of an all-regions equilibrium. In such a case, we can characterize the all-regions equilibrium by equating the wait time distributions.\footnote{In practice, we obtain the allocation equating wait times, i.e. solving $W_1(n) - W_2(N-n)=0$, which is equivalent to identifying the roots of the polynomial equation below:
\begin{eqnarray*}
-n^3 (\lambda_1 + \lambda_2) + n^2 (2N \lambda_1 + N \lambda_2) - n (N^2 \lambda_1 + 2t \lambda_1 \lambda_2) + Nt \lambda_1 \lambda_2 = 0
\end{eqnarray*}
By Descartes' rule of signs, this equation (i.e. the numerator) has potentially 3 positive roots. In the case of multiple roots, only the one that lies between the minimum points of the wait time curves where both curves are increasing is the symmetric equilibrium. See \Cref{proposition:focal}.
} $\blacksquare$

To Prove \Cref{prop:clustering}, we first introduce the following Lemma.
\begin{lemma}\label{lemma:proportional}
When (A1)-(A3) are satisfied and when drivers are allocated proportionally to demand, the proportional allocation lies between the  minimum wait times for the two regions: $n_1^{min} < \phi N < N-n_2^{min}$.
\end{lemma}

where $\phi$, as mentioned \cref{subsec: statements two regions}, is defined as $\phi=\frac{\lambda_1}{\lambda_1+\lambda_2}$. By our assumption  $\lambda_1>\lambda_2$, which came without loss of generality,  we  have $\phi>\frac{1}{2}$. In graphical terms represented by \cref{fig:eqA}, this lemma says the vertical dashed line  representing the proportional demand will fall between the troughs of the  two wait-time curves.

\begin{lproof}
First, we prove that the proportional allocation line lies between the two minima.
$n_1^{min}=\sqrt{\lambda_1 t}$ and $n_2^{min}=\sqrt{\lambda_2 t}$. Denote the total demand across both locations as $\Lambda = \lambda_1+\lambda_2$ and the fraction of demand in the (higher-demand) location $1$ to be $\phi = \frac{\lambda_1}{\Lambda} > \frac{1}{2}$.

For proportional allocation to be situated between the two minimums on the graph, the following conditions need to hold:
\begin{description}
\item[C1:] $\phi N > n_1^{min} = \sqrt{\lambda_1 t} = \sqrt{\phi \Lambda t}$ $\implies$ $N>\sqrt{\frac{\Lambda t}{\phi}}$
\item[C2:] $(1-\phi) N > n_2^{min} = \sqrt{\lambda_2 t} = \sqrt{(1-\phi) \Lambda t}$ $\implies$ $N>\sqrt{\frac{\Lambda t}{1-\phi}}$
\end{description}

Observe that since $\phi > \frac{1}{2}$, $C2 \implies C1$. Thus, when the demand is more skewed (higher $\phi$), we need to have a larger platform size for condition (C2) to be satisfied.

We now prove that assumption (A1) + (A3) $\implies$ (C2).
Observe that condition that shows up is the following:
Assumption (A3) implies
\begin{equation*}
2 \sqrt{\frac{t}{(1-\phi)\Lambda}} < \frac{N-\sqrt{(1-\phi)\Lambda t}}{\phi \Lambda} + \frac{t}{N-\sqrt{(1-\phi)\Lambda t}}
\end{equation*}
The second term on the RHS can be bounded as: $ \frac{t}{N-\sqrt{(1-\phi)\Lambda t}} < \frac{t}{\sqrt{\phi \Lambda t}}$, since $N>\sqrt{\phi \Lambda t} + \sqrt{(1-\phi)\Lambda t}$  by assumption (A1).

Thus assumption (A3) implies the following:
\begin{equation*}
\frac{N-\sqrt{(1-\phi)\Lambda t}}{\phi \Lambda}  > 2 \sqrt{\frac{t}{(1-\phi)\Lambda}} - \frac{t}{\sqrt{\phi \Lambda t}} \\
\implies
N  >\sqrt{\Lambda t} \left(2 \phi \sqrt{\frac{1}{1-\phi}} -\sqrt{\phi} +\sqrt{1-\phi}\right)
\end{equation*}

Next, we prove that the above inequality implies condition (C2), which stated that $N>\sqrt{\frac{\Lambda t}{1-\phi}}$. Thus, we need to prove the following:
\begin{align*}
&2 \phi \sqrt{\frac{1}{1-\phi}} -\sqrt{\phi} +\sqrt{1-\phi} > \sqrt{\frac{1}{1-\phi}} \Leftrightarrow \frac{2 \phi -1}{\sqrt{1-\phi}} -\sqrt{\phi} +\sqrt{1-\phi} >0\\
&\Leftrightarrow \sqrt{\phi}( \sqrt{\phi}-  \sqrt{1-\phi})  >0 
\end{align*}
Observe that the last inequality must be true given our assumption  that $\phi>\frac{1}{2}$, so (A1) + (A3) $\implies$ (C2). This finishes the proof of the lemma.\qed
\end{lproof}

\textbf{Proof of Proposition \ref{prop:clustering}.} 
By \Cref{lemma:proportional},  the demand-proportional allocation is in between the minimum wait times for both regions. We show the proportional allocation or any point to  the  left of it  (i.e., an alloction with $n_1\leq \phi N$) cannot be  an  all-regions equilibrium. This, combined with the assumption in  the  proposition that an all-regions equilibrium exists, implies that the all-regions equilibrium should be to the right of the proportional allocation. That is: $\frac{n_1^*}{\lambda_1}>\frac{n_2^*}{\lambda_2}$.

To see why no all-regions equilibrium  can be found weakly to the left of the proportional allocation, note  that  at proportional allocation $n=\phi N$, region 2 wait time is higher than region 1, i.e. $W_2((1-\phi)N) = \frac{N}{\Lambda} + \frac{t}{(1-\phi)N}  > W_1(\phi N)= \frac{N}{\Lambda} + \frac{t}{(\phi)N} $ since $\phi>\frac{1}{2}$ .  As we move left, region 2's wait time increases further, while region 1's wait time decreases until we reach the minimum wait time for region 1, $W(n_1^{min})$. Thus, the divergence between the two regions increases. For the wait time curves to intersect, it must be in region 1's decreasing wait time region. We know from \Cref{proposition:focal} that such an intersection will \textbf{not} be an all-regions equilibrium. \cref{fig:eqA}  panel (c) should help illustrate this  point. This completes the proof of the  proposition.
 $\blacksquare$

\begin{lemma}\label{lemma:existencescaling}
When an all-regions equilibrium exists for a ridesharing platform with $N$ drivers facing demand $\phi \Lambda$ and $(1-\phi)\Lambda$ in the two regions:
\begin{enumerate}
\item an all-regions equilibrium also exists when demand is unchanged and there are $N' = \gamma N$ drivers where $\gamma>1$.
\item an all-regions equilibrium also exists when both the demand and number of drivers are scaled up by a common factor $\gamma>1$ to $N' = \gamma N$ and $\Lambda' = \gamma \Lambda$.
\end{enumerate}
\end{lemma}

\begin{lproof}
Consider the equivalent conditions required for the existence of an all-regions equilibrium, characterized by assumptions (A1)-(A3). Below, we show that if the conditions are satisfied for a given $(N,\Lambda)$, then they must be satisfied for (a) $(N',\Lambda')=(\gamma N,\Lambda)$ as well as (b) $(N',\Lambda')=(\gamma N,\gamma\Lambda)$.

First, consider (A1). The proof of (a) is immediate. For (b), we observe that:
\begin{eqnarray*}
\gamma N > \sqrt{\gamma} \sqrt{\Lambda t} \left(\sqrt{\phi} + \sqrt{1-\phi} \right)
\end{eqnarray*}
holds since $\gamma>1$ and (A1) holds for $(N,\Lambda)$.

Next, we prove (A2). The proof of (A3) is similar to that of (A2) and is omitted.

For (A2), first we denote the following function $\phi$:
\begin{eqnarray*}
\psi(\rho) = \frac{\rho N - \sqrt{\phi \Lambda t}}{(1-\phi)\Lambda} + \frac{t}{\rho N - \phi \Lambda t} 
\end{eqnarray*}
We prove that $\psi$ is increasing in $\rho$, or $\frac{d\phi}{d\rho}>0$. Observe that:
\begin{eqnarray*}
\frac{d\psi}{d\rho} = N \left( \frac{1}{(1-\phi)\Lambda} - \frac{t}{\left(\rho N - \sqrt{\phi \Lambda t}\right)^2} \right)
\end{eqnarray*}
After some algebra and applying (A1), we obtain $\frac{d\psi}{d\rho} >0$.

Now, for part (a), observe that setting $\rho = \frac{N'}{N}>1$ implies that, in (A2), the RHS increases and the LHS does not change implying that (A2) still holds for $(N',\Lambda')=(kN,\Lambda)$.

Next, for (b), observe that applying (A2) with $(N',\Lambda')=(\gamma N,\gamma\Lambda)$ gives us:
\begin{eqnarray*}
2 \sqrt{\frac{t}{\gamma \phi \Lambda}} &< \frac{\gamma N - \sqrt{\phi \gamma \Lambda t}}{(1-\phi)\gamma\Lambda} + \frac{t}{\gamma N - \phi \gamma \Lambda t}  \Leftrightarrow
2 \sqrt{\frac{t}{ \phi \Lambda}} &< \frac{\sqrt{\gamma} N - \sqrt{\phi \Lambda t}}{(1-\phi)\Lambda} + \frac{t}{\sqrt{\gamma}N - \phi \Lambda t} 
\end{eqnarray*}
We need to prove the above holds whenever (A1)-(A3) hold.

Since we know that $\psi$ is an increasing function, we know that $\psi(\sqrt{\gamma})>\psi(1)$ when $\gamma>1$. But if we write out $\psi(\sqrt{\gamma})>\psi(1)$, it gives  us exactly the expression we needed to be true:

$$2 \sqrt{\frac{t}{ \phi \Lambda}} < \frac{\sqrt{\gamma} N - \sqrt{\phi \Lambda t}}{(1-\phi)\Lambda} + \frac{t}{\sqrt{\gamma}N - \phi \Lambda t} $$

Thus, when $(N',\Lambda')=(\gamma  N,\gamma\Lambda)$, we find that (A2) holds for $(N',\Lambda')$.

Thus, (A1)-(A3) hold under the conditions detailed in the Lemma. $\qed$
\end{lproof}

\textbf{Proof of Proposition \ref{prop_impactOfSize}.} The proof of existence of all-regions equilibrium  under the new  model primitives obtains from \Cref{lemma:existencescaling} above. To prove that the equilibrium  supply ratios tilts towards region 2, we first claim (but skip the straightforward proof) that if all the primitives of the model  $(\lambda,N,t)$ are multiplied by same scaling factor, the existence of an all-regions equilibrium  as well  as all of the $\frac{n^*_i}{n^*_j}$ ratios (and, by construction, all $\frac{\lambda_i}{\lambda_j}$ ratios) are preserved. Therefore, in this proof, instead of a multiplication of $N$ and $\lambda$ by a factor of $\gamma>1$, we focus on fixing $N$ and $\lambda$ and, instead, replacing  $t$ by $t\frac{1}{\gamma}$.

For the all-regions equilibrium  $(n^*_1,n^*_2)$, define $\alpha=\frac{n^*_1}{N}$. We know from \Cref{prop:clustering} that $\alpha>\phi$. The quilibrium condition, written in terms  of $\alpha$ will be:
\begin{eqnarray}
W^d(\alpha) \equiv W_1(\alpha N)-W_2((1-\alpha) N)=-\frac{(1-\alpha ) N}{(1-\phi ) \Lambda }+\frac{(\alpha)  N}{\phi  \Lambda
   }+\frac{t}{(1-\alpha) N}+\frac{t}{(\alpha)  N}=0
\end{eqnarray}

where $W^d$ represents the difference between  the total wait times between  the two  regions, which should  be zero at  the equilibrium. We now  use the implicit function theorem  to show that $\alpha$ increases as  we increase $t$, which would prove  the proposition.

\begin{eqnarray}
\dfrac{d \alpha}{d t} =-\dfrac{\dfrac{\partial W^d}{\partial t} }{\dfrac{\partial W^d}{\partial \alpha} } =\frac{(1-\alpha ) \alpha  (2 \alpha -1) (1-\phi) \phi  \Lambda }{(\alpha -1)^2 \alpha ^2 N^2+(2
   (\alpha -1) \alpha +1) (\phi -1) \phi  \Lambda  t}
\end{eqnarray}
The numerator is positive since $\alpha>\frac{1}{2}$. Thus, the sign of $\dfrac{d \alpha}{d t}$ is determined by the denominator. Below, we prove that the denominator is positive as well. The argument takes the following steps:
\begin{enumerate}
\item Define the denominator as $g(\alpha) = (\alpha -1)^2 \alpha ^2 N^2+(2 (\alpha -1) \alpha +1) (\phi -1) \phi  \Lambda  t$. 
\item Observe that $g'(\alpha) =- 2 (2 \alpha -1) \left((1-\alpha) \alpha  N^2+(1-\phi) \phi  \Lambda  t\right) < 0$, implying that $g(\alpha)$ is a decreasing function.
\item Since $\alpha \in\left[\phi,1-\frac{n_2^{min}}{N}\right]$, the inequality $g(\alpha) \geq g\left(1-\frac{n_2^{min}}{N}\right)$ has to hold.
\item We prove that $\min g(\alpha) = g\left(1-\frac{n_2^{min}}{N}\right)>0$. 
\begin{eqnarray}
g\left(1-\frac{n_2^{min}}{N}\right) &= \frac{\phi^2 \Lambda  t }{N^2}\left(N^2-2 N \sqrt{\phi  \Lambda  t}+(2 \phi -1) \Lambda  t\right)\\
&= \frac{\phi^2 \Lambda  t }{N^2} \left((N-\sqrt{\phi  \Lambda  t})^2-t \Lambda (1-\phi)\right)
\end{eqnarray}
where the term in parentheses is positive directly from assumption \textbf{(A1)}.
\end{enumerate}

Thus, we know that $\dfrac{d \alpha}{d t} >0$ implying that  as $t$ increases, the proportion of supply going to the higher-demand region is greater. $\blacksquare$

\textbf{Proof of Proposition \ref{prop_impactOfSize_SupplyOnly}.} Note that a scale-up in $N$ can be thought of as a scale-up in $(\lambda_1,\lambda_2,N)$, followed by a scale back down in $(\lambda_1,\lambda_2)$. From Proposition \ref{prop_impactOfSize}, we know that the first scale-up (i) preserves the existence of an all-regions equilibrium and also (ii) makes it strictly less under-supplied in region 2. Therefore, the proof of Proposition (\ref{prop_impactOfSize_SupplyOnly}) will be complete if we show that the second scale back down also preserves the existence of an all-regions equilibrium and makes it less under-supplied inn region 2.

To see this, suppose $(n_1^*,n_2^*)$ is the all-regions equilibrium under $(\lambda_1,\lambda_2,N,t)$. Let $\lambda_i'=\frac{\lambda_i}{\gamma}$ for $i\in\{1,2\}$ and some $\gamma>1$. We will now show that under $(\lambda_1',\lambda_2',N,t)$, there is an all-regions equilibrium with strictly less under-supply in region 2 than what is implied by $(n_1^*,n_2^*)$.

\begin{lemma}\label{lem_oldEQ1}
The following statements are true about the ``old'' equilibrium allocation $(n_1^*,n_2^*)$ under the ``new'' parameters $(\lambda_1',\lambda_2',N,t)$:
\begin{enumerate}
    \item The total wait function $W_2(n)$ is strictly increasing at $n=n_2^*$.
    \item At the allocation $(n_1^*,n_2^*)$, the wait time in region 1 is strictly higher than that in region 2. That is, $W_1(n_1^*)>W_2(n_2^*)$.
    \item The total wait function $W_1(n)$ is strictly increasing at $n=N\times\frac{\lambda_1'}{\lambda_1'+\lambda_2'}$.
    \item At the allocation proportional to demand, the wait time in region 2 is strictly larger than that in region 1. That is, if we set $n_i=N\times\frac{\lambda_i'}{\lambda_1'+\lambda_2'}$, then $W_2(n_2)>W_1(n_1)$.
\end{enumerate}

\end{lemma}

\textbf{Proof of Lemma \ref{lem_oldEQ1}.} We start by statement 1. To see this, first note that from the assumption that $(n_1^*,n_2^*)$ was the all-regions equilibrium under $(\lambda_1,\lambda_2,N,t)$, we know $n_2^*$ has to be strictly larger than where the old $W_2$ function reached its trough. That is, $n_2^*>\sqrt{\lambda_2t}$. Now, given $\lambda_2'<\lambda_2$, we it is also the case that $n_2^*>\sqrt{\lambda_2't}$. Therefore, the new $W_2$ function is also strictly increasing at $n=n_2^*$.

We now turn to statement 2. Given that $(n_1^*,n_2^*)$ was the all-regions equilibrium under the old parameters, the total wait times in the two regions were equal to each other. That is:

\begin{equation}\label{eq_lem_oldEQ1_1}
\frac{n_1^*}{\lambda_1}+\frac{t}{n_1^*}=\frac{n_2^*}{\lambda_2}+\frac{t}{n_2^*}    
\end{equation}

Given Proposition (\ref{prop:clustering}), we know that $n_1^*>n_2^*$, therefore: $\frac{t}{n_1^*}<\frac{t}{n_2^*}$. This latter inequality, combined with equality (\ref{eq_lem_oldEQ1_1}), implies $\frac{n_1^*}{\lambda_1}>\frac{n_2^*}{\lambda_2}$. The sign of this inequality is preserved if we multiply both sides of it by the positive number $\gamma-1$. That is: $(\gamma-1)\times\frac{n_1^*}{\lambda_1}>(\gamma-1)\times\frac{n_2^*}{\lambda_2}$. The size of the inequality is also preserved when we add equal numbers to both sides. Those equal numbers are the two sides of equation (\ref{eq_lem_oldEQ1_1}). This will give us:

\begin{equation}
(\gamma-1)\times\frac{n_1^*}{\lambda_1}+\frac{n_1^*}{\lambda_1}+\frac{t}{n_1^*}>(\gamma-1)\times\frac{n_2^*}{\lambda_2}+\frac{n_2^*}{\lambda_2}+\frac{t}{n_2^*}    
\end{equation}

Therefore:

\begin{equation}
\gamma\times\frac{n_1^*}{\lambda_1}+\frac{t}{n_1^*}>\gamma\times\frac{n_2^*}{\lambda_2}+\frac{t}{n_2^*}    
\end{equation}

which gives us:

\begin{equation}
\frac{n_1^*}{\frac{\lambda_1}{\gamma}}+\frac{t}{n_1^*}>\frac{n_2^*}{\frac{\lambda_2}{\gamma}}+\frac{t}{n_2^*}    
\end{equation}

which, by definition, means:

\begin{equation}
\frac{n_1^*}{\lambda_1'}+\frac{t}{n_1^*}>\frac{n_2^*}{\lambda_2'}+\frac{t}{n_2^*}    
\end{equation}

This proves statement 2.

Next, we turn to statement 3. The argument is similar to that for statement 1. $N\times\frac{\lambda_1}{\lambda_1+\lambda_2}$ was larger than the trough of $W_1$ under the old parameters. Given that the trough gets smaller under the new parameters, it will keep being smaller than $N\times\frac{\lambda_1}{\lambda_1+\lambda_2}$.

Finally, statement 4 is obvious from the proof of Proposition (\ref{prop:clustering}). $\square$

Now notice that Lemma (\ref{lem_oldEQ1}) completes the proof of the proposition. Given that under the allocation $(n_1^*,n_2^*)$, we have $W_1>W_2$, and under the allocation fully proportional to demand, we have $W_1<W_2$, and given that both $W_1$ and $W_2$ are continuous functions, there should be an allocation $(n_1^{*'},n_1^{*2})$ in between the two such that $W_1(n_1^{*'})=W_2(n_2^{*'})$. This was achieved by statements 2 and 4. Now by statement 1, $W_2$ is strictly increasing at $n=n_2^{*'}$ because $n_2^{*'}>n_2^*>\sqrt{\lambda_2't}$. Also, by statement 3, $W_1$ is increasing at $n=n_1^{*'}$ because $n_1^{*'}>N\frac{\lambda_1}{\lambda_1+\lambda_2}>\sqrt{\lambda_1't}$. This implies that $(n_1^{*'},n_2^{*'})$ is the all-regions equilibrium under parameters $(\lambda_1',\lambda_2',N,t)$. Now, given $n_1^{*'}<n_1^*$ and $n_2^{*'}>n_2^*$, it follows that:

$$\kappa^{*'}=\frac{\frac{n_1^{*'}}{\lambda_1'}}{\frac{n_2^{*'}}{\lambda_2'}}<\frac{\frac{n_1^*}{\lambda_1'}}{\frac{n_2^*}{\lambda_2'}}=\frac{\frac{n_1^*}{\lambda_1}}{\frac{n_2^*}{\lambda_2}}=\kappa^*$$

which finishes the proof of the proposition.$\blacksquare$


\textbf{Proof of Theorem (\ref{prop_extensionKregions}).} 
Before stating the induction hypothesis, we add one statement to the five statements of Theorem (\ref{prop_extensionKregions}). The inclusion of this statement and leveraging it in the induction process will be helpful for the proof. We call it statement 6.

\textit{Statement 6. Suppose an all region equilibrium $n^*=(n^*_1,...,n^*_I)$ exists under primitives $(\lambda,N,t)$ where $\lambda=(\lambda_1,...,\lambda_I)$. Then, demand arrival rates are scaled down, that is, under new primitives $(\frac{\lambda}{\gamma},N,t)$ with $\gamma>1$, we have:
    \begin{itemize}
        \item An all-regions equilibrium $n^{*'}=(n^{*'}_1,...,n^{*'}_I)$ exists.
        \item The new equilibrium $n^{*'}$ shows less geographical supply inequity than $n^*$ in the sense that for any $i<j$, we have $\frac{\frac{n^{*'}_i}{\lambda_i}}{\frac{n^{*'}_j}{\lambda_j}}\leq \frac{\frac{n^*_i}{\lambda_i}}{\frac{n^*_j}{\lambda_j}}$. The inequality is strict if and only if $\lambda_i>\lambda_j$.
    \end{itemize}} 

In words, this statement simply says the geographical supply inequity decreases if, all else fixed, all demand arrival rates proportionally decrease. The intuition is that this makes idle times relatively more important than pickup times. 

We can now state the strong induction hypothesis.

\textbf{Induction Hypothesis.} Take some natural number $I_0>2$. If all statements of Theorem (\ref{prop_extensionKregions}), including statement 6 added above, are correct for $I\in\{2,...,I_0-1\}$, then they are also all correct for $I=I_0$.

Now, in order to prove the theorem, we need to take two steps. First, we should prove the basis of the induction process. That is, we must show the theorem holds under $I=2$. Second, we need to prove the induction hypothesis. As for the first step, note that propositions (\ref{proposition:focal}) through (\ref{prop_impactOfSize_SupplyOnly}) do this job. The only statement that is not explicitly proven by those theorem is statement 6. However, the proof of statement 6 was the main building block of the proof of Proposition (\ref{prop_impactOfSize_SupplyOnly}).\footnote{Also, propositions (\ref{proposition:focal}) through (\ref{prop_impactOfSize_SupplyOnly}) assume that $\lambda_1>\lambda_2$ and, hence, leave out the case where $\lambda_1=\lambda_2$. But the proofs for the case where $I=2$ and $\lambda_1=\lambda_2$ are straightforward and we leave them to the reader.}

We now turn to the second and main step of this proof, which is to show that the induction hypothesis is correct (Note that some of the statements are not really proven based on the induction. Nevertheless, we present all of the proofs in this inductive framework since we believe having one induction as well as one non-induction section for the proof will just make it harder to read).

\textbf{Proof of Statement 1.} If the total wait time in region $i$ is strictly higher than that in region $j$, then given the continuity of these wait-time functions, a small enough mass of drivers can leave region $i$ for $j$ and strictly benefit from that, violating the equilibrium assumption. To see why they are increasing, suppose on the contrary, that at the equilibrium allocation, for region $i$, the total wait time is strictly decreasing in the number of drivers in that region. Since drivers are equal across all regions in equilibrium, drivers from any other region $j$ will have the incentive to relocate to region $i$, given that (i) currently region $i$ has the same total wait as they do; and (ii) once they move to region $i$, the total wait time of that region will decrease. This is a violation of the equilibrium assumption. Therefore it has to be that at the equilibrium, the wait times are all increasing in the number of drivers at all regions.$\square$

\textbf{Proof of Statement 2.} Suppose, on the contrary, that there are two different all-regions equilibria $n^*$ and $\Bar{n}$. Given the two vectors are different, there has to be a region $i$ such that $n^*_i\neq \Bar{n}_i$. Without loss of generality, assume $n^*_i< \Bar{n}_i$. Given that, from statement 1, we know the total wait time is increasing at $n^*_i$, and given the fact that the wait time function, once it becomes increasing, it remains strictly increasing, we can say $W_i(n^*_i)< W_i(\Bar{n}_i)$. 

Now, again from statement 1, we know two things. First, $\forall j:\quad W_j(n^*_j)=W_i(n^*_i)\quad\&\quad W_j(\Bar{n}_j)=W_i(\Bar{n}_i)$, which implies: $\forall j:\quad W_j(n^*_j)< W_j(\Bar{n}_j)$. Second, we know that the total wait time function at each region $j$ must be strictly increasing after it hits its trough (which happens weakly  before $n^*_j$). This implies that in order for $\forall j:\quad W_j(n^*_j)< W_j(\Bar{n}_j)$ to hold, it must be that $\forall j:\quad n^*_j< \Bar{n}_j$. Therefore: $$\Sigma_{j=1,...,I_0} \Bar{n}_j > \Sigma_{j=1,...,I_0} n^*_j$$ But this cannot be given that both of the sums should be equal to $N$.$\square$

\textbf{Proof of Statement 3.} Note that the definition of equilibrium is that no driver should have the incentive to relocate from one region to another. This definition, by construction, implies that if $n^*=(n^*_1,...,n^*_{I_0})$ is an equilibrium under $(\lambda,N,t)$, then once we fix $\Tilde{N}=n^*_i+n^*_j$ for some $i,j$ with $i<j$, then the allocation $(n^*_i,n^*_j)$ is itself an equilibrium of the two-region game with primitives $(\lambda_i,\lambda_j,\Tilde{N},t)$. Thus, by Proposition (\ref{prop:clustering}) (or alternatively, by the base of the induction), we know that if $\lambda_i>\lambda_j$, then $\frac{n^*_i}{\lambda_i}>\frac{n^*_j}{\lambda_j}$. Also in case $\lambda_i=\lambda_j$, it is fairly straightforward to verify that $\frac{n^*_i}{\lambda_i}=\frac{n^*_j}{\lambda_j}$. To see this, note that in that case, $\frac{n^*_i}{\lambda_i}=\frac{n^*_j}{\lambda_j}$ if and only if $n^*_i=n^*_j$. It is easy to see that $n^*_i=n^*_j$ is an equilibrium given that it gives the two regions the same total wait time and that at it, the total wait times must be increasing according to previous statements.$\square$

\textbf{Proof of Statement 4.} Before we start the proof of this statement, we note that, similar to the case of Proposition (\ref{prop_impactOfSize}), we can work with primitives $(\lambda, N,\frac{t}{\gamma})$ instead of $(\gamma\lambda,\gamma N,t)$. As a reminder, this is because there is a one-to-one and onto mapping between the equilibria under the two primitives, which preserves all of the $\frac{n^*_i}{\lambda_i}$ values.

We start by proving the first statement. That is, if an all-regions equilibrium exists under $(\lambda, N,t)$, then one does under $(\lambda, N,\frac{t}{\gamma})$ as well. To see this, let us assume that under the ``old'' primitives $(\lambda, N,t)$, the all-regions equilibrium allocation $n^*$ is such that $\forall i\in\{1,...,I_0\}:\quad W_i(n^*_i)=w$. We know this common $w$ must exist from statement 1, and we know it is unique from statement 2. 
We show existence of an equilibrium allocation under the new primitive by first describing two ``partial equilibrium'' allocations. We construct the first partial equilibrium allocation $\Bar{n}=(\Bar{n}_1,...,\Bar{n}_{I_0})$ by fixing $\Bar{n}_1=n^*_1$ and assuming the rest of values $(\Bar{n}_2,...,\Bar{n}_{I_0}$ to be the equilibrium allocation of drivers among regions 2 to $I_0$ under primitives $((\lambda_2,...,\lambda_{I_0}),N-n^*_1,\frac{t}{\gamma})$. In words, this allocation fixes the number of drivers in region 1 (i.e., the region with the highest demand arrival rate $\lambda_1$) at its value under the old primitives but allows the drivers of all other regions to reshuffle themselves among those regions. The second partial equilibrium allocation $\Tilde{n}$ fixes $\Tilde{n}_1=N\times\frac{\lambda_1}{\Sigma_{i=1,...,I_0}\lambda_i}$, and assumes the rest of values $(\Tilde{n}_2,...,\Tilde{n}_{I_0}$ to be the equilibrium allocation of drivers among regions 2 to $I_0$ under primitives $((\lambda_2,...,\lambda_{I_0}),N\times(1-\frac{\lambda_1}{\Sigma_{i=1,...,I_0}\lambda_i}),\frac{t}{\gamma})$. In words, this allocation fixes the total number of drivers in region 1 at the value it would take if drivers were to be allocated fully proportional to demand arrival rates. It then allows the rest of the drivers to reshuffle themselves among other areas under the new primitives. We will use these two partial equilibrium allocations to prove existence of an all-region equilibrium. But first we need to prove the existence of these partial equilibrium allocations themselves. Lemma \ref{lem_induction_partialEQ} below does this job.

\begin{lemma}\label{lem_induction_partialEQ}
Partial equilibrium allocations $\Tilde{n}$ and $\Bar{n}$ described above exist, are unique, and allocate a strictly positive number of drivers to each region.
\end{lemma}

\textbf{Proof of Lemma (\ref{lem_induction_partialEQ}).} We first start from $\Bar{n}$. Note that the assumption of $n^*$ being the equilibrium allocation under $(\lambda,N,t)$, by construction implies that $(n^*_2,...,n^*_{I_0})$ is the unique all-region equilibrium allocation under $((\lambda_2,...,\lambda_{I_0}),N-n^*_1,t)$. Now, given that by our induction assumption all results (including statement  4) hold for $I_0-1$ regions, if the primitives remain the same except that $t$ is divided by some $\gamma>1$, a unique all-region equilibrium will still exist. This is what we were denoting $\bar{n}_2$ through $\bar{n}_{I_0}$.

Next, we turn to $\tilde{n}$ and construct it from $\bar{n}$. We just showed that $(\bar{n}_2,...,\bar{n}_{I_0})$ is  the unique all-regions equilibrium under $((\lambda_2,...,\lambda_{I_0}),N-n^*_1,t)$. Also note that by statement 3, we know $n^*_1>\frac{\lambda_1}{\Sigma_{i=1,...,I_0}\lambda_i})$, which implies $N-n^*_1<N\times(1-\frac{\lambda_1}{\Sigma_{i=1,...,I_0}\lambda_i}))$. Therefore, primitives  $((\lambda_2,...,\lambda_{I_0}),N\times(1-\frac{\lambda_1}{\Sigma_{i=1,...,I_0}\lambda_i}),\frac{t}{\gamma})$ can be constructed from primitives $((\lambda_2,...,\lambda_{I_0}),N-n^*_1,t)$ by increasing the total number of drivers. Given that $\bar{n}$ was the unique all-regions equilibrium allocation under $((\lambda_2,...,\lambda_{I_0}),N-n^*_1,t)$, and given the induction assumption on statement 5 for $I=I_0-1$  regions, we can say that primitives  $((\lambda_2,...,\lambda_{I_0}),N\times(1-\frac{\lambda_1}{\Sigma_{i=1,...,I_0}\lambda_i}),\frac{t}{\gamma})$  also have a unique all-regions equilibrium allocation. This is exactly what was  denoted $\tilde{n}_2,...,\tilde{n}_{I_0}$. This completes the proof of the lemma.$\square$

We now use these two partial equilibrium allocations to show that a unique all-regions equilibrium allocation exists under primitives $(\lambda,N,\frac{t}{\gamma})$. Our next step is to prove the following useful lemma.

\begin{lemma}\label{lem_induction_existence}
At the partial equilibrium allocation $\bar{n}$, the total wait time in region 1 is larger than that in any other region. Conversely, at the partial equilibrium allocation $\tilde{n}$, the total wait time in region 1 is smaller than that in any other region.
\end{lemma}

\textbf{Proof of Lemma (\ref{lem_induction_existence}).} To see why the result holds for $\bar{n}$, note that under the old equilibrium $n^*$ and old primitives $(\lambda,N,t)$, all of the wait times were equal. This means for any $i>1$ we had 
$$\frac{n^*_1}{\lambda_1}+\frac{t}{n^*_1}=\frac{n^*_i}{\lambda_i}+\frac{t}{n^*_i}$$
But given that for all $i>1$ we have $n^*_1\geq n^*_i$ we get the following inequality under the new primitives $(\lambda,N,\frac{t}{\gamma})$:
$$\frac{n^*_1}{\lambda_1}+\frac{\frac{t}{\gamma}}{n^*_1}\geq\frac{n^*_i}{\lambda_i}+\frac{\frac{t}{\gamma}}{n^*_i}$$
Next, note that the main and only difference between allocations $n^*$ and $\bar{n}$ is that under $\bar{n}$, drivers reshuffle among regions 2 to $I_0$ in order to reduce their total wait times. Therefore, there has to be at least one region $j$ such that:
$$\frac{\bar{n}_j}{\lambda_j}+\frac{\frac{t}{\gamma}}{\bar{n}_j}\leq \frac{n^*_j}{\lambda_j}+\frac{\frac{t}{\gamma}}{n^*_j}$$
Combining the above two, we get:
$$\frac{\bar{n}_j}{\lambda_j}+\frac{\frac{t}{\gamma}}{\bar{n}_j}\leq \frac{n^*_1}{\lambda_j}+\frac{\frac{t}{\gamma}}{n^*_1}$$
But the total wait time under $(\lambda,N,\frac{t}{\gamma})$ is equal across regions 2 through $I_0$ under allocation $\bar{n}$. Therefore, the above inequality holds not only for a specific $j$ , but under any $j>1$. This proves the lemma for $\bar{n}$ given that $\bar{n}_1=n^*_1$.

Next, we prove the lemma for $\tilde{n}$. We first show that the wait time in region 1 is smaller than that in region 2 if both get drivers proportional to their demand arrival rates. We then show that the wait time in region 2 under $\tilde{n}_2$ is larger than the wait time in region 2 if region 2 were to get drivers proportional to its demand arrival rate. These two statements, combined, will prove our intended result. To see the first claim, note that the wait time in region 1, if it gets $N\times\frac{\lambda_1}  {\Sigma_{i\in\{1,...,I_0\}}\lambda_i}$  drivers, will be:
$$w_1=   \frac{ N\times\frac{\lambda_1}  {\Sigma_{i\in\{1,...,I_0\}}\lambda_i}}{\lambda_1}+\frac{\frac{t}{\gamma}}{ N\times\frac{\lambda_1}  {\Sigma_{i\in\{1,...,I_0\}}\lambda_i}}   $$which gives:

\begin{equation}
w_1=\frac{N}{\Sigma_{i\in\{1,...,I_0\}}\lambda_i}+
\frac{\frac{t}{\gamma}\times \Sigma_{i\in\{1,...,I_0\}}\lambda_i}
{N\lambda_1}
\end{equation}

Similarly, if region 2 were to get $N\times\frac{\lambda_2}  {\Sigma_{i\in\{1,...,I_0\}}\lambda_i}$ drivers, its total wait time will be:

\begin{equation}
w_2=\frac{N}{\Sigma_{i\in\{1,...,I_0\}}\lambda_i}+
\frac{\frac{t}{\gamma}\times \Sigma_{i\in\{1,...,I_0\}}\lambda_i}
{N\lambda_2}
\end{equation}

It is easy to see that the first terms of $w_1$ and $w_2$ are the same, and the second term is larger in $w_2$ given that $\lambda_1\geq\lambda_2$. Now note that under allocation $\tilde{n}$, the wait time in region 1 is indeed $w_1$. So, it remains to show that $W_2(\tilde{n}_2)\geq w_2$. To show this, we make two observations (and prove them both shortly). First, $\tilde{n}_2\geq N\times\frac{\lambda_2}  {\Sigma_{i\in\{1,...,I_0\}}\lambda_i}$. This simply says under $\tilde{n}$, region 2 is getting more drivers than it would if drivers were to be allocated to regions proportionally to their demand rates. Second, the total wait time function in region 2 is increasing between $N\times\frac{\lambda_2}  {\Sigma_{i\in\{1,...,I_0\}}\lambda_i}$ and $\tilde{n}_2$. Together, these two observations imply $W_2(\tilde{n}_2)\geq w_2$, as desired. Therefore, we have shown that $W_2(\tilde{n}_2)\geq w1$. But given that $(\tilde{n}_2,...,\tilde{n}_{I_0})$ was an all-regions equilibrium under $((\lambda_2,...,\lambda_{I_0}),N\times(1-\frac{\lambda_1}{\Sigma_{i=1,...,I_0}\lambda_i},\frac{t}{\gamma})$, we know that for any $i,j>1:\quad W_i(\tilde{n}_i)=W_j(\tilde{n}_j)$. This, combined with $W_2(\tilde{n}_2)\geq w_1$, completes the proof of the lemma, of course with the exception of the two observations made in this paragraph. We now turn to proving those observations and finish the proof of the lemma.

The first observation was that  $\tilde{n}_2\geq N\times\frac{\lambda_2}  {\Sigma_{i\in\{1,...,I_0\}}\lambda_i}$. To see why this is true, note that $\tilde{n}$ is the all-regions equilibrium under $((\lambda_2,...,\lambda_{I_0}),N\times(1-\frac{\lambda_1}{\Sigma_{i=1,...,I_0}\lambda_i}),\frac{t}{\gamma})$. Therefore, by our induction assumption on statement 3, region 2 will get disproportionately more drivers relative to all other regions, because it has the highest $\lambda_i$ amongst regions $2,...,I_0$. That is $\forall i>2:\quad \frac{\tilde{n}_2}{\lambda_2}\geq\frac{\tilde{n}_i}{\lambda_i}$. It is then easy to show that:
\begin{equation}\label{eq_induction_observation1}
\frac{\tilde{n}_2}{\lambda_2}\geq \frac{\Sigma_{i=2,...,I_0} \tilde{n}_i }  {\Sigma_{i=2,...,I_0} \lambda_i }
\end{equation}
But we know, from the primitives, that $\Sigma_{i=2,...,I_0} \tilde{n}_i=N\times(1-\frac{\lambda_1}{\Sigma_{i=1,...,I_0}\lambda_i})=N\times\frac{\Sigma_{i=2,...,I_0}\lambda_i}{\Sigma_{i=1,...,I_0}\lambda_i}$. Now, plugging this into (\ref{eq_induction_observation1}) and rearranging, we get  $\tilde{n}_2\geq N\times\frac{\lambda_2}  {\Sigma_{i\in\{1,...,I_0\}}\lambda_i}$, which is exactly our first observation.

We now turn to the proof of the second observation. That is, we want to show that the total wait time function in region 2 is increasing between $N\times\frac{\lambda_2}  {\Sigma_{i\in\{1,...,I_0\}}\lambda_i}$ and $\tilde{n}_2$. To see this, note that the wait time curve in region 2 takes the form that was depicted in figure (\ref{fig:equilibria}). In particular, it is a curve with only one trough; and once past the trough, the curve will remains strictly increasing indefinitely. Thus, to prove that the wait-time is increasing over the interval $[N\times\frac{\lambda_2}  {\Sigma_{i\in\{1,...,I_0\}}\lambda_i},\tilde{n}_2]$, it is sufficient to show that the smallest point in this interval is past the trough. One can show the trough happens at $n_2=\sqrt{\frac{t}{\gamma}\lambda_2}$. Therefore, what we need to show is:
\begin{equation}\label{eq_induction_existence_obs2}
    N\times\frac{\lambda_2}  {\Sigma_{i\in\{1,...,I_0\}}\lambda_i}\geq \sqrt{\frac{t}{\gamma}\lambda_2}
\end{equation}

In order to prove this, we first assume, to the contrary, that $N\times\frac{\lambda_2}  {\Sigma_{i\in\{1,...,I_0\}}\lambda_i}< \sqrt{\frac{t}{\gamma}\lambda_2}$; then we arrive at a contradiction with the result that $\tilde{n}$ is the all-regions equilibrium under $((\lambda_2,...,\lambda_{I_0}),N\times(1-\frac{\lambda_1}{\Sigma_{i=1,...,I_0}\lambda_i},\frac{t}{\gamma})$. Note that we are assuming, without loss of generality, $\lambda_2\geq\lambda_i$ for any $i>2$. Therefore, given that all $\lambda_i$ are positive, for any $i>2$, we have $\frac{\lambda_i}{\lambda_2}\leq \sqrt{\frac{\lambda_i}{\lambda_2}}$. Thus, if we multiply the left hand side of the inequality $N\times\frac{\lambda_2}  {\Sigma_{i\in\{1,...,I_0\}}\lambda_i}< \sqrt{\frac{t}{\gamma}\lambda_2}$ by $\frac{\lambda_i}{\lambda_2}$ and the right hand side by $\sqrt{\frac{\lambda_i}{\lambda_2}}$, then the direction of the inequality should not not change. Therefore, not only for region 2, but also for any region $i\geq 2$, we will have: 
$$N\times\frac{\lambda_i}  {\Sigma_{j\in\{1,...,I_0\}}\lambda_j}< \sqrt{\frac{t}{\gamma}\lambda_i}$$
Now, if we sum over all $i=2,...,I_0$ on both sides of the inequality above, we get:
$$N\times \Sigma_{i=2,...,I_0} \frac{\lambda_i}  {\Sigma_{j\in\{1,...,I_0\}}\lambda_j}< 
\Sigma_{i=2,...,I_0} \sqrt{\frac{t}{\gamma}\lambda_i}$$
Rearranging, we get:
$$N\times (1- \frac{\lambda_1}  {\Sigma_{j\in\{1,...,I_0\}}\lambda_j})< 
\Sigma_{i=2,...,I_0} \sqrt{\frac{t}{\gamma}\lambda_i}$$

But $N\times (1- \frac{\lambda_1}  {\Sigma_{j\in\{1,...,I_0\}}\lambda_j})$ is the total number of drivers in regions $2$ through $I_0$. That is: $N\times (1- \frac{\lambda_1}  {\Sigma_{j\in\{1,...,I_0\}}\lambda_j})=\Sigma_{j=2,...,I_0}\tilde{n}_j$. Therefore, we get:

$$\Sigma_{j=2,...,I_0}\tilde{n}_j< 
\Sigma_{i=2,...,I_0} \sqrt{\frac{t}{\gamma}\lambda_i}$$
which implies there should be at least one $j\geq 2$ such that $\tilde{n}_j<\sqrt{\frac{t}{\gamma}\lambda_j}$. But this means that for that region $j$, the wait time function is decreasing at $n=\tilde{n}_j$ contradicting the result that $\tilde{n}_j$ is part of an all-regions equilibrium. This completes the proof of the second observation, and hence that of lemma (\ref{lem_induction_existence}).$\square$

Next, we use lemma (\ref{lem_induction_existence}) to construct an all-regions equilibrium under primitives $(\lambda,N,\frac{t}{\gamma})$. This will be a constructive proof to the existence portion of statement 4. To this end, we start from the first partial equilibrium $\bar{n}$, gradually shifting drivers from region 1 to other regions until we are left with $\tilde{n}$ drivers in region 1. That is, for any $\hat{n}_1\in[\tilde{n}_1,\bar{n}_1]$ we consider the partial equilibrium $\hat{n}=(\hat{n}_1,...,\hat{n}_{I_0})$ such that the $(\hat{n}_2,...,\hat{n}_{I_0})$ is the all-regions equilibrium allocation under primitives $(\lambda_2,...,\lambda_{I_0}),N-\hat{n}_1,\frac{t}{\gamma})$. The argument for why such partial equilibrium exists for any $\hat{n}_1<\bar{n}_1$ is similar the argument given in proof of lemma (\ref{lem_induction_partialEQ}) for $\tilde{n}_1$.

Now, note that by lemma (\ref{lem_induction_existence}), the total wait time in region 1 is  larger than that in other regions when  $\hat{n}_1=\bar{n}_1$ and it is smaller in region 1 than it is in other regions when $\hat{n}_1=\tilde{n}_1$. Therefore, there should be some $\hat{n}_1\in[\tilde{n}_1,\bar{n}_1]$ for which the total wait time in region 1 is equal to the total wait time in all of the other regions, which themselves are equal to each other by $\hat{n}$ being a partial equilibrium the way defined above.\footnote{Note that in order to make this argument we also need to know that as we move $\hat{n}_1$ within $[\tilde{n}_1,\bar{n}_1]$, the total wait time in region 1 as well as the common total wait time in the other regions both move continuously. This is true by construction for region 1, since the total wait time function is continuous. For other regions, this needs to be shown that as we add drivers to the collection of these regions, the equilibrium total wait time moves continuously. We skip the proof of this claim here, but can provide it upon request.} We claim such allocation $\hat{n}$ is the all-regions equilibrium of the whole market (that is, under primitives $(\lambda,N,\frac{t}{\gamma})$ ). The proof for this claim is as follows:

We know that under allocation $\hat{n}$ all regions have the same total wait time. We also know, by $(\hat{n}_2,...,\hat{n}_{I_0})$ being the all-regions equilibrium under primitives $(\lambda_2,...,\lambda_{I_0}),N-\hat{n}_1,\frac{t}{\gamma})$, that the total wait time in each region $i>1$ is increasing at $n=\hat{n}_i$. Thus, the only thing that remains to be shown is that for $i=1$ too the total wait time curve is increasing at $n=\hat{n_1}$. To this end, as argued before in a similar case, we need to show that $\hat{n}_1\geq \sqrt{\frac{t}{\gamma}\lambda_1}$. Note that given $\hat{n}_1\geq\tilde{n}_1$, it would suffice to show  $\tilde{n}_1\geq\sqrt{\frac{t}{\gamma}\lambda_1}$. We show this latter inequality by borrowing from what we already did in the proof of the last observation we made as part of proof of lemma (\ref{lem_induction_existence}). There, we proved inequality (\ref{eq_induction_existence_obs2}) holds. Now, given that we have been assuming (without loss of generality) that $\lambda_1\geq\lambda_2$, and given that all $\lambda_i$ are positive numbers, we get: $\frac{\lambda_1}{\lambda_2}\geq \sqrt{\frac{\lambda_1}{\lambda_2}}$. Therefore, if we multiply the left-hand side of equation (\ref{eq_induction_existence_obs2}) by $\frac{\lambda_1}{\lambda_2}$ and the right hand side by $\sqrt{\frac{\lambda_1}{\lambda_2}}$, the sign of the inequality should not change. This operation gets us:

$$
    N\times\frac{\lambda_1}  {\Sigma_{i\in\{1,...,I_0\}}\lambda_i}\geq \sqrt{\frac{t}{\gamma}\lambda_1}
$$
which is exactly what we were after. This shows that $\hat{n}$ is the all-regions equilibrium, completing the proof of the first part of statement 4 in the theorem.$\square$

Now that we have shown the all-region equilibrium $\hat{n}$ under primitives $(\lambda,N,\frac{t}{\gamma})$ exists, we show that it indeed shows less geographical supply inequity than the old equilibrium $n^*$. As the first step towards this goal, note that for any $j>i>1$, we can show the result holds based on our induction assumption. More precisely, we know that $(n^*_2,...,n^*_{I_0})$ is the all-regions equilibrium under primitives $((\lambda_2,...,\lambda_{I_0}),N-n^*_1,t)$. We also know that $(\hat{n_2},...,\hat{n}_{I_0})$ is the all regions equilibrium under primitives $((\lambda_2,...,\lambda_{I_0}),N-\hat{n}_1,\frac{t}{\gamma})$. The move from primitives $((\lambda_2,...,\lambda_{I_0}),N-n^*_1,t)$ to primitives $((\lambda_2,...,\lambda_{I_0}),N-\hat{n}_1,\frac{t}{\gamma})$ involves two steps. The first step is to divide $t$ by some $\gamma>1$. The second step is to add $n^*_1-\hat{n}_1$ drivers. Based on our induction assumption, both statements 4 and 5 of Theorem (\ref{prop_extensionKregions}) hold for $I_0-1$ regions. Therefore, for any $j>i>1$ we have:

$$\frac{\frac{\hat{n}_i}{\lambda_i}}{\frac{\hat{n}_j}{\lambda_j}}\leq \frac{\frac{n^*_i}{\lambda_i}}{\frac{n^*_j}{\lambda_j}}$$

with the inequality strict if $\lambda_i>\lambda_j$. Now the only thing that remains to show is that we can say the same not only for $j>i>1$, but also for $j>i=1$. In order to show this, we consider three cases.

\textbf{Case 1: for every $j>1$, we have $\hat{n}_j>n^*_j$.} In this case, the result is becomes trivial given that we know $\hat{n}_1\leq n^*_1$.

\textbf{Case 2: for at least two distinct $j,j'>1$, we have $\hat{n}_j\leq n^*_j$ and $\hat{n}_{j'}\leq n^*_{j'}$.} We start with $j$ and note that the allocation of drivers in all regions other than j --i.e., allocation $(\hat{n}_1,...,\hat{n}_{j-1},\hat{n}_{j+1},...,\hat{n}_{I_0})$ is the all-region equilibrium under primitives $((\lambda_1,...,\lambda_{j-1},\lambda_{j+1},...,\lambda_{I_0}),N-\hat{n}_j,\frac{t}{\gamma})$. Also note that allocation $(n^*_1,...,n^*_{j-1},n^*_{j+1},...,n^*_{I_0})$ is the all-region equilibrium under primitives $((\lambda_1,...,\lambda_{j-1},\lambda_{j+1},...,\lambda_{I_0}),N-n^*_j,\frac{t}{\gamma})$. Note that the former primitives can be obtained from the latter by two moves. First, going from $t$ to $\frac{t}{\gamma}$ for some $\gamma>1$; and second, changing the total number of drivers from $N-n^*_j$ to the (by assumption) larger number of $N-\hat{n}_j$. Based on our induction assumptions, we know that both of these moves reduce the geographical supply inequity. Therefore, now we can claim the following for any $i\neq j$:

$$\frac{\frac{\hat{n}_1}{\lambda_1}}{\frac{\hat{n}_i}{\lambda_i}}\leq \frac{\frac{n^*_1}{\lambda_1}}{\frac{n^*_i}{\lambda_i}}$$

with the inequality strict if $\lambda_1\neq \lambda_i$. This covers all of the comparisons that we needed with the exception of the comparison between region 1 and region $j$ itself. But we can prove the inequality for that case as well, by going through the exact same process as above, except excluding region $j'$ this time instead of region $j$. This finishes the proof of statement 4 of the theorem under case 2.

\textbf{Case 3: for exactly one region $j>1$, we have $\hat{n}_j\leq n^*_j$.} In this case, we can go through the same process as that described in case 2, to show for any $i\neq j$:

$$\frac{\frac{\hat{n}_1}{\lambda_1}}{\frac{\hat{n}_i}{\lambda_i}}\leq \frac{\frac{n^*_1}{\lambda_1}}{\frac{n^*_i}{\lambda_i}}$$

with the inequality strict if $\lambda_1\neq \lambda_i$. This time, however, we are not able to use a similar argument for the to show the result holds between regions 1 and $j$. The following lemmas, however, demonstrate a different way to prove the result for this specific comparison.

\begin{lemma}\label{lem_induction_sqrt1}
Under the conditions of case 3, we have $\hat{n}_1\geq \frac{n^*_1}{\sqrt{\gamma}}$ and $\hat{n}_j\geq \frac{n^*_j}{\sqrt{\gamma}}$.
\end{lemma}

\textbf{Proof of Lemma (\ref{lem_induction_sqrt1}).} We only show $\hat{n}_1\geq \frac{n^*_1}{\sqrt{\gamma}}$. The argument for $\hat{n}_j\geq \frac{n^*_j}{\sqrt{\gamma}}$ is the same.

We start by observing that the total wait time $w_1$ in region 1 under $n_1=\frac{n^*_1}{\sqrt{\gamma}}$ is given by:

$$w_1=\frac{n_1}{\lambda_1}+\frac{\frac{t}{\gamma}}{n_1}$$
$$=\frac{\frac{n^*_1}{\sqrt{\gamma}}}{\lambda_1}+\frac{\frac{t}{\gamma}}{\frac{n^*_1}{\sqrt{\gamma}}}$$
$$=(\frac{n^*_1}{\lambda_1}+\frac{t}{n^*_1})\frac{1}{\sqrt{\gamma}}$$
\begin{equation}\label{eq_lem_inclusion_sqrt_1_1}
  =\frac{w^*}{\sqrt{\gamma}}  
\end{equation}

where $w^*$ is the common total wait time among all regions under primitives $(\lambda,N,t)$ and the all-regions equilibrium $n^*$ given those primitives.

Next, we show that under the \textit{new primitives} $(\lambda,N,\frac{t}{\gamma})$, but at the \textit{old equilibrium allocation $n^*$}, the total wait-time in any region $i$ is weakly larger than $\frac{w^*}{\sqrt{\gamma}}$. To see this, we write out one such total wait time:
$$\frac{n^*_i}{\lambda_i}+\frac{\frac{t}{\gamma}}{n^*_i}$$

Note that because $n^*$ is the all-region equilibrium under the old primitives, it must be that for all $i:\quad n^*_i\geq \sqrt{t\lambda_i}$. This gives $\frac{n^*_i}{\lambda_i}\geq\frac{t}{n^*_i}$, or, alternatively: $\frac{t}{n^*_i}\leq\frac{1}{2}(\frac{n^*_i}{\lambda_i}+\frac{t}{n^*_i})=\frac{w^*}{2}$.

Therefore, we can write:

$$\frac{n^*_i}{\lambda_i}+\frac{\frac{t}{\gamma}}{n^*_i}= (\frac{n^*_i}{\lambda_i}+\frac{t}{n^*_i})-\frac{t}{n^*_i}(1-\frac{1}{\gamma})$$
$$=w^*-\frac{t}{n^*_i}(1-\frac{1}{\gamma})$$
$$\geq w^*(1-\frac{1}{2}(1-\frac{1}{\gamma}))$$
$$=w^*\times(\frac{1+\frac{1}{\gamma}}{2})$$
$$> w^*\times(\sqrt{1\times \frac{1}{\gamma}})$$
\begin{equation}\label{eq_lem_inclusion_sqrt_1_2}
    =\frac{w^*}{\sqrt{\gamma}}
\end{equation}

Equations (\ref{eq_lem_inclusion_sqrt_1_1}) and (\ref{eq_lem_inclusion_sqrt_1_2}), together, tell us that for any $i$, we have $\frac{n^*_i}{\lambda_i}+\frac{\frac{t}{\gamma}}{n^*_i}> w_1$. Now notice that the total wait time under the new primitives at the old equilibrium alloction in any region is increasing. This is simply because $\forall i:\quad n^*_i\geq\sqrt{t\lambda_i}>\sqrt{\frac{t}{\gamma}\lambda_i}$. This, combined with the fact that there is at least one region $i$ with $\hat{n}_i>n^*_i$,\footnote{This is true because there are at least three regions; and besides regions 1 and $j$, case 3 assumes $\hat{n}_i>n^*_i$ for all $i$.} tells us:
$$\hat{w}\equiv\frac{\hat{n}_i}{\lambda_i}+\frac{\frac{t}{\gamma}}{\hat{n}_i}
>\frac{n^*_i}{\lambda_i}+\frac{\frac{t}{\gamma}}{n^*_i}>w_1$$

where $\hat{w}$ is defined as the common total wait time among all regions under the new primitives and new equilibrium allocation.

What $\hat{w}>w_1$ tells us is that if we reduce the number of drivers in region 1 to $n_1=\frac{n^*_1}{\sqrt{\gamma}}$, the total wait time in region 1 falls below the equilibrium total wait time. But this means it has to be that $\hat{n}_1>n_1=\frac{n^*_1}{\sqrt{\gamma}}$. To see why, consider two scenarios. First, if $n_1<\sqrt{\frac{t}{\gamma}\lambda_1}$, then by $\hat{n}_1\geq \sqrt{\frac{t}{\gamma}\lambda_1}$, we get $\hat{n}_1>n_1$. Next, if $n_1\geq\sqrt{\frac{t}{\gamma}\lambda_1}$, then the wait time curve is strictly increasing when moving up from $n_1$, which means at some point past $n_1$, it hits the higher wait time $\hat{w}>w_1$. That point would be $\hat{n}_1$. Thus, the lemma has been proven for region 1. The proof for region $j$ is exactly the same. $\square$

We now present the another useful lemma which helps us better understand what happens to the two regions 1 and $j$.

\begin{lemma}\label{lem_induction_sqrt2}
Consider a market with two regions 1 and 2 only. Allocation $n^*$ is an equilibrium in this market under primitives $(\lambda_1,\lambda_2,N,t)$ if and only if allocation $\frac{n^*}{\sqrt{\gamma}}$ is an equilibrium under primitives $(\lambda_1,\lambda_2,\frac{N}{\sqrt{\gamma}},\frac{t}{\gamma})$.
\end{lemma}
\textbf{Proof of Lemma (\ref{lem_induction_sqrt2}).} Follows directly from definitions.$\square$

Now, lemmas (\ref{lem_induction_sqrt1}) and (\ref{lem_induction_sqrt2}) show us a clear way to complete the last piece of the inductive proof of statement 4 in the theorem. Based on lemma (\ref{lem_induction_sqrt1}), we know $\hat{n}_1+\hat{n}_j>\frac{n^*_1+n^*_j}{\sqrt{\gamma}}$. Now define $N^*=n^*_1+n^*_j$ and $\hat{N}=\hat{n}_1+\hat{n}_j$. We know that $(n^*_1,n^*_j)$ was the all-regions equilibrium under primitives $(\lambda_1,\lambda_j,N^*,t)$. Thus, by lemma (\ref{lem_induction_sqrt2}) we can claim that $(\frac{n^*_1}{\sqrt{\gamma}},\frac{n^*_j}{\sqrt{\gamma}})$ is the all-regions equilibrium under primitives $(\lambda_1,\lambda_j,\frac{N^*}{\sqrt{\gamma}},\frac{t}{\gamma})$. 

On the other hand, we know that $(\hat{n}_1,\hat{n}_j)$ is the all-regions equilibrium under primitives $(\lambda_1,\lambda_j,\hat{N},\frac{t}{\gamma})$. Given that we showed $\hat{N}>\frac{N^*}{\sqrt{\gamma}}$, and given that by our strong induction assumption statement 5 is correct for all two-region cases, we can write:

$$\frac{\frac{\hat{n}_1}{\lambda_1}}{\frac{\hat{n}_j}{\lambda_j}}
\geq \frac{\frac{ \frac{n^*_1}{\sqrt{\gamma}}}{\lambda_1}}{\frac{ \frac{n^*_j}{\sqrt{\gamma}}}{\lambda_j}}
= \frac{\frac{n^*_1}{\lambda_1}}{\frac{n^*_j}{\lambda_j}} $$

with the inequality strict whenever $\lambda_1>\lambda_j$. This completes the proof of case 3, and hence finishes the inductive proof of statement 4 of Theorem  (\ref{prop_extensionKregions}) with the exception of the last claim about $\gamma\rightarrow\infty$, which we turn to next.

To see why for any $i<j$ we have $\frac{\frac{n^{*'}_i}{\lambda_i}}{\frac{n^{*'}_j}{\lambda_j}}\rightarrow 1$ as $\gamma\rightarrow\infty$, assume first on the contrary, that this claim  is not true. That is $\exists\, i<j$ such that $\frac{\frac{n^{*'}_i}{\lambda_i}}{\frac{n^{*'}_j}{\lambda_j}}$ does not approach 1 as $\gamma\rightarrow\infty$. We use this assumption to get a contradiction. Note that given the other claims in statement 4 of the theorem, we  know that $\frac{\frac{n^{*'}_i}{\lambda_i}}{\frac{n^{*'}_j}{\lambda_j}}$ monotonically decreases as $\gamma$ increases.  Therefore, the only possibility for it to not approach 1, is for it to  approach a number  strictly above one. Denote that number by $\kappa>1$.

Also note that as $\gamma\rightarrow\infty$, the equilibrium number of drivers in none of the regions tends to zero.  This is because (i) as immediately implied by the other claims in statement 4, the number of drivers in the lowest demand region $n^{*'}_I$ is increasing in $\gamma$; and (ii) the number of drivers in any other region is always weakly larger than $n^{*'}_I$. This, along with the fact $\gamma\rightarrow\infty$ is equivalent to $t\rightarrow 0$, means that the total wait time in each region $k$ will tend to the idle time in that region. Therefore, $\frac{\frac{n^{*'}_i}{\lambda_i}}{\frac{n^{*'}_j}{\lambda_j}}\rightarrow \kappa>1$ implies regions $i$ and $j$ have different limiting total wait times at the equilibrium, which contradicts statement 1. This finishes the proof of the statement. $\square$

\textbf{Proof of Statement 6.} The steps of this proof closely (almost exactly) follow the steps of the proof of statement 4. We skip it but can provide the detailed proof upon request.$\square$

\textbf{Proof of Statement 5.} Similar to the corresponding two-region case (i.e., proof of Proposition (\ref{prop_impactOfSize_SupplyOnly}). This statement can be proven in  a straightforward manner once we have proven statements  4 and 6. To be more precise, if we know that geographical supply inequity decreases in the sense defined in the statement of  the theorem both (i) when we proportionally scale-up $N$ and the vector $\lambda$ and (ii) when we scale down the vector $\lambda$, it follows that the geographical supply inequity also decreases when we only scale up $N$, which is a certain combination of (i) and (ii).$\square$

The above proofs show that (i) the theorem holds for $I=2$ and that (ii) the theorem holds for any $I_0>2$ if it holds for all $I\in\{2,...,I_0-1\}$. This means our proof is complete.$\blacksquare$

\subsection{Proofs under the assumption that region size is homogeneous}\label{subsec: proofs when region size homogeneous}

\textbf{Proof of \cref{prop: driver behavior EQ description}.}
\textbf{Proof of Statement 1.} To see that an equilibrium always exists, note that all $N$ drivers being in one region $i$ is always an equilibrium. This is because under this allocation, pickup times in all other regions are extremely large, incentivizing drivers to remain in $i$.

To see uniqueness of equilibrium allocation $n^*$ conditional on  the set of served areas $J=\{i:n^*_i>0\}$, note that allocation $n^*_J$ (i.e., restriction of $n^*$ to regions in $J$) is an all-regions equilibrium under primitives $(\lambda_J,N,t)$. We know from \cref{prop_extensionKregions} that this equilibrium is unique. $\square$

\textbf{Proof of Statement 2.} This also follows immediately from \cref{prop_extensionKregions} and the observation that $n^*_J$ is an all-regions equilibrium under $(\lambda_J,N,t)$. $\square$

\textbf{Proof of Statement 3.} Suppose $J=\{i:n^*_i>0\}$ and $J'=\{i:n^{*'}_i>0\}$. Also suppose $J\subsetneq J'$. Given that the total number of drivers under $n^*$ and $n^{*'}$ are the same, and given $J\subsetneq J'$, there has to be one region $i\in J'$ such that $n^*_i>n^{*'}_i$.

Observe that $n^*_J$ is the all-regions equilibrium under $(\lambda_J,N,t)$ and $n^{'*}_{J'}$ is the all-regions equilibrium under $(\lambda_{J'},N,t)$. Therefore, by \cref{prop_extensionKregions}, we know that in region $i$, total wait time is strictly increasing in $n_i$ for any $n_i>n^{*'}_i $. This, combined with $n^*_i>n^{*'}_i$, implies that the total wait time in region $i$ is strictly higher under $n^*$ relative to $n^{*'}$. Given the equivalence of toatl wait time across regions with positive supply, this means the total wait time in any served region under $n^*$ is strictly higher than that in any served region  under $n^{*'}$.$\square$

This finishes the proof of the proposition (albiet still under the assumption that $t$ is uniform across regions). $\blacksquare$

\textbf{Proof of \cref{prop: driver behavior econ of density}.} Denote the set up regions that get positive supply under $n^*$by $J$. Statement 1 directly follows from \cref{prop_extensionKregions} once we note that $n^*_J$ is the all-regions equilibrium under primitives $(\lambda_J,N,t)$. Statement 2 follows directly from statement 1 given that under uniform prices and wages, equilibrium wait times are uniform across regions as shown by \cref{prop_extensionKregions}. More precisely, we know $A_i(n^*_i)=\frac{n^*_i}{W_i(n^*_i)\lambda_i}$ and the same holds for region $j$. Now, given $W_i(n^*_i)=W_j(n^*_j)$, statement 2 directly follows from statement 1. $\blacksquare$

\textbf{Proof of \cref{prop:MarketThickness_fixedN}.} Again, the proofs of all sections directly follow from \cref{prop_extensionKregions} and the observation that once we restrict attention to the set $J=\{i:n^*_i>0\}$ of regions, then (i) $n^*_J$ is an all regions equilibrium, and $(\lambda_J',N',t')$ is still a one (or two) sided thickening of $(\lambda_J,N,t)$. $\blacksquare$

\textbf{Proof of \cref{prop:Platform vs EQ}.} Before proving the proof of the proposition, we prove the following lemma.

\begin{lemma}\label{lem:platformIdealLemma}
$n^{**}_1> n^{**}_2$.
\end{lemma}

\textbf{Proof of \cref{lem:platformIdealLemma}.} Suppose, on the contrary, that $n^{**}_i< n^{**}_j$. In this case, we can show the platform will be better off swapping the allocation of drivers between the two regions. To see this, note that $n^{**}_i< n^{**}_j$ implies:

$$(\frac{1}{\lambda_i}+\frac{t}{n^{**^2}_j})\times (\frac{1}{\lambda_j}+\frac{t}{n^{**^2}_j})<
(\frac{1}{\lambda_i}+\frac{t}{n^{**^2}_i})\times (\frac{1}{\lambda_j}+\frac{t}{n^{**^2}_i})$$

By $\lambda_i>\lambda_j$ we get:

$$\frac{\frac{1}{\lambda_j}-\frac{1}{\lambda_i} }{(\frac{1}{\lambda_i}+\frac{t}{n^{**^2}_j})\times (\frac{1}{\lambda_j}+\frac{t}{n^{**^2}_j})}>
\frac{\frac{1}{\lambda_j}-\frac{1}{\lambda_i} }{(\frac{1}{\lambda_i}+\frac{t}{n^{**^2}_i})\times (\frac{1}{\lambda_j}+\frac{t}{n^{**^2}_i})}$$

Therefore:
$$\frac{1}{\frac{1}{\lambda_i}+\frac{t}{n_j^{**^2}}}-
\frac{1}{\frac{1}{\lambda_j}+\frac{t}{n_j^{**^2}}}>
\frac{1}{\frac{1}{\lambda_i}+\frac{t}{n_i^{**^2}}}-
\frac{1}{\frac{1}{\lambda_j}+\frac{t}{n_i^{**^2}}}$$

$$\Rightarrow\frac{1}{\frac{1}{\lambda_i}+\frac{t}{n_j^{**^2}}}+
\frac{1}{\frac{1}{\lambda_j}+\frac{t}{n_i^{**^2}}}>
\frac{1}{\frac{1}{\lambda_i}+\frac{t}{n_i^{**^2}}}+
\frac{1}{\frac{1}{\lambda_j}+\frac{t}{n_j^{**^2}}}$$

$$\Rightarrow\frac{n_j^{**}}{\frac{n_j^{**}}{\lambda_i}+\frac{t}{n_j^{**}}}+
\frac{n_i^{**}}{\frac{n_i^{**}}{\lambda_j}+\frac{t}{n_i^{**}}}>
\frac{n_i^{**}}{\frac{n_i^{**}}{\lambda_i}+\frac{t}{n_i^{**}}}+
\frac{n_j^{**}}{\frac{n_j^{**}}{\lambda_j}+\frac{t}{n_j^{**}}}$$

But this last statement precisely says that the platform can strictly increase the total number of rides given by swapping the allocation of drivers between regions $i$ and $j$ from $(n_i^{**},n_j^{**})$ to $(n_j^{**},n_i^{**})$. Given that prices and wages are spatially uniform, this means the platform will also strictly increase its profit through this action, which is a contradiction. Therefore, it has to be that $n^{**}_1\geq n^{**}_2$.

Now we also show that $n^{**}_1= n^{**}_2$ is not the case either. To see this, note that the total platform profit from these two regions is given by: 

$$(p-c)\times (
\frac{n_i^{**}}{\frac{n_i^{**}}{\lambda_i}+\frac{t}{n_i^{**}}}+
\frac{n_j^{**}}{\frac{n_j^{**}}{\lambda_j}+\frac{t}{n_j^{**}}})  $$

Subject to a constant $n^{**}_i+n^{**}_j$, maximizing this profit through the first order condition will give:

\begin{equation}\label{eq:Lem_Platform_Ideal_1}
  \sqrt{n_i^{**}}W_i=\sqrt{n_j^{**}}W_j  
\end{equation}

where $W_i$ is a suppressed notation for $W_i(n_i^{**})$, and same for $W_j$. If $n^{**}_1= n^{**}_2$, then we will have $W_i=W_j$ which means the platform optimal allocation coincides with the driver equilibrium. But in that case, we know for sure that $n^{**}_1\neq n^{**}_2$ due to the fact that supply is skewed toward region $i$ This finishes the proof of the lemma. $\square$

Now with \cref{lem:platformIdealLemma} in hand, we prove the main statements in the proposition. We first show $$\frac{A_j(n^*_j)}{A_i(n^*_i)}<\frac{A_j(n^{**}_j)}{A_i(n^{**}_i)}$$.

To see this, note that by \cref{eq:Lem_Platform_Ideal_1} and \cref{lem:platformIdealLemma}, we have $W_i(n_i^{**})<W_j(n_j^{**})$. This means if we start from the driver equilibrium allocation which sets $W_i(n_i^{*})=W_j(n_j^{*})$, we will have to reallocate some drivers from region $i$ to region $j$ in order to get to $(n_1^{**},n_j^{**})$. Therefore, $n_j^{**}>n_j^{*}$ and $n_i^{**}<n_i^{*}$. Based on this, it is easy to see that $A_j(n^{**}_j)>A_j(n^{*}_j)$ and $A_j(n^{**}_i)<A_j(n^{*}_i)$. It follows that $\frac{A_j(n^*_j)}{A_i(n^*_i)}<\frac{A_j(n^{**}_j)}{A_i(n^{**}_i)}$.

Now we show $$\frac{A_j(n^{**}_j)}{A_i(n^{**}_i)}<1.$$ To see this, note that, by \cref{eq:Lem_Platform_Ideal_1}, we have:

$$\sqrt{n^{**}_i}\times (\frac{n_i^{**}}{\lambda_i}+\frac{t}{n_i^{**}})=
\sqrt{n^{**}_j}\times (\frac{n_j^{**}}{\lambda_j}+\frac{t}{n_j^{**}})$$

$$\Rightarrow \frac{\sqrt{n_i^{**^3}}}{\lambda_i}+\frac{t}{\sqrt{n_i^{**}}}=
\frac{\sqrt{n_j^{**^3}}}{\lambda_j}+\frac{t}{\sqrt{n_j^{**}}}$$

Which, by \cref{lem:platformIdealLemma}, implies 

\begin{equation}\label{eq:Lem_platformIdeal_2}
  \frac{\sqrt{n_i^{**^3}}}{\lambda_i}>
\frac{\sqrt{n_j^{**^3}}}{\lambda_j}  
\end{equation}

Now note that:

$$A_i(n^{**}_i)=\frac{n^{**}_i}{W_i \lambda_i}=
\frac{1}{1+\frac{t\lambda_i}{n^{**^2}_i}}=
\frac{1}{1+\frac{t\lambda_i}{\sqrt{n^{**^3}_i}\sqrt{n^{**}_i}}}
$$

By \cref{eq:Lem_platformIdeal_2} and \cref{lem:platformIdealLemma}, we get:

$$A_i(n^{**}_i)<
\frac{1}{1+\frac{t\lambda_j}{\sqrt{n^{**^3}_j}\sqrt{n^{**}_j}}}=A_j(n^{**}_j)
$$
 
 which completes the proof of the proposition. $\blacksquare$

\subsection{Proofs under the assumption that region size is heterogeneous}\label{subsec: full proofs of driver behavior}

We have shown so far that all of the results we proposed under fixed and uniform prices and wages hold when there is no size heterogeneity across regions. We now show that all of these four propositions still hold even if we relax that assumption and allow each region to have its own size $t_i$.\footnote{We would like to emphasize that we believe this two step approach (first working with homogeneous $t$ and then considering its heterogeneity) was not necessary. We believe it is likely that the proofs we provided so far would work with heterogeneous $t$ with minimal changes in the main steps. Nevertheless, we decided to take the two step approach for two reasons. First, the proofs are already involved and directly dealing with heterogeneity in $t$  would have led to much less ``clean'' proofs. The second reason has to do with the progress on the proofs. Our original proofs were based on homogeneous $t$ but later we noticed that the homogeneity assumption was not necessary in any way.} The key to this extension is the following lemma.

\begin{lemma}\label{lem: heterogeneous t}
Suppose that there is a ``quantum region size'' $\Delta$. Consider market primitives $(\lambda,N,t)$ such that each region's size $t_i$ can be written as a multiplier of this quantum size. That is: $t_i=x_i \Delta$ where $x_i\in\mathbb{N}$. Also, consider market primitives $(\hat{\lambda},N,\hat{t})$ such that 
$$\hat{t}=(
\underbrace{\Delta,...,\Delta}_{x_1 \text{ times}},
\underbrace{\Delta,...,\Delta}_{x_2 \text{ times}},
...,
\underbrace{\Delta,...,\Delta}_{x_I \text{ times}})$$

and 
$$\hat{\lambda}=(
\underbrace{\frac{\lambda_1}{x_1},...,\frac{\lambda_1}{x_1}}_{x_1 \text{ times}},
\underbrace{\frac{\lambda_2}{x_2},...,\frac{\lambda_2}{x_2}}_{x_2 \text{ times}},
...,
\underbrace{\frac{\lambda_I}{x_I},...,\frac{\lambda_I}{x_I}}_{x_I \text{ times}})$$

Then:

\begin{enumerate}
    \item allocation $n$ is an equilibrium under $(\lambda,N,t)$ if and only if allocation

$$\hat{n}=(
\underbrace{\frac{n_1}{x_1},...,\frac{n_1}{x_1}}_{x_1 \text{ times}},
\underbrace{\frac{n_2}{x_2},...,\frac{n_2}{x_2}}_{x_2 \text{ times}},
...,
\underbrace{\frac{n_I}{x_I},...,\frac{n_I}{x_I}}_{x_I \text{ times}})$$

is an equilibrium under primitives $(\hat{\lambda},N,\hat{t})$.

    \item allocations $\hat{n}$ that take the above form (i.e., with equal number of drivers among all regions $j$ with the same $\hat{\lambda}_j$) are the only possible equilibria under $(\hat{\lambda},N,\hat{t})$ and the only possible allocations that can maximize the platform profit under $(\hat{\lambda},N,\hat{t})$.
    
    \item platform profit from allocation $n$ under market primitives $(\lambda,N,t)$ is equal to its profit from allocation $\hat{n}$ under market primitives $(\hat{\lambda},N,\hat{t})$.
    
    \item Access to ride in any region $i$ under allocation $n$ and primitives $(\lambda,N,t)$ is equal to access to rides in any region $j$ with $\hat{\lambda}_j=\frac{\lambda_i}{x_i}$ under allocation $\hat{n}$ and primitives $(\hat{\lambda},N,\hat{t})$.
\end{enumerate}

\end{lemma}

\textbf{Proof of \cref{lem: heterogeneous t}.} The proof is straightforward. It is immediate that $\hat{n}$ sums up to $N$. It is also immediate that in each region $j$ with demand arrival rate  $\hat{\lambda}_j=\frac{\lambda_i}{x_i}$, the total wait time is given by:

$$\hat{W_j}(\hat{n}_j)=\frac{\hat{n}_j}{\hat{\lambda}_j}+\frac{\hat{t}_j}{\hat{n}_j}=
\frac{\frac{n_i}{x_i}}{\frac{\lambda_i}{x_i}}+
\frac{\Delta}{\frac{n_i}{x_i}}=
\frac{n_i}{\lambda_i}+
\frac{\Delta x_i}{n_i}=
\frac{n_i}{\lambda_i}+
\frac{t_i}{n_i}=W_i(n_i)$$

It is also immediate to see that $\hat{W_j}(\cdot)$ is increasing at $\hat{n}_j$ if and only if $W_i(\cdot)$ is increasing at $n^*_i$. Finally, by \cref{prop_extensionKregions}, it can be seen that no equilibrium under $(\hat{\lambda},N,\hat{t})$ can give two different volumes of drivers to two regions of the same size that have the same arrival rate of demand. 

Next, note that by \cref{prop:Platform vs EQ}, if two regions have the same size and the same demand density, then the platform would also optimally want the same number of drivers in those regions. 

Additionally, given what we established regarding the wait-time equivalence between the two regions, it is straightforward to verify that platform profits are equal between $n$ and $\hat{n}$.

Finally, regarding access, note that $$\hat{A}_j(\hat{n}_j)=\frac{\hat{n}_j}{\hat{\lambda}_j\hat{W}(\hat{n}_j)}
\frac{\frac{n_i}{x_i}}{\frac{\lambda_i}{x_i}W(n_i)}=A_i(n_i)
$$

This establishes the lemma. $\square$

This lemma is powerful in that it shows any market primitives with heterogeneous $t$ can, with a small enough $\Delta$, be approximated with some market primitives with homogeneous $t$ without loss of the equilibria. In other words, the lemma shows there is a 1-to-1 mapping between the equilibria under $(\hat{\lambda},N,\hat{t})$ and $(\lambda,N,t)$. This allows us to complete the proofs of \cref{prop: driver behavior EQ description} through \cref{prop:Platform vs EQ} --which were stated for heterogeneous $t$ but so far proven only for homogeneous $t$. Here, we only provide the basic steps of the proof and avoid going through all the steps, which are notationally cumbersome but conceptually straightforward (full steps would available upon request). The basic idea for proving \cref{prop: driver behavior EQ description} through \cref{prop:Platform vs EQ} for general $(\lambda,N,t)$ with heterogeneous $t$ is as follows: If all $t_i$ are dividable by a common $\Delta$, then use that $\Delta$ and take the following steps:

\begin{enumerate}
    \item Using \cref{lem: heterogeneous t}, ``transform'' the allocations and primitives in the proposition from the $(\lambda,N,t)$ space to the $(\hat{\lambda},N,\hat{t})$ space. 
    \item Given $\hat{t}$ is homogeneous, apply the proposition. 
    \item Again, using the provisions of \cref{lem: heterogeneous t}, show that the proposition holding under $(\hat{\lambda},N,\hat{t})$ means it also holds under $(\lambda,N,t)$.
\end{enumerate}

If there is no common $\Delta$ by which all $t_i$ are dividable, then each $x_i$ will be the quotient in the division of $t_i$ by $\Delta$. This means the proposition holding for $(\hat{\lambda},N,\hat{t})$ will not imply it also holds for $(\lambda,N,t)$. It will, rather, hold for an approximation of $(\lambda,N,t)$. But the approximation will get more and more accurate as we make $\Delta$ smaller. Thus, the proof will be obtained as $\Delta\rightarrow0$.

\section{Proofs for results with endogenous platform strategy}\label{appx: proof 2}
Before starting these proofs, we make an observation. As mentioned in the main text of the paper, we have made the total number of drivers $N$ endogenous in this section. Although in the first glance this adds another layer of complexity on top of prices and wages becoming endogenous, a closer look shows this in fact makes the proofs more manageable. This is because with an endogenous $N$, the problem in each region will be, in some sense, separate from other regions. As such, induction in the number of regions will not be a required step any more. The analysis of the platform optimal behavior does present a challenge, however. We employ monotone comparative static techniques from the literature in order to address those challenges and prove our propositions.

\textbf{Proof of \cref{prop:endogenous Wages}.} We start by proving the statements about economies of density. Then we will move on to statements regarding the role of market thickness.

\textbf{Proof of Statements 1 and 2 on Economies of Density.}
To prove the result for the whole market, we can prove it for every region $i$ separately. As such, in what follows, we suppress the index $i$. As such, in this proof, $n^*$ does not stand for the vector of all driver presence volumes in all regions, but just the number of drivers in region $i$. We will also suppress some equilibrium notations. For instance, we denote the equilibrium total wait time for drivers by $W$ instead of $W_i(n^*_i)$.

Note that in the equilibrium, the hourly revenue for drivers has to be equal to the reservation value $\bar{c}$. Therefore, we have:

\begin{equation}\label{eq:ReservationValue}
\frac{c}{W}=\bar{c}    
\end{equation}

If we replace $W$ in the above with the expression in terms of $n^*$, and then solve for $n^*$ we get two solutions:

$$n^*=\frac{\frac{c}{\bar{c}}\mp\sqrt{(\frac{c}{\bar{c}})^2-\frac{4t}{\lambda(p)}}}
{\frac{2}{\lambda(p)}}$$

Note that only the larger solution (i.e., the one with the $+$ sign) is an equilibrium because at the lower solution, further driver entry will lead to lower overall wait time, increasing driver revenue. As such, we have:

\begin{equation}\label{eq:ReservationValue2}
n^*=\frac{\frac{c}{\bar{c}}+\sqrt{(\frac{c}{\bar{c}})^2-\frac{4t}{\lambda(p)}}}
{\frac{2}{\lambda(p)}} 
\end{equation}

Now note that platform profit in the region is given by $\pi=(p-c)\times \frac{n^*}{W}$. 

Replacing for $W$ and $n^*$ respectively from \cref{eq:ReservationValue} and \cref{eq:ReservationValue2}, we get:

\begin{equation}\label{eq:Profit_Flexible_N}
    \pi=(p-c)\times\frac{\lambda(p)}{2}\times [1+\sqrt{1-\frac{4t}{\lambda(p)}(\frac{\bar{c}}{c})^2}]
\end{equation}

Next, we make two observations. First, note that if the exogenous price $p$ is such that

\begin{equation}\label{eq:ReservationValue3}
1-\frac{4t}{\lambda(p)}(\frac{\bar{c}}{p})^2<0 
\end{equation}

then it means \cref{eq:ReservationValue2} will have a real valued root only if the platform sets $c>p$. That is, the platform can only attract drivers to the region by offering them a wage higher than price $p$, thereby running a net loss itself. Thus, the platform's optimal action will be to set a wage that will attract no driver if and only if \cref{eq:ReservationValue3} holds. Note that \cref{eq:ReservationValue3} is equivalent to:

$$\frac{\lambda(p)}{t}<4(\frac{\bar{c}}{p})^2$$

Given that $p$ is fixed in this proposition, then $f(p)$ is also fixed. Therefore, by $\lambda(p)=\bar{\lambda}f(p)$, we have:

\begin{equation}\label{eq: Density Threshold}
  \frac{\bar{\lambda}}{t}<\frac{4}{f(p)}(\frac{\bar{c}}{p})^2  
\end{equation}

which exactly says the region will not get any drivers if the density of potential demand falls short of some cutoff as statement 2 of the proposition says.

Our second observation is about the uniqueness of the equilibrium wage when the region does get a positive number of drivers. To ease the notation, denote $\frac{4t\bar{c}^2}{\lambda(p)}$ simply by $a$. This gives

$$\pi=(p-c)\times \frac{\lambda(p)}{2}\times[1+\sqrt{1-\frac{a}{c^2}}]$$

writing the first order condition in terms of $c$ and rearranging yields

$$\sqrt{c^2-a}=\frac{ap}{c^2}-c$$

This equation can have only one root given that the left hand side is strictly increasing in $c$ while the right hand side is strictly decreasing. Thus, the optimal wage $c^*$ is unique. We know by \cref{eq:ReservationValue2} that this means $n^*$ is unique too. This finishes the proofs of statements 1 and 2 of the proposition. $\square$

\textbf{Proof of Statement 3 on Economies of Density.} Given that the problem in each region is separate from others, this statement is equivalent to the statement that $A(n^*)$ is strictly increasing in $\frac{\bar{\lambda}}{t}$ in each region (Note that we are keeping suppressing the region subscripts). In the rest of the proof, we also suppress the notation on $A(n^*)$ and write it simply as $A$. Recall that $$A=\frac{r}{\bar{\lambda}}=\frac{n}{W\bar{\lambda}}$$ 

Writing out $A$ in terms of the primitives of the model as well as platform strategy parameters $p$ and $c$, we get:

$$A=\frac{\frac{c}{\bar{c}}+\sqrt{(\frac{c}{\bar{c}})^2-\frac{4t}{\lambda(p)}}}
{\frac{2}{\lambda(p)}}\times \frac{1}{\bar{\lambda}\times \frac{c}{\bar{c}}}$$

Simplifying, we get:

\begin{equation}\label{eq:AccessOnlyFunctionOfDensity}
  A=\frac{1+\sqrt{1-\frac{4t}{\bar{\lambda}f(p)}\times (\frac{\bar{c}}{c})^2}}
{\frac{2}{f(p)}}  
\end{equation}

This equation shows that $A$ depends on $\frac{\bar{\lambda}}{t}$ but not on either $\bar{\lambda}$ or $t$ separately. Therefore, in order to show $A$ increases as $\frac{\bar{\lambda}}{t}$, we do not need to show that $A$ is both increasing in $\bar{\lambda}$ and decreasing in $t$. Only one of the two would be sufficient. We choose to focus on the comparative static in $t$. We will make the same choice in other parts of the proof of this proposition as well as the other remaining propositions.

Our next step is to show that $A$ is strictly decreasing in $t$. Note that given the uniqueness result in statement 1, there is a 1-to-1 relationship between access $A$ and wage $c$ in the region. Therefore, the platform could be thought of as optimally choosing $A$ as opposed to optimally choosing $c$. This reduces the problem to a monotone comparative static problem of showing the optimal $A$ is decreasing in the value of $t$. This allows us to use standard monotone comparative statics theorems a la Milgrom and Shannon. We need to prove that the profit function is strictly submodular in $A$ and $t$. That is: $$\frac{\partial^2 \pi}{\partial A \times \partial t}<0$$

To show this, we first write $\pi$  out in terms of $A$ instead of $c$. Recall that platform profit in the region is given by:

\begin{equation}\label{eq: platform profit}
    \pi=(p-c)\frac{n}{W}
\end{equation}

We need to rewrite so that dependence of $\pi$ on $c$ (both direct and through $n$ and $W$) is expressed through dependence on $A$. First note that $A=\frac{n}{W\bar{\lambda}}=\frac{n}{\frac{c}{\bar{c}}\bar{\lambda}}$. Thus 

\begin{equation}\label{eq: n in terms of A}
 n=A\frac{c}{\bar{c}} \bar{\lambda}  
\end{equation}

Also, writing $W$ out \cref{eq:ReservationValue} in terms of $n$ and replacing for $n$ from \cref{eq: n in terms of A}, we get:

$$\frac{c}{\bar{c}}=\frac{A \frac{c}{\bar{c}}}{f(p)}+\frac{t}{A\frac{c}{\bar{c}}\bar{\lambda}}$$

Solving for $c$, we get:

\begin{equation}\label{eq: wage in terms of access}
    c=\bar{c}\sqrt{\frac{\frac{t}{\bar{\lambda}}f(p)}{Af(p)-A^2}}
\end{equation}

Replacing from \cref{eq: n in terms of A} and \cref{eq: wage in terms of access} into \cref{eq: platform profit}, we get:

\begin{equation}\label{eq: profit in terms of A}
\pi = (p-\bar{c}\sqrt{\frac{\frac{t}{\bar{\lambda}}f(p)}{Af(p)-A^2}})A\bar{\lambda}    
\end{equation}

Now that we have the right expression, we want to show $\frac{\partial^2 \pi}{\partial A \times \partial t}<0$. Or equivalently, we want do show that $\frac{\partial \pi}{ \partial t}$ is strictly decreasing in $A$. Taking the partial derivative with respect to $t$ and simplifying yields:

\begin{equation}\label{eq: partial pi partial t}
    \frac{\partial \pi}{ \partial t}=-\frac{\bar{c}}{2t}
    \sqrt{\frac{A\bar{\lambda}tf(p)}
    {f(p)-A}}
\end{equation}

Given that $\bar{\lambda}$, $f(p)$, $\bar{c}$, and $t$ are all positive constants, it would suffice to show $-\frac{A}{f(p)-A}$ is strictly decreasing in $A$. This is immediate by noticing the negative sign, the fact that the numerator is positive and strictly increasing in $A$, and that the denominator is positive and strictly decreasing in $A$. This finishes the proof of statement 3 of the proposition. $\square$

\textbf{Proof of Statement 4 on Economies of Density.} Next, we will show that regions with higher density of potential demand $\frac{\bar{\lambda}_i}{t_i}$ will get lower wages $c_i$ when prices are fixed and spatially uniform. Similar to the previous statements, this one can also be proven  through a comparative static result focusing on each region. We show that if a region's density of potential demand increases, then the optimal wage chosen by the platform for that region decreases. Again, we will be suppressing notations on region indices as well as equilibrium ``$^*$'' notations. In the proof for this statement, we take a similar approach to the one we took for the previous statement. We write the profit function fully in terms of wage $c$ by replacing for $n$ from \cref{eq:ReservationValue2}. Then we show that 

\begin{equation}\label{eq: supermodularity wage density}
\frac{\partial^2 \pi}{\partial c \times \partial t}>0
\end{equation}

which means the optimal wage $c^*$ will be strictly increasing in $t$ and, hence, strictly decreasing in density $\frac{\bar{\lambda}}{t}$. 

Replacing from \cref{eq:ReservationValue2} into the \cref{eq: platform profit} and rearranging yields:

$$\pi=(p-c)\frac{1+\sqrt{1-\frac{4t}{\bar{\lambda}f(p)}(\frac{\bar{c}}{c})^2}}
{\frac{2}{\bar{\lambda}f(p)}}$$

Differentiating with respect to $t$ and rearranging terms yields:

$$\frac{\partial \pi}{\partial t}=
-\frac{(p-c)(\frac{\bar{c}}{c})^2}
{\sqrt{(\frac{c}{\bar{c}})^2-\frac{4t}{\bar{\lambda}f(p)}}}$$

It is easy to see that $\frac{\partial \pi}{\partial t}$ is strictly increasing in $c$. This is because the two terms in the numerator are strictly decreasing in $c$, the term in the denominator is strictly increasing in $c$, and there is a negative sign. This shows that $\pi$ is a strictly supermodular function in $c$ and $t$, thereby showing that if $t$ increases, so does the optimal level of $c$. This finishes the proof of this statement. $\square$

Next we turn to those statements in the proposition that have to do with the role of market thickness.

\textbf{Proof of Statement 1 on the Role of Market Thickness.} This is straightforward given the proof statement 2 on economies of density. By \cref{eq: Density Threshold}, we have 

$$n^*_i>0\Rightarrow \frac{\bar{\lambda}_i}{t_i}\geq \frac{4}{f(p)}(\frac{\bar{c}}{p})^2  $$

$$\Rightarrow \frac{\gamma \bar{\lambda}_i}{t_i}\geq \frac{4}{f(p)}(\frac{\bar{c}}{p})^2  \Rightarrow n^{*'}_i>0. \quad \square$$

\textbf{Proof of Statement 2 on the Role of Market Thickness.} First note that the second part of the inequality in this statement (i.e., $\frac{\bar{\lambda}_i}{t_i}> \frac{\bar{\lambda}_j}{t_j} \Rightarrow 
    \frac{A_j(n^{*'}_j)}{A_i(n^{*'}_i)}< 1$) is  implied by statement 3 on economies of density, which we have already proved. We, hence, will only prove: $$\frac{\bar{\lambda}_i}{t_i}> \frac{\bar{\lambda}_j}{t_j} \Rightarrow \frac{A_j(n^{*}_j)}{A_i(n^{*}_i)}<
    \frac{A_j(n^{*'}_j)}{A_i(n^{*'}_i)}$$

In order to prove this statement, we first state and prove a lemma. Recall, again, that given under flexible $N$ the optimization problems in different regions are separate from each other, then the equilibrium access $A_i(n^{*}_i)$ is only a function of $\bar{\lambda}_i,t_i,p_i,c_i$ rather than those at other regions. Therefore, the notation $i$ may be suppressed. In the proofs of some of the last statements we have even suppressed the notation on $n^*$ and denote the equilibrium access in region $i$ simply by $A$. In our next lemma, we will keep suppressing those notations. We also note that access has been an implicit function of model parameters, in particular $\frac{\bar{\lambda}f(p)}{t}$ (we showed that equilibrium access is not separately dependent on $\bar{\lambda}$, $t$, and $f(p)$ otherwise). Denote $\frac{\bar{\lambda}f(p)}{t}$ by $d$. In the proof of this lemma, we make the exposition of the dependence of equilibrium access $A$ on parameter $d$ explicit by denoting the equilibrium access as $A(d)$. In other words, $A(d)$ is the equilibrium access in a region with density $d$ when the platform has set the optimal wage $c^*$ given density $d$ and price $p$ and drivers respond accordingly. 

\begin{lemma}\label{lem: If B concave}
Define function $B(\tau)$ as 
$$B(\tau)\equiv \log(A(e^\tau))$$
Then statement 2 of \cref{prop:endogenous Wages} on the role of market thickness holds if $B$ is a strictly convcave function whenever defined.
\end{lemma}

\textbf{Proof of \cref{lem: If B concave}.} First notice that $B$ is defined whenever $A>0$ which is whenever $\tau$ is larger than some cutoff. 

Next, take two regions $i,j$ with $\frac{\bar{\lambda}_i}{t_i}> \frac{\bar{\lambda}_j}{t_j}$. Set $\tau_i=\log(\frac{\bar{\lambda}_i f(p)}{t_i})$ and
$\tau_j=\log(\frac{\bar{\lambda}_j f(p)}{t_j})$. Also set $\beta=\log(\gamma)$ where $\gamma$ is the scalar larger than one in the statement of the proposition. By strict concavity of $B$, we have:

$$B(\tau_i+\beta)-B(\tau_i)<B(\tau_j+\beta)-B(\tau_j)$$

This implies:

$$\log[A(e^{\tau_i+\beta})]-\log[A(e^{\tau_i})]<\log[A(e^{\tau_j+\beta})]-\log[A(e^{\tau_j})]$$

$$\Rightarrow\log[\frac{A(e^{\tau_i+\beta})}{A(e^{\tau_i})}]<\log[\frac{A(e^{\tau_j+\beta})}{A(e^{\tau_j})}]$$

$$\Rightarrow\frac{A(\frac{\gamma\bar{\lambda}_i f(p)}{t_i})}{A(\frac{\bar{\lambda}_i f(p)}{t_i})}<\frac{A(\frac{\gamma\bar{\lambda}_j f(p)}{t_j})}{A(\frac{\bar{\lambda}_j f(p)}{t_j})}$$

$$\Rightarrow\frac{A(\frac{\bar{\lambda}_j f(p)}{t_j})}
{A(\frac{\bar{\lambda}_i f(p)}{t_i})}
<\frac{A(\frac{\gamma\bar{\lambda}_j f(p)}{t_j})}{
A(\frac{\gamma\bar{\lambda}_i f(p)}{t_i})}$$

which is exactly the statement we needed to prove. This completes the proof of the lemma. $\square$

Therefore, the only thing we need to prove is strict concavity of the function $B$. Our next lemma does this.

\begin{lemma}\label{lem: B concave}
Function $B$ as defined in the previous lemma is strictly concave.
\end{lemma}

\textbf{Proof of \cref{lem: B concave}.} Take $\tau$ and $\tau+2\beta$ for some positive $\beta$ such that $B(\tau)$ and $B(\tau+2\beta)$ are both defined (i.e., the corresponding $A$ values are both positive). The proof of the lemma will be complete if we can show:

\begin{equation}\label{eq: B concave}
    B(\tau+\beta)>\frac{B(\tau)+B(\tau+2\beta)}{2}
\end{equation}

Denote $d=e^\tau$ and $\gamma=e^\beta$. Also denote $\hat{A}=A(d)$. Additionally, from the previous parts of this proposition, we know $A(d\gamma^2)>A(d)$. Therefore, if we write $A(d\gamma^2)=A(d)\zeta^2$ for some positive $\zeta$, then it has to be that $\zeta>1$. At this point, one can observe that \cref{eq: B concave} is equivalent to

\begin{equation}\label{eq: A log concave}
    A(d\gamma)>\sqrt{A(d)A(d\gamma^2)}=A(d)\zeta=\hat{A}\zeta
\end{equation}

To establish the above inequality, we first write out the platform profit function in terms of $A$ (again treating $A$ is the decision variable) and, then, characterize the first order condition which gives $A(d)$ for any $d$ (recall that $A(d)$ is the optimal $A$ under $d$). This is done in \cref{eq: profit in terms of A}. If we differentiate \cref{eq: profit in terms of A} with respect to $A$, we get:

\begin{equation}\label{eq: d pi/ d A}
    \frac{\partial \pi}{\partial A}=
    \bar{\lambda}[p-\frac{\bar{c}}{2}
    \sqrt{ \frac{\frac{t}{\bar{\lambda}}}{A}
    (\frac{f(p)}{f(p)-A})^3}]
\end{equation}

Equating $\frac{\partial \pi}{\partial A}$ to zero and   rearranging yields the following first order condition:

\begin{equation}\label{eq: Access FOC}
    \frac{2p}{\bar{c}\sqrt{f(p)^3}}-\sqrt{ \frac{\frac{t}{\bar{\lambda}}}{A}
    (\frac{1}{f(p)-A})^3}=0
\end{equation}

Replacing for $\frac{t}{\bar{\lambda}}$ with $\frac{1}{d}$, we get:

\begin{equation}\label{eq: Access FOC rewritten}
    \frac{2p}{\bar{c}\sqrt{f(p)^3}}-\sqrt{ \frac{1}{dA}
    (\frac{1}{f(p)-A})^3}=0
\end{equation}

Next, define the function $g(d,A)$ by:

$$g(d,A)\equiv \sqrt{ \frac{1}{dA}
    (\frac{1}{f(p)-A})^3}$$
    
Therefore, the optimality of $\hat{A}$ under $d$ and the optimality of $\hat{A}\zeta^2$ under $d\gamma^2$ mean:

\begin{equation}\label{eq: Access FOC in terms of g}
    g(d,\hat{A})=\frac{2p}{\bar{c}\sqrt{f(p)^3}}=g(d\gamma^2,\hat{A}\zeta^2)
\end{equation}

In order to prove \cref{eq: A log concave}, we need to show that under $d\gamma$ and at $A=\hat{A}\zeta$ we have $\frac{\partial \pi}{\partial A}>0$ (because this would mean the first order condition would hold at some $A>\hat{A}\zeta$, implying exactly \cref{eq: A log concave}). In other words, we need need to show that from \cref{eq: Access FOC in terms of g} one can conclude:

\begin{equation}\label{eq: A log concave in terms of g}
    g(d\gamma,\hat{A}\zeta)<\frac{2p}{\bar{c}\sqrt{f(p)^3}}
\end{equation}

To prove this, it would be sufficient to show 

$$\log(g(d\gamma,\hat{A}\zeta))<\frac{\log(g(d,\hat{A}))+\log(g(d\gamma^2,\hat{A}\zeta^2))}{2}$$

Expanding the above expression in terms of $g$, we get:

$$-\frac{1}{2}\log(dA)-\frac{1}{2}\log(\zeta\gamma)-\frac{3}{2}\log(f(p)-A\zeta)<
$$

$$\frac{-\frac{1}{2}\log(dA)-\frac{3}{2}\log(f(p)-A)
-\frac{1}{2}\log(dA)-\log(\zeta\gamma)-\frac{3}{2}\log(f(p)-A\zeta^2)}
{2}$$

$$\Leftrightarrow-\frac{3}{2}\log(f(p)-A\zeta)<
$$

$$\frac{-\frac{3}{2}\log(f(p)-A)
-\frac{3}{2}\log(f(p)-A\zeta^2)}
{2}$$

$$\Leftrightarrow\log(f(p)-A\zeta)>
\frac{\log(f(p)-A)+\log(f(p)-A\zeta^2)}
{2}$$

This last inequality holds if we can show that the function $\log(f(p)-Ae^\nu)$ is strictly concave in $\nu$. To see this, note that:

$$\frac{\partial}{\partial \nu} \log(f(p)-Ae^\nu)=
\frac{-Ae^\nu}{f(p)-Ae^\nu}=
1-\frac{f(p)}{f(p)-Ae^\nu}$$

which is strictly decreasing in $\nu$. This finishes the proof of this statement of the proposition. $\square$

\textbf{Proof of Statement 3 on the Role of Market Thickness.} This statement is straightforward is is left to the reader. $\square$

This finishes the proof of \cref{prop:endogenous Wages}. $\blacksquare$

\textbf{Proof of \cref{prop:endogenous Prices}.}

\textbf{Proof of Statements 1 and 2 on Economies of Density.} The steps of this proof are similar to the corresponding steps in the previous theorem.  A given region will have a positive number of drivers in equilibrium if it can attract drivers when the price in the region is equal to the fixed wage $c$. Based on \cref{eq:Profit_Flexible_N}, this means:

\begin{equation}\label{eq:Positive drivers as func of density}
1-\frac{4t}{\bar{\lambda}(1-\alpha c)}(\frac{\bar{c}}{c})^2\geq0 
\end{equation}

or equivalently, when:

$$\frac{4}{(1-\alpha c)}(\frac{\bar{c}}{c})^2\leq \frac{\bar{\lambda}}{t}$$

which proves statement 2. Also, \cref{eq:ReservationValue2} holds in this case too, which implies uniqueness of $n^*$ if optimal price $p$ is unique. We now show that optimal price $p$ is indeed unique. More precisely, we want to show that there is a unique $p$ maximizing $\pi$ in \cref{eq:Profit_Flexible_N} when $\lambda(p)\equiv\bar{\lambda}(1-\alpha p)$. Alternatively, we can show that there is a unique $p$ maximizing $\log(\pi)$. If we write the first order condition for $\log(\pi)$ in terms of $p$, we get:

$$\frac{\partial \log(\pi)}{\partial p}=\frac{\partial \log(p-c)}{\partial p}+\frac{\partial \log(\bar{\lambda})}{\partial p}+$$

$$\frac{\partial \log(1-\alpha p)}{\partial p}+\frac{\partial 
\log(1+\sqrt{1-\frac{4t}{\bar{\lambda}W^2}\frac{1}{1-\alpha p})}}{\partial p}=0$$

Some expanding, simplifying, and rearranging yields:

$$\frac{1}{p-c}+\frac{-\alpha}{1-\alpha p}+
\frac{-4t\alpha}{\bar{\lambda}W^2}
\frac{1}{1-\alpha p}
\frac{1}{\sqrt{1-\alpha p}+\sqrt{1-\alpha p-\frac{4t}{\bar{\lambda}W^2}}}
\frac{1}{2\sqrt{1-\alpha p-\frac{4t}{\bar{\lambda}W^2}}}
=0$$

The first order condition can only have one solution given that every one of the three terms in the left hand side expression is strictly decreasing in $p$. This finishes the proofs of statements 1 and 2 regarding economies of density. $\square$

\textbf{Proof of Statements 3 on Economies of Density.} Similar to the case of the previous proposition, it will suffice to show that the optimal $p$ is strictly increasing in the density of the region. This immediately follows from observing that, in the proof of the previous statement of this theorem, $\frac{\partial \log(\pi)}{\partial p}$ is strictly decreasing in $t$, meaning $\log(\pi)$ is strictly sub-modular in $t$ and $p$. This implies that the $p$ that maximizes $\log(\pi)$ (and hence $\pi$ itself) is strictly decreasing in $t$. That is, the optimal $p$ is strictly increasing in density of potential demand $\frac{\bar{\lambda}}{t}$. $\square$

\textbf{Proof of Statement 1 on the Role of Market Thickness.} Similar to the equivalent statement in \cref{prop:endogenous Wages}, this statement follows directly from the proofs of statements 1 and 2 on economies of density. $\square$

\textbf{Proof of Statement 2 on the Role of Market Thickness.} Similar to the equivalent statement in \cref{prop:endogenous Wages}, the proof of this statement is straightforward and is left to the reader. $\square$

This completes the proof of the proposition. $\blacksquare$

\textbf{Proof of \cref{prop:endogenous Wages and Prices}.} This proposition is substantially more complex than \cref{prop:endogenous Wages} and \cref{prop:endogenous Prices} because the endogeneity of prices and wages at the same time makes things intractable. On the other hand, we show that prices and wages both being endogenous also leads to an advantage. It reduces the set of the input parameters of the model to  $(\bar{\lambda},t,\bar{c},\alpha)$. That is, $p$ or $w$ is not part of the model parameters any more. We prove a series of lemmas that show under these parameters, the proposition can be reduced to a numerical problem. That is, we first show that the proposition holds for a general $(\bar{\lambda},t,\bar{c},\alpha)$ if and only if it holds for $(\bar{\lambda},1,1,1)$. Showing that the proposition holds for $(\bar{\lambda},1,1,1)$ is essentially numerical, which we carry out accordingly.

\begin{lemma}\label{lem: equivalance demand reservation wage}
Assume $\gamma\in\mathbb{R}$ is a strictly positive number.  Under primitives $(\bar{\lambda},t,\bar{c},\alpha)$, vectors $p^*$, $w^*$, and $n^*$ constitute an equilibrium if and only if under primitives $(\bar{\lambda}',t',\bar{c}',\alpha')=(\gamma^2 \bar{\lambda},t,\gamma\bar{c},\alpha)$, vectors $p^{*'}=p^*$, $c^{*'}=c^*$, and $n^{*'}=\gamma n^*$ constitute an equilibrium  (note that here, unlike the previous two propositions, we are \textit{not} suppressing the $^*$ notation for equilibria or the regional indices). Additionally, access to rides in all regions are equal under these two equilibria: $\forall i: A_i^{*'}=A_i^*$.
\end{lemma}

\textbf{Proof of \cref{lem: equivalance demand reservation wage}} Let us expand the notation on platform profits and equilibrium numbers of drivers across the market in a way that allows for explicitly tracking the dependence of all of these on market primitives. To be more specific, we denote the equilibrium number of drivers in region $i$ under primitives $(\bar{\lambda},t,\bar{c},\alpha)$ and platform strategy $(p,c)$ by:

$$n^*_i(p,c|\bar{\lambda},t,\bar{c},\alpha)$$

Similarly, we denote the platform profit by:

$$\pi_i(p,c|\bar{\lambda},t,\bar{c},\alpha)$$

We now turn to the proof of the lemma, starting with a claim:

\begin{claim}\label{claim: lem equivalence 1: pi proportional}
For any given $(p,c)$, we have:

$$\pi_i(p,c|\gamma^2\bar{\lambda},t,\gamma\bar{c},\alpha)=\gamma^2 \pi_i(p,c|\bar{\lambda},t,\bar{c},\alpha).$$

\end{claim}

\textit{Proof of \cref{claim: lem equivalence 1: pi proportional}.} To see this, note that under $(\bar{\lambda},t,\bar{c},\alpha)$, region $i$ will attract no drivers --hence will generate zero profit-- if any only if:

$$(\frac{c}{\bar{c}})^2-\frac{4t}{\bar{\lambda}(1-\alpha p)}<0$$

The corresponding condition under $(\gamma^2\bar{\lambda},t,\gamma\bar{c},\alpha)$ will be:

$$(\frac{c}{\gamma\bar{c}})^2-\frac{4t}{\gamma^2\bar{\lambda}(1-\alpha p)}<0$$

The above two conditions are equivalent. This finishes the proof of the claim. $\square$

Now let us assume $p$ and $c$ are such that region $i$ does get a positive number of drivers. There, from \cref{eq:ReservationValue2}, the number of drivers present in $i$ under $(\bar{\lambda},t,\bar{c},\alpha)$ will be uniquely give by:

$$n^*_i(p,c|\bar{\lambda},t,\bar{c},\alpha)=
\frac{\frac{c_i}{\bar{c}}+
\sqrt{(\frac{c_i}{\bar{c}})^2-\frac{4t_i}{\bar{\lambda}_i(1-\alpha p_i)}}}
{\frac{2}{\bar{\lambda}_i(1-\alpha p_i)}}
$$

The same quantity under $(\gamma^2\bar{\lambda},t,\gamma \bar{c},\alpha)$ will be uniquely give by:

$$n^*_i(p,c|\gamma^2\bar{\lambda},t,\gamma\bar{c},\alpha)=
\frac{\frac{c_i}{\gamma\bar{c}}+
\sqrt{(\frac{c_i}{\gamma\bar{c}})^2-\frac{4t_i}{\gamma^2\bar{\lambda}_i(1-\alpha p_i)}}}
{\frac{2}{\gamma^2\bar{\lambda}_i(1-\alpha p_i)}}
$$

$$=\gamma \frac{\frac{c_i}{\bar{c}}+
\sqrt{(\frac{c_i}{\bar{c}})^2-\frac{4t_i}{\bar{\lambda}_i(1-\alpha p_i)}}}
{\frac{2}{\bar{\lambda}_i(1-\alpha p_i)}}=\gamma n^*_i(p,c|\bar{\lambda},t,\bar{c},\alpha)$$

Also note that the equilibrium driver wait time $W_i$ in region $i$ under $(\bar{\lambda},t,\bar{c},\alpha)$ has to be $\frac{c_i}{\bar{c}}$ in order to equate driver earnings in the region with the reservation value. Similarly, under $(\gamma^2\bar{\lambda},t,\gamma\bar{c},\alpha)$, the driver wait time in region $i$ will be $\frac{c_i}{\gamma \bar{c}}$.

Next, note that for any $p,c$:

$$\pi_i(p,c|\bar{\lambda},t,\bar{c},\alpha)=(p_i-c_i)\times\frac{n^*_i(p,c|\bar{\lambda},t,\bar{c},\alpha)}{\frac{c_i}{\bar{c}}}$$

Also:

$$\pi_i(p,c|\gamma^2\bar{\lambda},t,\gamma\bar{c},\alpha)=(p_i-c_i)\times\frac{n^*_i(p,c|\gamma^2\bar{\lambda},t,\gamma\bar{c},\alpha)}{\frac{c_i}{\gamma\bar{c}}}$$

$$=(p_i-c_i)\times\frac{\gamma n^*_i(p,c|\bar{\lambda},t,\bar{c},\alpha)}{\frac{c_i}{\gamma\bar{c}}}=\gamma^2 \pi_i(p,c|\bar{\lambda},t,\bar{c},\alpha)$$

Therefore, the profit functions under $(\bar{\lambda},t,\bar{c},\alpha)$ and $(\gamma^2\bar{\lambda},t,\gamma\bar{c},\alpha)$ are fully proportional in each region, which mean they will have the exact same maximizers $p^*$ and $c^*$. Given these maximizers, we will have the number of drivers in each region $i$ under $(\gamma^2\bar{\lambda},t,\gamma\bar{c},\alpha)$ will be given by:

$$n^{*'}_i\equiv n^*_i(p^*,c^*|\gamma^2\bar{\lambda},t,\gamma\bar{c},\alpha)=\gamma n^*_i(p^*,c^*|\bar{\lambda},t,\bar{c},\alpha)\equiv \gamma n^*_i$$

Finally:

$$A^{*'}_i=\frac{n^{*'}_i}{(\frac{c^*_i}{\gamma\bar{c}})\times\gamma^2\bar{\lambda}_i}=\frac{\gamma n^{*}_i}{(\frac{c^*_i}{\gamma\bar{c}})\times\gamma^2\bar{\lambda}_i}=\frac{ n^{*}_i}{(\frac{c^*_i}{\bar{c}})\times\bar{\lambda}_i}=A^*_i$$

The proof of the lemma is now complete. $\square$

\begin{lemma}\label{lem: equivalance reservation wage price sensitivity}
 Assume $\gamma\in\mathbb{R}$ is a strictly positive number. Under primitives $(\bar{\lambda},t,\bar{c},\alpha)$, vectors $p^*$, $w^*$, and $n^*$ constitute an equilibrium if and only if under primitives $(\bar{\lambda}',t',\bar{c}',\alpha')=( \bar{\lambda},t,\gamma\bar{c},\frac{\alpha}{\gamma})$, vectors $p^{*'}=\gamma p^*$, $c^{*'}=\gamma c^*$, and $n^{*'}=n^*$ constitute an equilibrium  Additionally, access to rides in all regions are equal under these two equilibria: $\forall i: A_i^{*'}=A_i^*$.
\end{lemma}

\textbf{Proof of \cref{lem: equivalance reservation wage price sensitivity}.} The steps of the proof for this lemma are very similar to those in the proof of \cref{lem: equivalance demand reservation wage}. Hence, we skip the details. $\square$

We just point out that there is a simple intuition for this result: a change of primitives from $(\bar{\lambda},t,\bar{c},\alpha)$ to $( \bar{\lambda},t,\gamma\bar{c},\frac{\alpha}{\gamma})$ should not be expected to change the equilibria because it is effectively a ``currency change.'' The only two places where ``money'' becomes involved in the model are where demand is determined based on the price of a ride and where supply is determined based on wage. If the reservation hourly revenue is multiplied by $\gamma$ and, at the same time, the price sensitivity of passengers is divided by the same factor $\gamma$, it is as if we are looking at the same market but with a new currency one unit of which is worth $\frac{1}{\gamma}$ times one unit of the old currency. This should not change the equilibrium supply, demand, and access. It should only multiply the equilibrium prices and wages by $\gamma$.

\begin{lemma}\label{lem: equivalance demand region size}
 Assume $\gamma\in\mathbb{R}$ is a strictly positive number. Under primitives $(\bar{\lambda},t,\bar{c},\alpha)$, vectors $p^*$, $w^*$, and $n^*$ constitute an equilibrium if and only if under primitives $(\bar{\lambda}',t',\bar{c}',\alpha')=(\gamma \bar{\lambda},\gamma t,\bar{c},\alpha)$, vectors $p^{*'}= p^*$, $c^{*'}= c^*$, and $n^{*'}=\gamma n^*$ constitute an equilibrium  Additionally, access to rides in all regions are equal under these two equilibria: $\forall i: A_i^{*'}=A_i^*$.
\end{lemma}

\textbf{Proof of \cref{lem: equivalance demand region size}.} Again, the steps of this proof are very similar to those in the previous two lemmas. As such, we skip the details. $\square$.

Our next lemma combines the previous three.

\begin{lemma}\label{lem: equivalance full}
 Under primitives $(\bar{\lambda},t,\bar{c},\alpha)$, vectors $p^*$, $c^*$, and $n^*$ constitute an equilibrium if and only if under primitives $(\tilde{\lambda},1,1,1)$ where $\tilde{\lambda}_i=\frac{\bar{\lambda}_i}{(\bar{c}\alpha)^2t_i}$ for each $i$, the following vectors constitute an equilibrium: $p^{*'}=\alpha p^*$, $c^{*'}=\alpha c^*$, and $n^{*'}=\frac{1}{(\bar{c}\alpha)t} n^*$.  Additionally, access to rides in all regions are equal under these two equilibria: $\forall i: A_i^{*'}=A_i^*$.
\end{lemma}

\textbf{Proof of \cref{lem: equivalance full}.} Apply \cref{lem: equivalance demand reservation wage} with $\gamma=\frac{1}{\bar{c}\alpha}$ on primitives $(\bar{\lambda},t,\bar{c},\alpha)$. On the resulting primitives, apply \cref{lem: equivalance reservation wage price sensitivity} with $\gamma=\alpha$. Finally, on the resulting primitives, apply \cref{lem: equivalance demand region size} with $\gamma=\frac{1}{t}$. This will establish the equivalence claimed in this lemma. $\square$

With this result in hand, we now turn to the main two lemmas that prove \cref{prop:endogenous Wages and Prices}.

\begin{lemma}\label{lem: prop holds if it does for lambda 111}
\cref{prop:endogenous Wages and Prices} holds under all generic primitives $(\bar{\lambda},t,\bar{c},\alpha)$ if it holds under primitives in the form of $(\tilde{\lambda},1,1,1)$.
\end{lemma}

\textbf{Proof of \cref{lem: prop holds if it does for lambda 111}.} Suppose \cref{prop:endogenous Wages and Prices} holds for all primitives that take the form of $(\tilde{\lambda},1,1,1)$. We show it holds for a general $(\bar{\lambda},t,\bar{c},\alpha)$. To this end, choose vector $\tilde{\lambda}$ such that $\forall i:\, \tilde{\lambda}_i= \frac{\bar{\lambda}_i}{(\bar{c}\alpha)^2t_i}$. Next, we prove each statement of \cref{prop:endogenous Wages and Prices} for general primitives.

\textit{Statement 1 on economies of density.} There is a unique equilibrium under $(\tilde{\lambda},1,1,1)$. Denote it $\tilde{n}$. By \cref{lem: equivalance full}, there is also a unique equilibrium $n^*$ under $(\bar{\lambda},t,\bar{c},\alpha)$, which is given by $\tilde{n}_i(\bar{c}\alpha t_i)$ for each $i$. Also, for any $i$ for which $\tilde{n}_i>0$, we know there are unique equilibrium price $\tilde{p}_i$ and wage $\tilde{w}_i$. This happens if and only if $n^*_i>0$. In this case, again according to \cref{lem: equivalance full}, the price and wage are also unique under $(\bar{\lambda},t,\bar{c},\alpha)$.

\textit{Statement 2 on economies of density.} By assumption know there is a $\mu$ such that:

$$\forall i:\, \frac{\tilde{\lambda}_i}{1}<\mu \Leftrightarrow \tilde{n}_i=0$$

Note that by construction, $\tilde{\lambda}_i<\mu$ is equivalent to $\frac{\bar{\lambda}_i}{t_i}<\mu (\bar{c}\alpha)$. Also note that $n^*_i=0$ if and only if $\tilde{n}_i=0$. Therefore, we have:

$$\forall i:\, \frac{\bar{\lambda}_i}{t_i}<\mu(\bar{c}\alpha) \Leftrightarrow n^*_i=0$$

which finishes the proof of statement 2 on economies of density.

\textit{Statement 3 on economies of density.} By assumption, this statement holds under $(\tilde{\lambda},1,1,1)$. By the construction of $\tilde{\lambda}$, it is immediate that:

$$\tilde{\lambda}_i>\tilde{\lambda}_j\Leftrightarrow
\frac{\bar{\lambda}_i}{t_i}>
\frac{\bar{\lambda}_j}{t_j}$$

Also by \cref{lem: equivalance full}, we know that equilibrium access to rides in each region $i$ is the same under $(\tilde{\lambda},1,1,1)$ and $(\bar{\lambda},t,\bar{c},\alpha)$. It immediately follows that statement 3 holds under  $(\bar{\lambda},t,\bar{c},\alpha)$.

\textit{Statement 4 on economies of density.} Note that by construction of $\tilde{\lambda}$ and by \cref{lem: equivalance full}, for any $i,j$ the following hold:

\begin{enumerate}
    \item $\tilde{\lambda}_i\geq\tilde{\lambda}_j\Leftrightarrow
\frac{\bar{\lambda}_i}{t_i}\geq 
\frac{\bar{\lambda}_j}{t_j}$ and a similar statement when both inequalities are strict.
    \item $\tilde{c}_i\leq \tilde{c}_j \Leftrightarrow c^*_i\leq c^*_j$ and a similar statement when both inequalities are strict.
    \item $\tilde{p}_i\leq \tilde{p}_j \Leftrightarrow p^*_i\leq p^*_j$ and a similar statement when both inequalities are strict.
    \item $\tilde{p}_i-\tilde{p}_j\leq \tilde{c}_i-\tilde{c}_j \Leftrightarrow p^*_i-p^*_j\leq c^*_i-c^*_j$ and a similar statement when both inequalities are strict.
\end{enumerate}

These four statements, together with the assumption that statement 4 of the proposition on economies of density holds under $(\tilde{\lambda},1,1,1)$ imply that it also holds under $(\bar{\lambda},t,\bar{c},\alpha)$.

Next, we prove the statements of the proposition that are about the role of market thickness. Key to proving all of those results is the observation that the same equivalence established between $(\tilde{\lambda},1,1,1)$ and $(\bar{\lambda},t,\bar{c},\alpha)$ by \cref{lem: equivalance full} may also be established between $(\gamma\tilde{\lambda},1,1,1)$ and $(\gamma\bar{\lambda},t,\bar{c},\alpha)$ for any positive real $\gamma$. Even the multipliers $\frac{1}{(\bar{c}\alpha)^2t_i}$ remain the same. With this observation, we turn to the proofs.

\textit{Statement 1 on the role of market thickness.} By \cref{lem: equivalance full}, we know $n^*_i>0\Rightarrow\tilde{n}_i>0$. By the proposition being true for $(\tilde{\lambda},1,1,1)$, we know that $\tilde{n}_i>0\Rightarrow\tilde{n}'_i>0$ where $\tilde{n}'$ is the equilibrium driver allocation under $(\gamma\tilde{\lambda},1,1,1)$. Finally, again by \cref{lem: equivalance full}, we have $\tilde{n}'_i>0\Rightarrow n^{*'}_i>0$. This finishes the proof of this statement.

\textit{Statement 2 on the role of market thickness.} \cref{lem: equivalance full} shows that access to rides in all regions is preserved under the transformation from $(\bar{\lambda},t,\bar{c},\alpha)$ to $(\tilde{\lambda},1,1,1)$ and the transformation from $(\gamma\bar{\lambda},t,\bar{c},\alpha)$ to $(\gamma\tilde{\lambda},1,1,1)$.  This, together with $\tilde{\lambda}_i\geq\tilde{\lambda}_j\Leftrightarrow
\frac{\bar{\lambda}_i}{t_i}\geq 
\frac{\bar{\lambda}_j}{t_j}$ and the assumption that the proposition holds under $(\tilde{\lambda},1,1,1)$ gives the proof of this statement under $(\gamma\bar{\lambda},t,\bar{c},\alpha)$.

\textit{Statement 3 on the role of market thickness.} The logic for the proof of this statement is identical to that for the proof of the previous one.

The proof of \cref{lem: prop holds if it does for lambda 111} is now complete. $\square$

Our last step in proving \cref{prop:endogenous Wages and Prices} is to show that it does indeed hold under $(\tilde{\lambda},1,1,1)$. The convenient feature of this remaining step is that it can be done fully numerically. The following lemma takes on this task.

\begin{lemma}\label{lem: prop holds for lambda 111}
\cref{prop:endogenous Wages and Prices} holds if $(\bar{\lambda},t,\bar{c},\alpha)$ takes the specific form of $(\tilde{\lambda},1,1,1)$.
\end{lemma}

\textbf{Proof of \cref{lem: prop holds for lambda 111}.} We start by proving statement 2 and then move to statement 1 on economies of density.

\textit{Statement 2 on economies of density.} Here, we are looking for conditions on potential demand $\bar{\lambda}_i$ such that the platform can get a positive number of drivers into region $i$ without sustaining any loss. Note that from our analysis before, we know the region can get a positive number of drivers if:

$$(\frac{c_i}{\bar{c}})^2\geq \frac{4t_i}{\bar{\lambda}_i(1-\alpha p_i)}$$

From $\bar{c}=t_i=\alpha=1$ and $\bar{\lambda}_i=\tilde{\lambda}_i$, this turns into:

$$c_i^2\geq \frac{4}{\tilde{\lambda}_i(1- p_i)}$$

Or, equivalently: $$\tilde{\lambda}_i\geq \frac{4}{c_i^2(1- p_i)}$$

As such, the minimum requirement volume for $\tilde{\lambda}_i$ depends on the maximum possible value for $c_i^2(1- p_i)$ subject to the constraint that the platform does not run any losses. This happens when $p_i=c_i$. Therefore, the region will get a positive number of drivers if and only if:

$$\tilde{\lambda}_i\geq \frac{4}{\max_{c_i} c_i^2(1- c_i)}=27$$

This says each region $i$ gets a positive number of drivers if and only if $\tilde{\lambda}_i\geq 27$ which proves statement 2 of the proposition on economies of density.

\begin{figure}[H]
\centering
\caption{Numerical results that are necessary to prove \cref{lem: prop holds for lambda 111}.}\label{fig:prop 7}
\centering
\vspace{-0.2in}
\subfigure[Response of wage to density]{\includegraphics[width=0.48\textwidth]{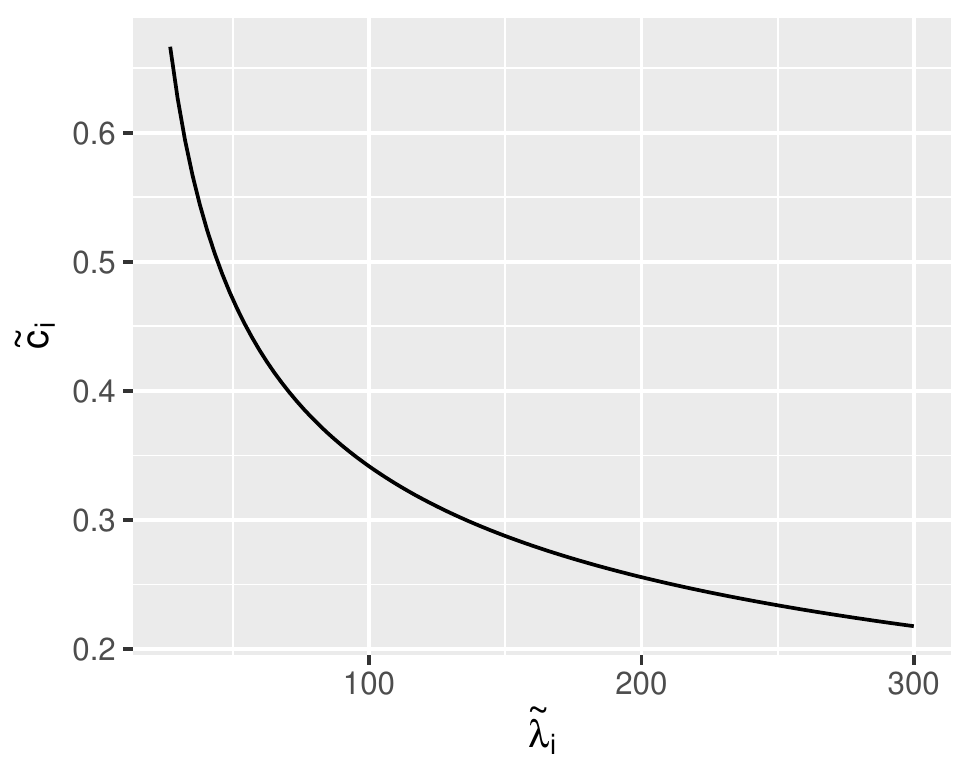}}\label{fig:prop7Wages}
\subfigure[Response of price to density]{\includegraphics[width=0.48\textwidth]{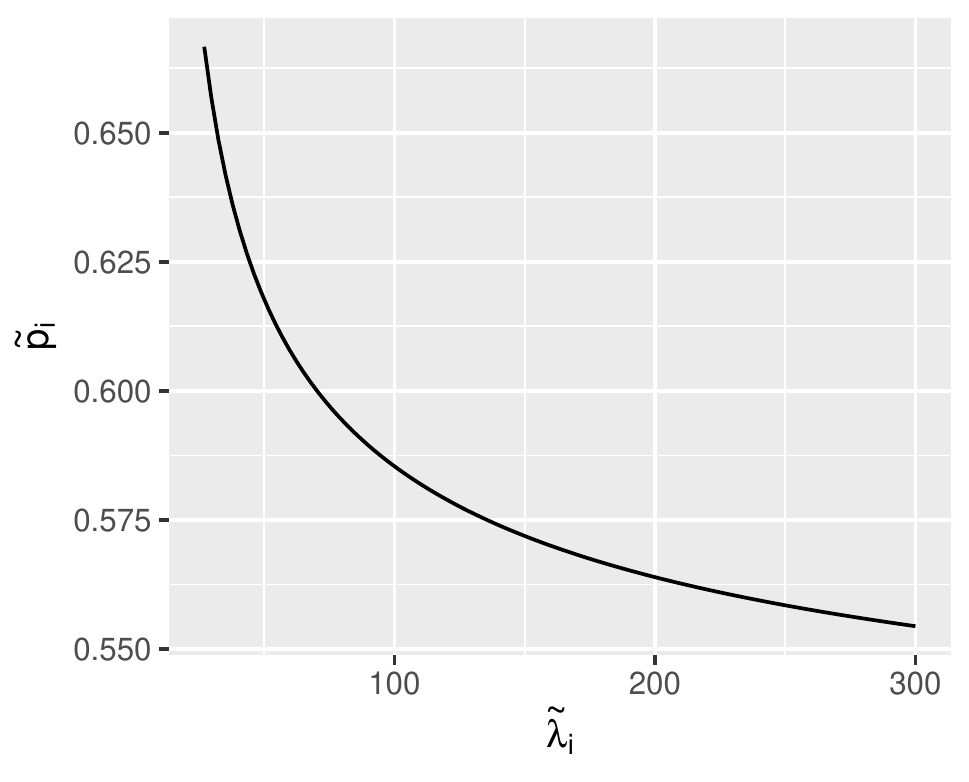}}\label{fig:prop7Prices}\\
\vspace{0.2in}
\subfigure[Response of access to density]{\includegraphics[width=0.48\textwidth]{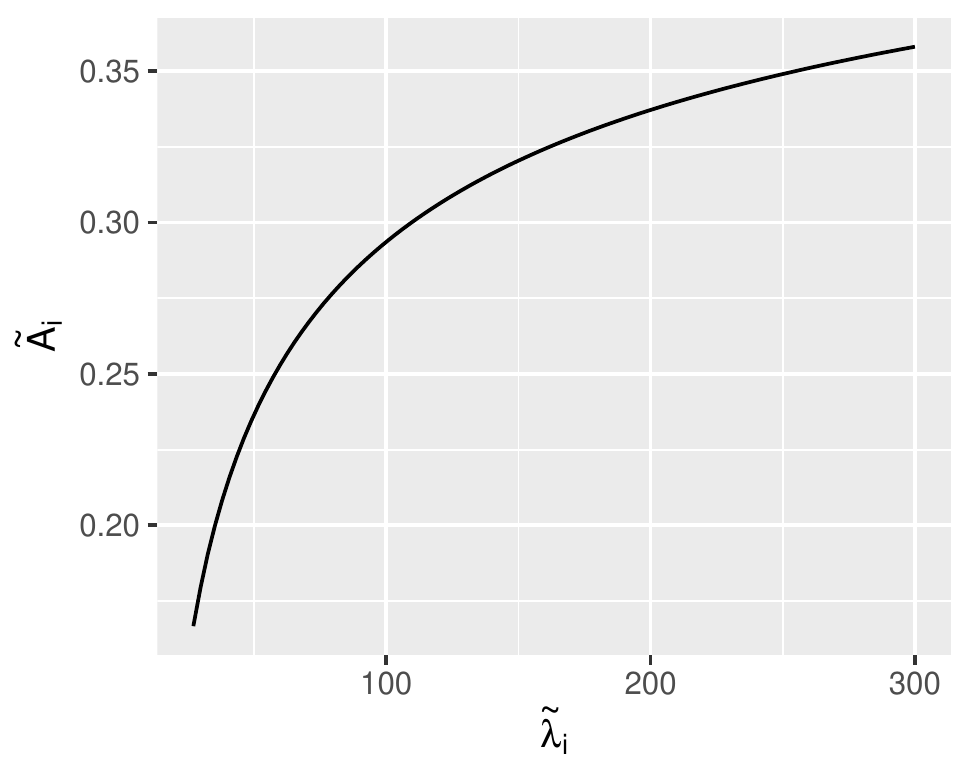}}\label{fig:prop7Access}
\subfigure[Response of margin (price less wage) to density]{\includegraphics[width=0.48\textwidth]{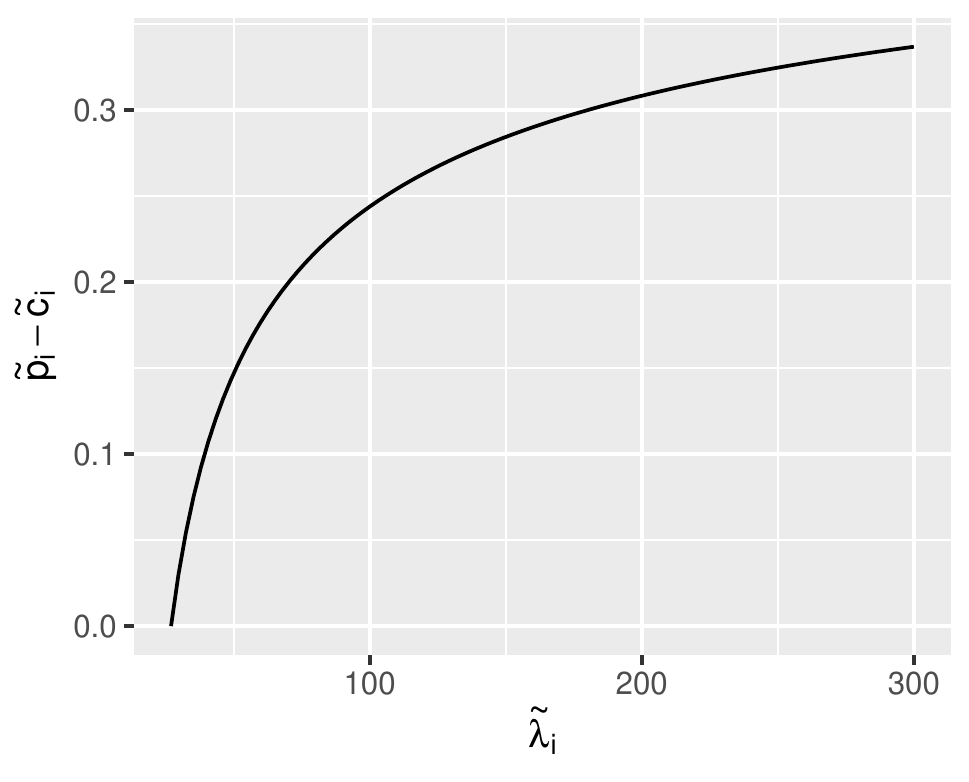}}\label{fig:prop7Margin}\\
\subfigure[Response of log access to log density]{\includegraphics[width=0.48\textwidth]{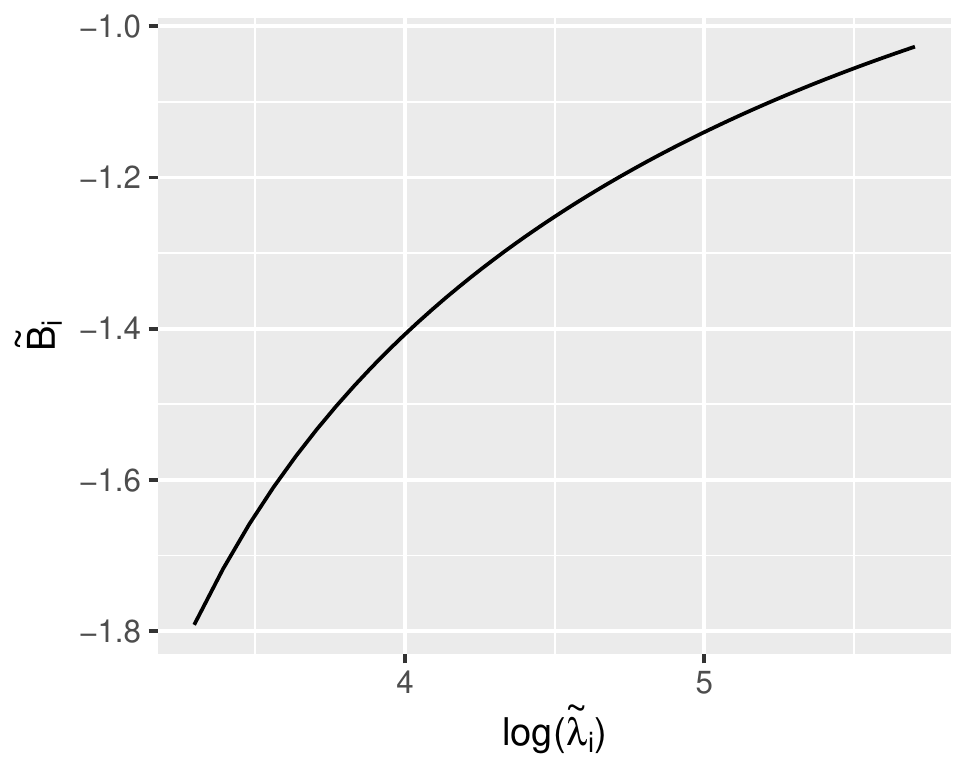}}\label{fig:prop7B}

\end{figure}

Next, we prove the rest of the statements numerically. To do this, we need to study the equilibrium price $\tilde{p}_i$, equilibrium wage $\tilde{c}_i$, equilibrium number of drivers $\tilde{n}_i$, and equilibrium access $\tilde{A}_i$ in a given region $i$ as a function of $\tilde{\lambda}_i$. The key is that all of the other parameters are equal to 1 and $\tilde{\lambda}_i$ is the only parameter changing. Therefore, for any value that $\tilde{\lambda}_i$ assumes, we have a fully numerical problem that can be solved using a software such as R. Below, we are providing graphs of these equilibrium quantities as functions of $\tilde{\lambda}_i$. \cref{fig:prop 7} provides these results. The figure gives the unique optimal price and wage computed for  $\tilde{\lambda}_i\geq 27$. As shown in the figure, optimal price $\tilde{p}_i$ and optimal wage $\tilde{c}_i$ both strictly decrease in density $\tilde{\lambda}_i$. This means if two regions $i,j$ are such that $\tilde{\lambda}_i>\tilde{\lambda}_j$, then region $j$ will have a higher price and a higher wage in the equilibrium. Also margin $\tilde{p}_i-\tilde{c}_i$ is strictly increasing in density. This means if two regions $i,j$ are such that $\tilde{\lambda}_i>\tilde{\lambda}_j$, then region $j$ will have a lower margin in the equilibrium. Additionally, access $\tilde{A}_i$ is strictly increasing in density. This means if two regions $i,j$ are such that $\tilde{\lambda}_i>\tilde{\lambda}_j$, then region $j$ will have a lower access to rides in the equilibrium. These results show that the statements on economies of density hold.

As for statements on the role of market thickness, statement 1 follows directly from statement 2 on economies of density. Statement 2 on the role of market thickness can be seen from the last panel in \cref{fig:prop 7} which shows that log of access is concave in log of density. Recall from \cref{lem: If B concave} that this would imply statement 2 on the role of market thickness. Finally, statement 3 on the role of market thickness is straightforward and left to the reader (the key would be to note that as the market gets sufficiently large, pickup times in all regions become negligible; hence the pricing problem in all regions boils down to that in a model without economies of density, giving all regions the same price, wage, and access to rides).

This finishes the proof of \cref{lem: prop holds for lambda 111} and, hence, \cref{prop:endogenous Wages and Prices}.$\blacksquare$

\textbf{Proof of Proposition \ref{prop: inter-region equivalence}.} First suppose $n^*$ is \textit{not} a driver equilibrium in the version of the problem with no inter-region rides. This, by definition, means that either a small number $\delta_0$ of new drivers can join one of the regions $i$ from outside and strictly improve their payoffs, of a small $\delta_0$ can relocate from $i$ to $i'$ or to the outside option and have a strictly positive improvement. In that case, one can find the corresponding small perturbation in $\rho$ that would in a given amount of time, change $n^*$ in a way that would increase the payoff in those regions. Similarly, if we assume $(n^*,\rho^*)$ with (for a $\rho^*$ that guarantees stability) is not an equilibrium of the version of the problem with cross-region rides, one can use the small profitable perturbation in $\rho^*$ and use them to construct changes in $n^*$ that yield strictly improved payoffs in some regions in the original version. This finishes the proof of the proposition. $\blacksquare$

\textbf{Proof of \cref{prop: testable implications 1}.} The proof is straightforward:
$$\frac{A_1}{A_2}=\frac{\frac{r_1}{\bar{\lambda}_1}}{\frac{r_2}{\bar{\lambda}_2}}$$

But we know that $\frac{r_i}{\bar{\lambda}_i}=\frac{r_{i,3-i}}{\bar{\lambda}_{3-i}}$ for $i=1,2$. Therefore, the above equality yields:

$$\frac{A_1}{A_2}=\frac{\frac{r_{12}}{\bar{\lambda}_{12}}}{\frac{r_{21}}{\bar{\lambda}_{21}}}$$

From balancedness, we have $\bar{\lambda}_{12}=\bar{\lambda}_{21}$. Thus:

$$\frac{A_1}{A_2}=\frac{r_{12}}{r_{21}}=RO_1.\blacksquare$$

\textbf{Proof of \cref{prop: testable implications 2}.} This proof is similar to that of the previous proposition.

$$\frac{A_1/A_2}{A'_1/A'_2}=\frac{\frac{r_1/\bar{\lambda}_1}{r_2/\bar{\lambda}_2}}{\frac{r'_1/\bar{\lambda}'_1}{r'_2/\bar{\lambda}'_2}}$$

$$=\frac{\frac{r_{12}/\bar{\lambda}_{12}}{r_{21}/\bar{\lambda}_{21}}}{\frac{r'_{12}/\bar{\lambda}'_{12}}{r'_{21}/\bar{\lambda}'_{21}}}=\frac{r_{12}/r_{21}}{r'_{12}/r'_{21}}\times \frac{\bar{\lambda}'_{12}/\bar{\lambda}'_{21}}{\bar{\lambda}_{12}/\bar{\lambda}_{21}}$$

By similar unbalancedness, the second term in the latter expression is equal to 1. Thus, we get:

$$\frac{A_1/A_2}{A'_1/A'_2}=\frac{r_{12}/r_{21}}{r'_{12}/r'_{21}}=\frac{RO_1}{RO'_1}.\blacksquare$$

\section{Proof of the Proposition in the Empirical Section}\label{appx: proof empirical}
\textbf{Proof of \cref{prop: regression equivalance 1}.} The basic idea behind this proof is that the difference between $RO_{ikd}$ and $\frac{\bar{\lambda}^\rightarrow_{ikd}}{\bar{\lambda}^\rightarrow_{ikd}}$ ``gets absorbed'' by the $Q$-level fixed effects, keeping intact all of the other coefficients as well as the error terms. Here, we provide a proof for the case where fixed effects embedded in controls $X_{ikd}$ are exactly at the $Q$ level. We will skip the proof for the case where the fixed effects are at a level finer than $Q$ but $\frac{\bar{\lambda}^\rightarrow_{ikd}}{\bar{\lambda}^\rightarrow_{ikd}}$ can be determined fully based on $Q$. That proof is conceptually the same (because finer fixed effects will absorb the differences more easily) but the notations will be much less clean.

Now, assume the matrix of control variables  $X$ is given by:

$$X=[F^Q\quad \bar{X}]$$

where $F^Q$ is the set of columns giving the $Q$-level fixed effects and $\bar{X}$ consists of the rest of the controls. That is, each column in $F^Q$ corresponds to one  possible value $q$ that $Q_{ikd}$ can assume; and each row $ikd$ of $F^Q$ in the column corresponding to $q$ is 1 if $Q_{ikd}=q$ and zero otherwise.

Consequently, also assume $\beta=[\beta_F\quad \beta_{\bar{X}}]$. We now introduce the following lemma.

\begin{lemma}\label{lem: regression equivalence}
Coefficients $\alpha$ and $\beta=[\beta_F\quad \beta_{\bar{X}}]$ and error terms $\epsilon_{ikd}$ satisfy \cref{eq:Regression Test econ of density} if and only if coefficients $\alpha$, $\tilde{\beta}$, and error terms $\epsilon_{ikd}$ satisfy \cref{eq:Regression Test econ of density ideal data}, where $\tilde{\beta}$ is defined as $[\beta_G\quad \beta_{\bar{X}}]$, and $\beta_G$ is a vector constructed from $\beta_F$ by subtracting $\log(g(q))$ from any element of $\beta_F$ from the column in $F^Q$ that corresponds to value $q$.
\end{lemma}

\textbf{Proof of \cref{lem: regression equivalence}.}

$$\log(RO_{ikd})=\alpha \log(D^\leftarrow_{ikd})+\beta X_{ikd}+\epsilon_{ikd}\Leftrightarrow$$

$$\log(RO_{ikd})=\alpha \log(D^\leftarrow_{ikd})+\beta_F\times F^Q_{ikd}+\beta_{\bar{X}}\times\bar{X}_{ikd} +\epsilon_{ikd}\Leftrightarrow$$

$$\log(RO_{ikd})-\log(\frac{\bar{\lambda}^\rightarrow_{ikd}}{\bar{\lambda}^\leftarrow_{ikd}})=\alpha \log(D^\leftarrow_{ikd})+
(\beta_F\times F^Q_{ikd}-\log(\frac{\bar{\lambda}^\rightarrow_{ikd}}{\bar{\lambda}^\leftarrow_{ikd}}) )+
\beta_{\bar{X}}\times\bar{X}_{ikd} +\epsilon_{ikd}\Leftrightarrow$$

$$\log(RO_{ikd})-\log(\frac{\bar{\lambda}^\rightarrow_{ikd}}{\bar{\lambda}^\leftarrow_{ikd}})=\alpha \log(D^\leftarrow_{ikd})+
(\beta_F\times F^Q_{ikd}-\log(g(Q_{ikd})) )+
\beta_{\bar{X}}\times\bar{X}_{ikd} +\epsilon_{ikd}\Leftrightarrow$$

$$\log(RO_{ikd})-\log(\frac{\bar{\lambda}^\rightarrow_{ikd}}{\bar{\lambda}^\leftarrow_{ikd}})=\alpha \log(D^\leftarrow_{ikd})+
\beta_G\times F^Q_{ikd}+
\beta_{\bar{X}}\times\bar{X}_{ikd} +\epsilon_{ikd}\Leftrightarrow$$

$$\log(RO_{ikd})-\log(\frac{\bar{\lambda}^\rightarrow_{ikd}}{\bar{\lambda}^\leftarrow_{ikd}})=\alpha \log(D^\leftarrow_{ikd})+
\tilde{\beta} X_{ikd} +\epsilon_{ikd}\Leftrightarrow$$

$$\log(\frac{r^\rightarrow_{ikd}}{r^\leftarrow_{ikd}})-\log(\frac{\bar{\lambda}^\rightarrow_{ikd}}{\bar{\lambda}^\leftarrow_{ikd}})=\alpha \log(D^\leftarrow_{ikd})+
\tilde{\beta} X_{ikd} +\epsilon_{ikd}\Leftrightarrow$$

$$\log(\frac{r^\rightarrow_{ikd}}{\bar{\lambda}^\rightarrow_{ikd}})-\log(\frac{r^\leftarrow_{ikd}}{\bar{\lambda}^\leftarrow_{ikd}})=\alpha \log(D^\leftarrow_{ikd})+
\tilde{\beta} X_{ikd} +\epsilon_{ikd}\Leftrightarrow$$

$$\log(A^\rightarrow_{ikd})-\log(A^\leftarrow_{ikd})=\alpha \log(D^\leftarrow_{ikd})+
\tilde{\beta} X_{ikd} +\epsilon_{ikd}\Leftrightarrow$$

$$\log(\frac{A^\rightarrow_{ikd}}{A^\leftarrow_{ikd}})=\alpha \log(D^\leftarrow_{ikd})+
\tilde{\beta} X_{ikd} +\epsilon_{ikd}$$

which finishes the proof of the lemma. $\square$

Note that according to \cref{lem: regression equivalence}, each error term $\epsilon_{ikd}$ remains intact under this equivalence. It follows immediately that the unique estimated value for $\alpha$, which is part of estimated parameters minimizing the GMM error in \cref{eq:Regression Test econ of density}, will also be the unique estimated value for $\alpha$ in \cref{eq:Regression Test econ of density ideal data} as well. $\blacksquare$

\section{Relative Outflows at the Borough Level}\label{sec:Appx:RelOutfows Borough}
In this section, we present relative outflows at the borough level for Uber, Lyft, and Via. The observed patterns are similar to those at the zones level: (i) busier boroughs have higher relative outflows; and (ii) the relative-outflow gap is wider for smaller platforms. 
This figure should also help with the understanding of some of the less intuitive patterns in the zones-level relative outflows in \cref{figure:relOutflows1718_zones} in the main text. In \cref{figure:relOutflows1718_zones}, one can see that some relative outflows in parts of Staten Island and Queens are higher than some in Manhattan. Here, we would like to point out why these patterns emerge and explain why the potential bias they cause is in fact against our results (meaning the magnitudes of our results would be even larger if we ``corrected'' for these patterns).

\begin{figure}[H]
\centering
 
\subfigure[]{\includegraphics[width=0.48\textwidth]{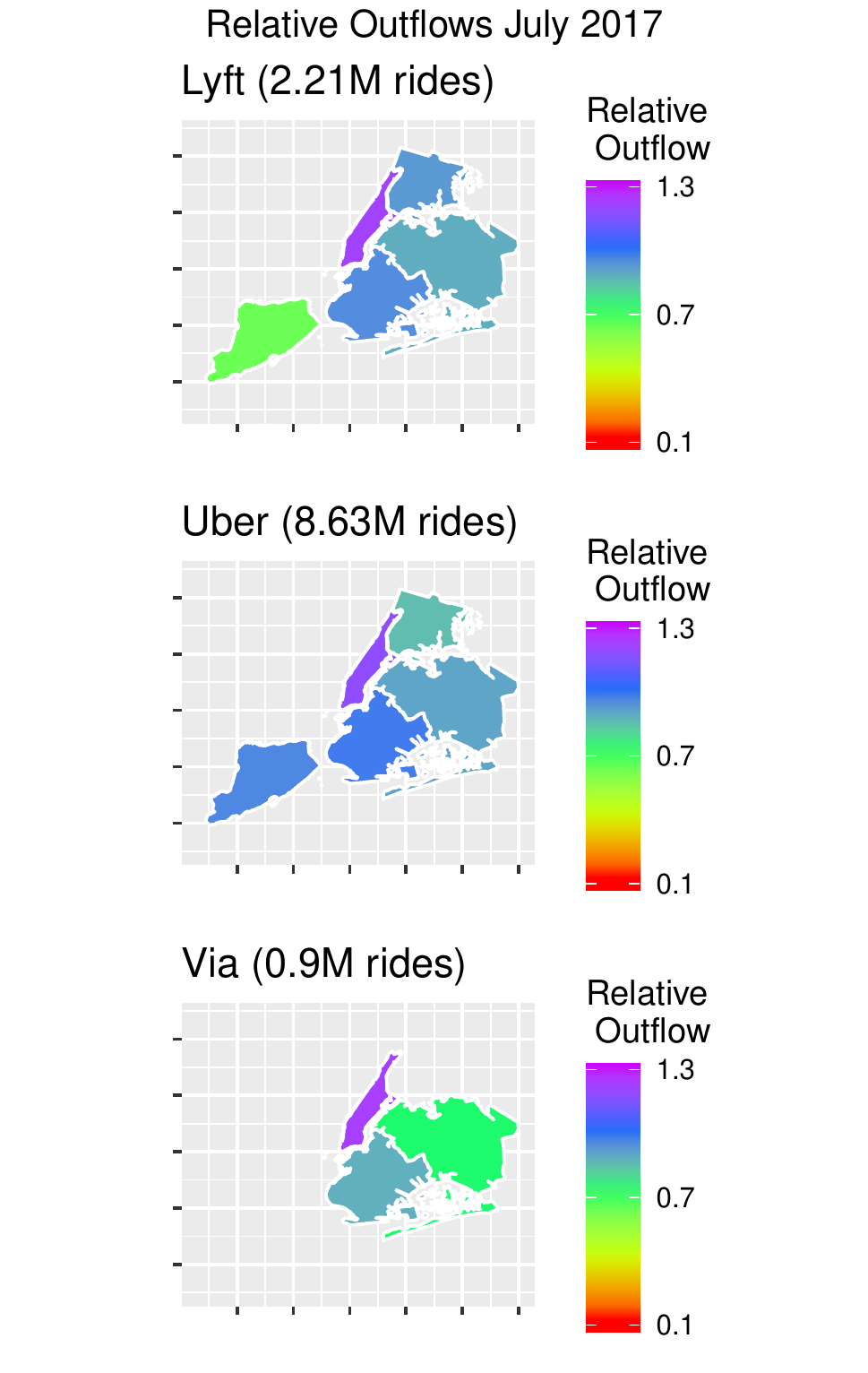}} 
\subfigure[]{\includegraphics[width=0.48\textwidth]{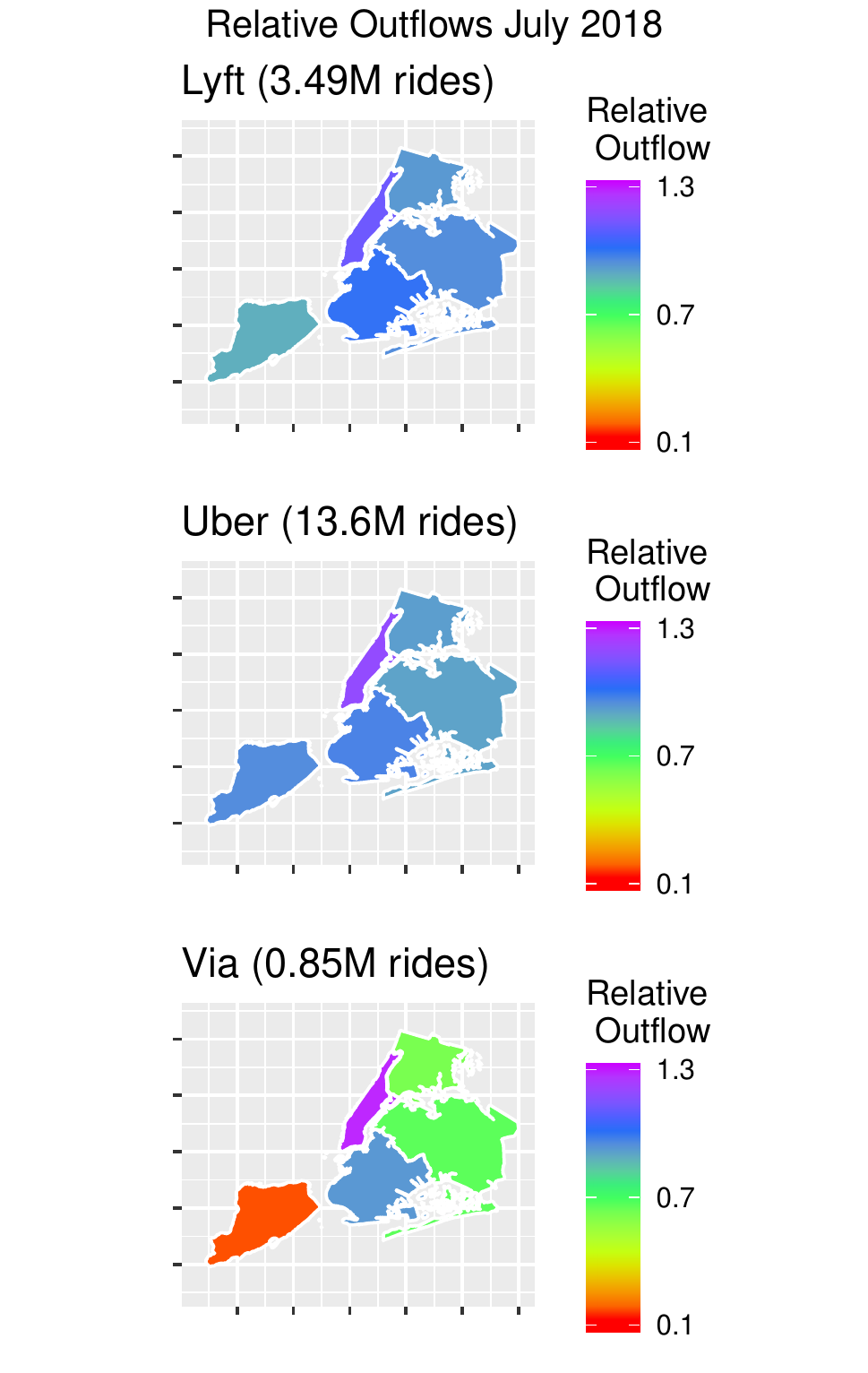}} 
\caption{\scriptsize{Relative outflows for Lyft, Uber, and Via at the borough level. Similar patterns to those at the zone level can be observed.}}\label{figure:relOutflows1718_boroughs}
\end{figure}

We believe the large relative outflows in Staten Island and Queens arise from two sources. First, in Queens, the areas with larger relative outflows tend to be close to airports. We believe this is likely a consequence of the fact that rideshare companies have historically been more restricted when it comes to picking up passengers at airports. Therefore, drivers who drop off passengers at airports may need to look for new passengers nearby which creates economies of density in those regions. The second, and more important, reason for larger relative outflows in some ``unexpected'' regions has more to do with the nature of intra-city transportation. As an extreme case, suppose all rides took place within boroughs, meaning each ride's origin was located in the same borough as its destination. In that case, one would expect the highest relative outflow in each borough to be above 1 and the lowest relative outflow in each borough to be below 1 (this is because rides are balanced overall). Therefore, each borough's highest relative outflow would be higher than any other borough's lowest relative outflow, \textit{no matter how the densities of the two boroughs compare.} Of course the assumption that all rides are within-borough is an extreme assumption but if a reasonably large fraction of rides happen within boroughs, we should expect similar relative outflows patterns. Consistent with our hypothesis about this explanation, these patterns are not observed when we look at the borough-level relative outflows graphs in \cref{figure:relOutflows1718_boroughs}. 

Note that these patterns (especially the second one) can be reasonably controlled for using borough fixed effects. We would also like to note that any potential bias these patterns may cause in our results would be against our hypothesis that $\alpha$ in \cref{eq:Regression Test econ of density} and \cref{eq:Regression Test econ of density ideal data} is positive. This is in line with the observation that once we control for boroughs, the estimated $\alpha$ increases.

\section{Discussion: Evidence of Economies of Density in the Taxicab Market?}\label{section:discussion}

We conjecture that economies of density also exist in the taxicab market and that they lead to agglomeration of supply in busier regions. Our theoretical model, however, does not directly speak to the taxicab market. Our model assumes a central dispatch structure for the matching of drivers to riders; and one of the key forces behind our results is the pickup time. In the taxi market, on the other hand, there are search frictions; but once a cab and a passenger find each other, they are not far apart (i.e., no pickup time).

\begin{figure}[H]
\centering
 \caption{\scriptsize{ Dropoff densities  in panel (a) and relative outflows in panel (b) by taxi zone across NYC for Yellow Cab rides during January 2009 (before Green Taxi was launched to serve the outer boroughs). The two quantities seem spatially highly correlated. We interpret this to indicate concentration of supply especially in mid and lower Manhattan due to economies of density.}}\label{fig: yellow taxi flows}
\subfigure[]{\includegraphics[width=0.48\textwidth]{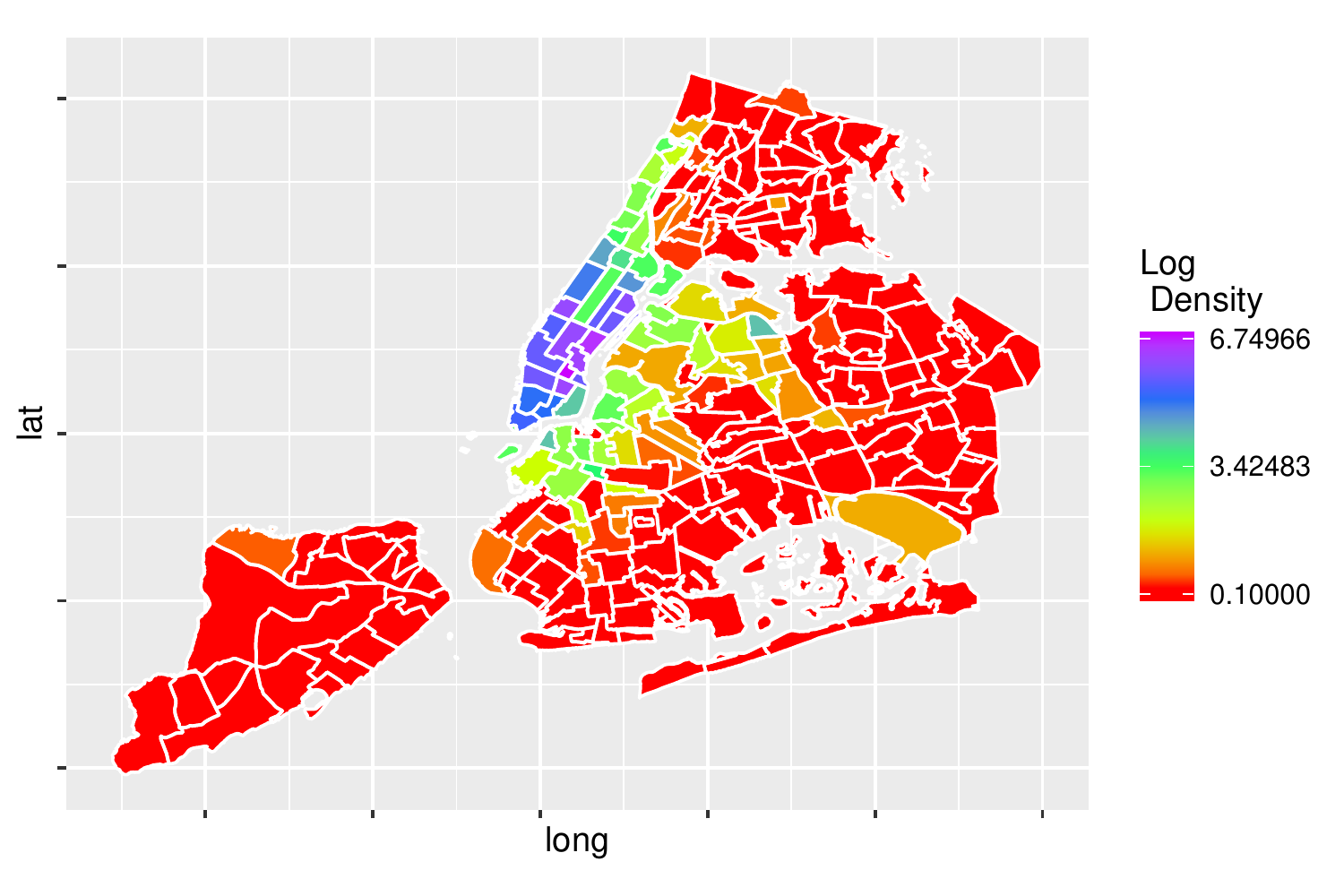}} 
\subfigure[]{\includegraphics[width=0.48\textwidth]{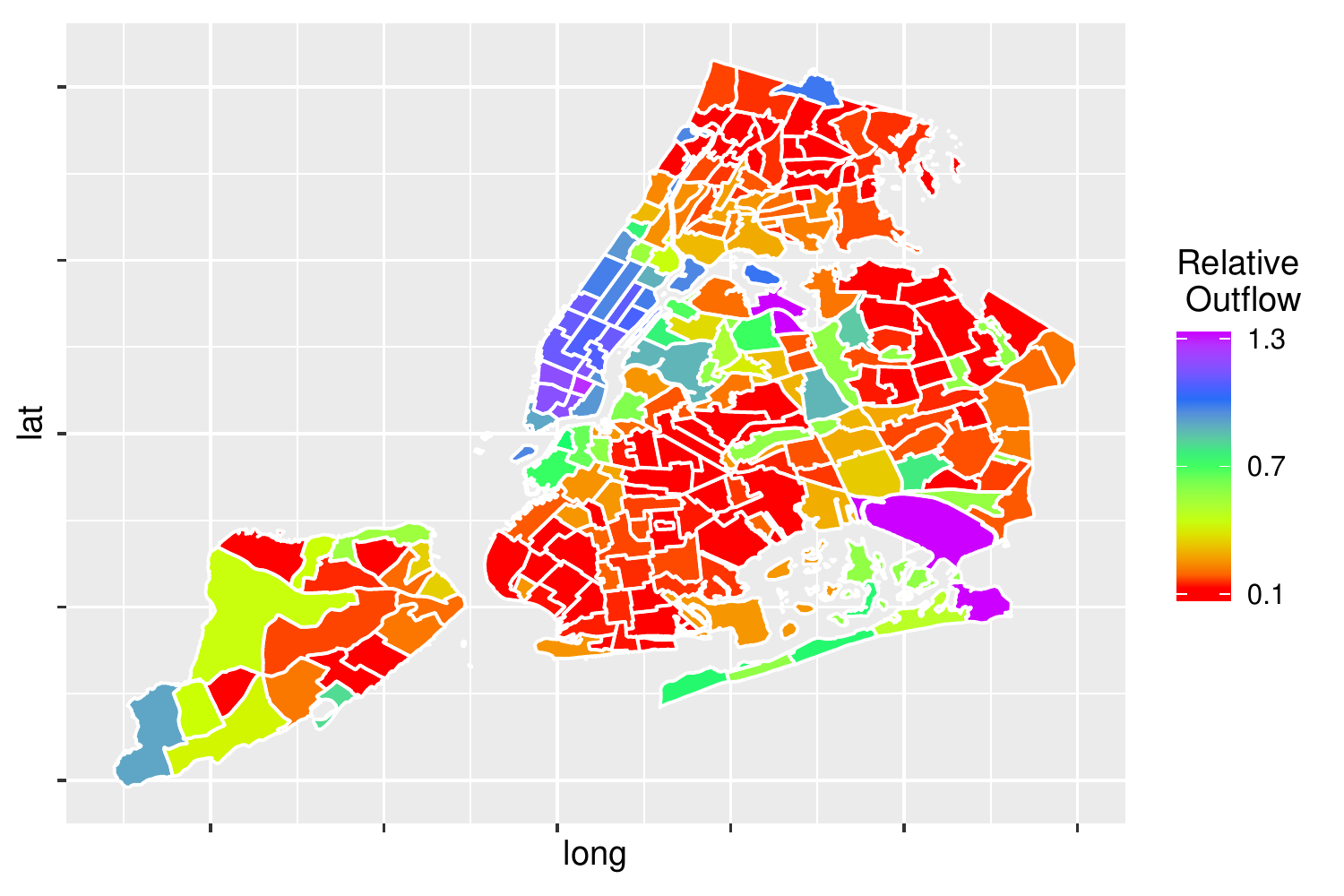}} 

\end{figure}

Unlike our theory model, our empirical approach (i.e., relative outflows analysis) can be applied regardless of the type of the matching mechanism in the market. Thus, we perform it  on data on Yellow Taxi rides in NYC during January 2009 (which is the first month for which data on taxicab rides are available from the TLC website) to gain insight into whether economies of density are at work there. \cref{fig: yellow taxi flows} suggests a strong positive association between relative outflows $RO_i$ and dropoffs densities $D^\leftarrow_i$ across regions $i$. Strikingly, numerous zones in the outer boroughs have relative outflows that are below 10\%. This means in those regions (which are regions with smaller dropoff rates per square mile), more than 90\% of those who enter using a taxi ride end up exiting using a different transportation option.

Next, we carry out the same regression analysis as in \cref{eq:Regression Test econ of density} for Yellow Taxi rides. The results, under all regression specifications with regards to borough and zone type come back positive and significant. Also the magnitude is much higher compared to the results in \cref{tab:zoneLevelRegression}, suggesting a stronger  economies of density effect in the taxicab market compared to the rideshare market.

\begin{table}[!htbp] \centering 
  \caption{\scriptsize{Results from regressing relative outflows on dropoff densities for yellow taxi rides in January 2009 (before the green taxi program started). The regression specification is the same as \cref{eq:Regression Test econ of density}. Similar to the results from the same regression on rideshare data, here too the coefficient $\alpha$ on dropoff density is positive and significant and this is robust to different fixed effects specifications. The magnitude of the estimated $\alpha$ is much higher here compared to rideshare, suggesting more pronounced economies of density in the taxicab market. }} 
  \label{} 
\begin{tabular}{@{\extracolsep{5pt}}lcccc} 
\\[-1.8ex]\hline 
\hline 
 & \multicolumn{4}{c}{\textit{Dependent variable: log relative outflow}} \\ 
\cline{2-5} 
& (1) & (2) & (3) & (4)\\ 
\hline \\[-1.8ex] 
 log dropoff density & 0.184$^{***}$ & 0.172$^{***}$ & 0.183$^{***}$ & 0.165$^{***}$ \\ 
  & (0.024) & (0.039) & (0.024) & (0.041) \\ 
 
  Fixed Effects$^{\dagger}$ & Constant & B &  Z & B$\times$Z \\ 
\hline \\[-1.8ex] 
Observations & 194 & 194 & 194 & 194 \\ 
R$^{2}$ & 0.233 & 0.833 & 0.792 & 0.839 \\ 
Adjusted R$^{2}$ & 0.229 & 0.828 & 0.786 & 0.819 \\ 

\hline 
\hline \\[-1.8ex] 
\textit{Note:}  & \multicolumn{4}{r}{$^{*}$p$<$0.1; $^{**}$p$<$0.05; $^{***}$p$<$0.01} \\ 
\multicolumn{5}{l}{$\dagger$: B:Borough, Z:Zone-type}\\
\end{tabular} 
\end{table}

Our results in this section suggest that economies of density, and the agglomeration that it leads to, have an important role in shaping the spatial distribution of supply in the taxicab market (Note that January 2009 was before ``Green Taxis'' were established. Thus, the lower access to yellow taxis in the outer boroughs cannot be because of the presence of green taxis). This points to a future research direction on the modeling and empirical analysis of transportation markets with decentralized matching systems such as taxi cabs: the incorporating of economies of density and its implications.
Much of the literature on transportation markets with decentralized matching (e.g., \cite{lagos2000alternative,lagos2003analysis,buchholz2018spatial,brancaccio2019geography,brancaccio2019efficiency}) is built on the assumption that the ``matching function'' between passengers and drivers is homogeneous of degree one. That is, the number of rides originated in region $i$ doubles if both the number of passengers searching for drivers and the number of drivers searching for passengers in $i$ double.  This abstracts away from economies of density. This would likely lead to an attribution of the relative outflows patterns in Figure \ref{fig: yellow taxi flows} to substantially lower need for cab rides entering Manhattan than rides exiting it.\footnote{If the matching-function-inversion approach from this literature is applied to recover search volumes for rides across regions, we would indeed conjecture that the result would point to fewer searches for rides from the outer regions. We would like to emphasize, however, that fewer searches do not necessarily indicate lower potential demand. In fact, based on our relative outflows analysis, those fewer searches would at least in part indicate that potential riders in outer regions forgo searches for taxicabs in anticipation of their likely failure.} 

Capturing economies of density in decentralized transportation markets (i.e., allowing the number of matches in a region to \textit{more than double} if demand and supply both double) will also have policy implications. This is because the policy maker may have an incentive to encourage supply in sparer areas in an effort to build economies of density there, whereas individual taxi drivers will not have the same incentive (they do not internalize the density-related externality they leave on others). This incentive mis-alignment in the taxi market will require remedies, which will likely take a similar form to the theoretical results in our rideshare model, and are worth further studying both theoretically and empirically.

\section{Relative Outflows Analysis in Austin}\label{appx: austin RO}

In this section, we carry out relative outflows analyses using our Austin data. More specifically, we do this at two levels. First, we calculate densities of rides and relative outflows at a geographically granular but time-wise less granular level. That is, our geographical units will be 1-mile by 1-mile tiles from a city grid published by the city of Austin,\footnote{See https://data.austintexas.gov/Locations-and-Maps/City-Grid/ty45-9ayu} while our time windows will be one month long each. We take regressions to examine the relationship between the density of incoming rides and the relative outflow at each tile during each month, both with and without city district fixed effects (see Figure \ref{fig: austin scatter} for Austin's city districts). Our second analysis makes the time window narrower (a day instead of a month) but uses the 10 districts of the city as the geographical units instead of 1 square mile tiles. We run this second analysis both with and without monthly fixed effects. 

More precisely, the regression we consider is as follows:

$$log(RO_{it})=\alpha_0+\alpha_1 \times log(D^\leftarrow_{it})+\epsilon_{it}$$

where $i$ is the region (tile or district) and $t$ is the time (month or day respectively). The above equation depicts the regression with no fixed effects. But as mentioned above, two of our four specifications have fixed effects. Table \ref{tab: Austin Relative Outflow Regression} presents the results.

\begin{table}[!htbp] \centering 
  \caption{\scriptsize{Relative outflow regression analysis results in Austin. Results, both at the tile-month level and at the district-day level, show that the coefficient of interest $\alpha_1$ is positive and significant.}} 
  \label{tab: Austin Relative Outflow Regression} 
\begin{tabular}{@{\extracolsep{5pt}}lcc|cc} 
\\[-1.8ex]\hline 
\hline \\[-1.8ex] 
 & \multicolumn{4}{c}{\textit{Dependent variable: $\log(RO)$}} \\ 
\cline{2-5} 
\\[-1.8ex] & \multicolumn{2}{c}{tile-month level} & \multicolumn{2}{c}{district-day level} \\ 
\\[-1.8ex] & (1) & (2) & (3) & (4)\\ 
\hline \\[-1.8ex] 
 inflow density & 0.034$^{***}$ & 0.027$^{***}$ & 0.052$^{***}$ & 0.028$^{***}$ \\
  & (0.006) & (0.007) &  &  \\

 Constant & $-$0.199$^{***}$ & - & $-$0.136$^{***}$ & - \\ 
  & (0.025) &  & (0.010) &  \\ 
  
  Fixed Effects$^{\dagger}$ & N & D & N & YM \\
  
\hline \\[-1.8ex] 
Observations & 3,388 & 3,388 & 3,084 & 3,084 \\ 
R$^{2}$ & 0.011 & 0.014 & 0.054 & 0.097 \\ 
Adjusted R$^{2}$ & 0.011 & 0.011 & 0.054 & 0.094 \\ 
\hline 
\hline \\[-1.8ex] 
\textit{Note:}  & \multicolumn{4}{r}{$^{*}$p$<$0.1; $^{**}$p$<$0.05; $^{***}$p$<$0.01} \\ 
\multicolumn{5}{l}{$\dagger$: D: District, YM: Year-month, N: none}\\
\end{tabular} 
\end{table} 

As can be seen from these results, the coefficient of interest is always positive and significant as expected. Note, again, that this positive association between $\log(RO)$ and $\log(D^\leftarrow)$ exists in spite of the fact that incoming rides appear with negative sign in the former and with positive sign in the latter.

\begin{figure}
\centering
\caption{\scriptsize{Panel (a) dropoff densities in different 1-mile by 1-mile squares across the city. Panel (b) does the same for relative outflows. Both quantities are larger in the middle and drop as we move towards outer areas. }}\label{fig: driver start date}
\subfigure[]{\includegraphics[width=0.48\textwidth]{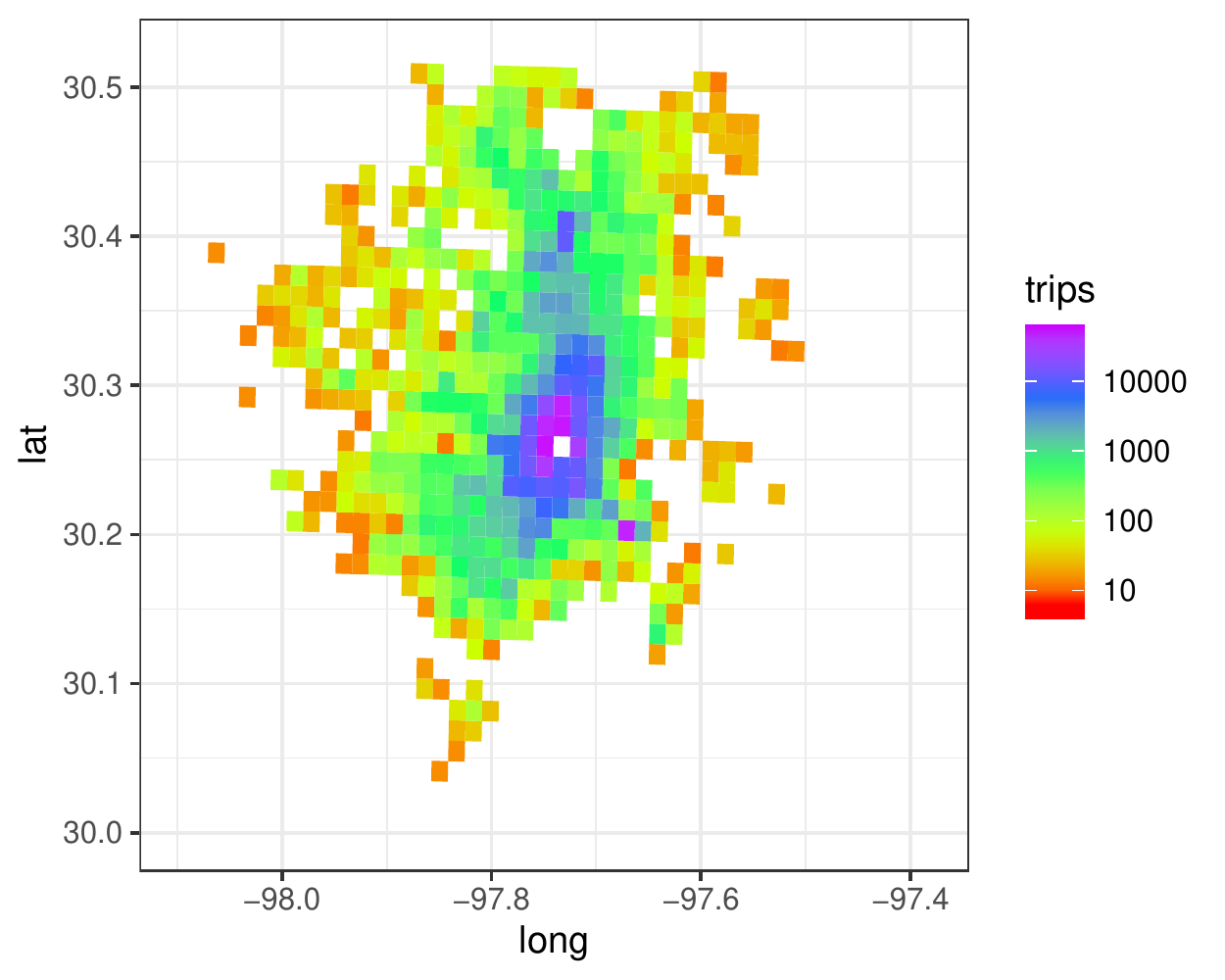}} 
\subfigure[]{\includegraphics[width=0.48\textwidth]{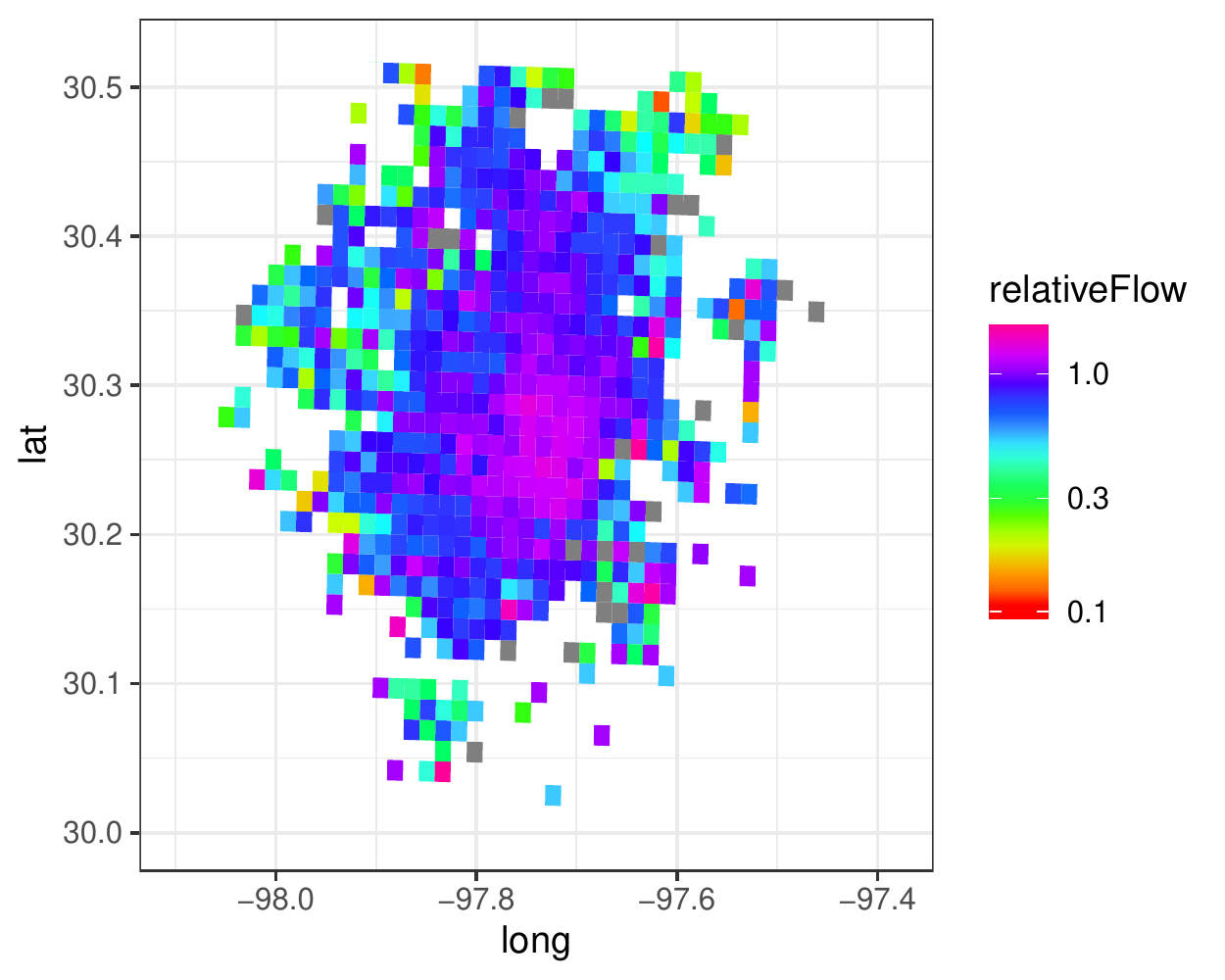}} 
\label{fig: austin RO}
\end{figure}

Finally, before closing this section, we show this positive association visually, by presenting heat maps of relative outflows and dropoff densities at the 1-mile by 1-mile tile level aggregated over the duration of our data. Figure \ref{fig: austin RO} shows these quantities. As can be seen there, central areas of the city have both higher dropoff rates and higher relative outflows.

\section{Further Analysis of Individual Driver Behavior in Austin}\label{appx: further Austin}

This section supplements the empirical analysis of individual driver app-turnoff behavior in Austin in the main text. Specifically, we provide the following analyses. First, we study what happens after app turnoffs and show two main things: (i) drivers are more likely to exit the market for the day upon an app turnoff if the turnoff is taking place in the outer regions of the city, and (ii) drivers are more likely to turn their apps off and relocate to busier regions of the city than the other way around.

Second, we track the movements of drivers between rides \textit{while their apps are on} and show that, when idle, drivers move toward the city center and search for rides.

\subsection{App turnoff, market exit, and location}
Here, we run the following logit regression:

\begin{equation}
    E_\eta=a_0+a_1\times C_\eta+\epsilon_eta
\end{equation}

where each $\eta$ is a ride at the end of which the driver turned off their app. Also, $E_\eta$ is whether the driver exited the supply for the day upon turrning off the app, and $C_\eta$ is a binary measure of whether the droppoff after which the app was turned off was centrally located in the city. It takes the value of 1 if ride $\eta$ was closer than the median ride to the geo locations of downtown Austin (-97.7444 in longitude and 30.2729 in lattitude) and 0 otherwise. We expect the $a_1$ coefficient to be negative and significant indicating that an app turnoff in the central areas of the city is more likely to be a simple break, whereas one in the outer regions is more likely to be an exit. Here are the regression results, confirming our hypothesis.

\begin{table}[!htbp] \centering 
  \caption{\scriptsize{The result shows that drivers are more likely to exit the market when they end up dropping off a passenger in outer regions.}} 
  \label{tab: app turnoff and exit} 
\begin{tabular}{@{\extracolsep{5pt}}lc} 
\\[-1.8ex]\hline 
\hline \\[-1.8ex] 
 & \multicolumn{1}{c}{\textit{Dependent variable:} driver exit} \\ 

\hline \\[-1.8ex] 
 $a_1$ & $-$0.044$^{***}$ \\ 
  & (0.001) \\ 
  
 $a_0$ & 0.332$^{***}$ \\ 
  & (0.001) \\ 

\hline \\[-1.8ex] 
Log Likelihood & $-$444,688.200 \\ 
Akaike Inf. Crit. & 889,380.400 \\ 
\hline 
\hline \\[-1.8ex] 
\textit{Note:}  & \multicolumn{1}{r}{$^{*}$p$<$0.1; $^{**}$p$<$0.05; $^{***}$p$<$0.01} \\ 
\end{tabular} 
\end{table} 

\subsection{Drivers' move across the city upon app turnoff}

Here, we examine drivers who turn off their apps and do not exit the market for the day. In particular, we study whether such drivers are more likely to move to busier regions of the city and restart operating or the other way around. To this end, we perform an analysis that is akin to our relative outflows study of rides, but this time we carry it out for inactive drivers. More specifically, we take different threshold distance values $\tau$ and for each $\tau$ we calculate the $\tau$-level driver-relative-outflow as follows:

\begin{enumerate}
    \item We divide the total number of times a driver turns their app off within the $\tau$-mile distance of Austin downtown and then restarts operating \textit{in the same day} outside of the $\tau$-mile radius of downtown. We call this the $\tau$-level outflow.
    \item We calculate the $\tau$-level inflow by doing the opposite of the above.
    \item We then calculate the relative outflow by dividing the outflow by the inflow.
\end{enumerate}

We expect this driver level relative outflow to be below 1 for different values of $\tau$. That is, we expect the number of drivers who turn their apps off in busy regions and restart operating in less busy ones to be smaller than those who do the opposite. We expect the wedge between these two flows to be bigger (hence, the relative outflow to be farther away from 1) for larger values of $\tau.$ Figure \ref{fig:driver RO} below confirms this expectation:

\begin{figure}[H]
    \centering
    \includegraphics{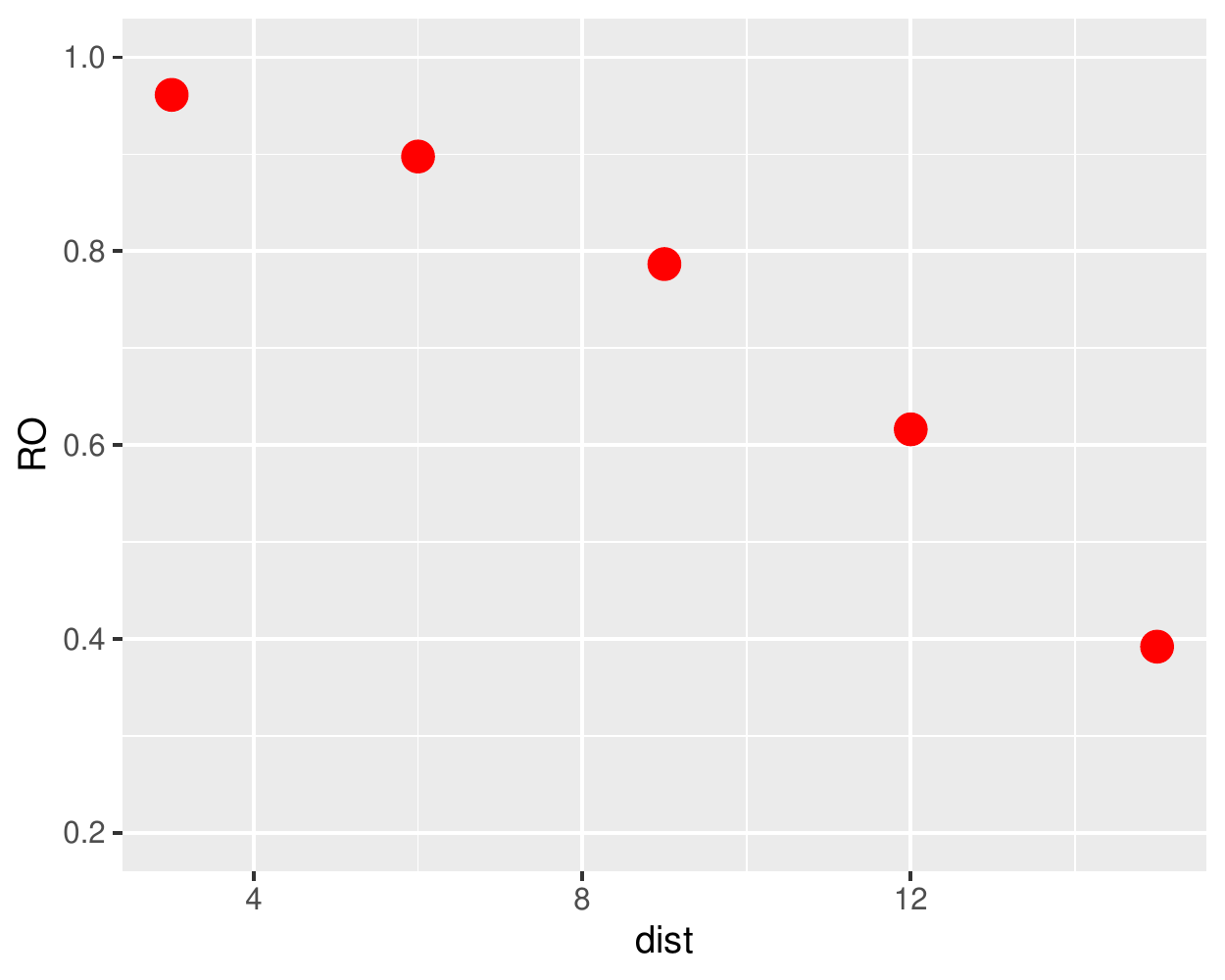}
    \caption{\scriptsize{Driver-level relative outflows analysis. The horizontal axis depicts $\tau$, in miles, as defined above and the vertical axis shows the $\tau$-level driver relative outflow. As expected, the results are always below 1.}}
    \label{fig:driver RO}
\end{figure}

\subsection{Driver idle movement while app is on}
In this section, we show evidence that drivers tend to move toward the city center while they are idle and searching for rides and their app is on. To this end, for each ride $\eta$ after whose completion the driver did \textit{not} turn their app off, we define the following: $\Delta^\leftarrow_\eta$ is defined to be the  how far the destination of $\eta$ was from Austin downtown, subtracted from how far the pickup location for the ride after $\eta$ was from downtown. A positive value for $\Delta^\leftarrow_\eta$ indicates that the driver moved closer to the city center while idle and searching for rides. We perform a simple $t$-test to examine whether the mean of $\Delta^\leftarrow_\eta$ is positive and statistically significant. The mean is given at 0.265 miles with the 95\% CI at $(0.26,0.27)$. This number has a considerable magnitude, given that the average ride in the city moves away from downtown by a distance slightly below 0.2 miles. (Note that the direction of the average ride should not be surprising given economies of density.)

\end{document}